\documentclass[preprint,review,10pt]{elsarticle}

\usepackage{amssymb}
\usepackage{amsmath,fullpage}
\usepackage{microtype}
\usepackage{graphicx}
\usepackage{color}
\usepackage{float}
\usepackage{subfigure}
\usepackage{booktabs} 

\usepackage{hyperref}

\usepackage{algorithmic}
\usepackage{multirow}
\usepackage{hyperref}
\usepackage[ruled,vlined]{algorithm2e}

\allowdisplaybreaks

\usepackage{amsmath}
\usepackage{amssymb}
\usepackage{mathtools}
\usepackage{amsthm}

\newcommand{\bx}{\mathbf{x}}

\theoremstyle{plain}
\newtheorem{theorem}{Theorem}[section]

\newtheorem{lemma}[theorem]{Lemma}

\theoremstyle{definition}

\theoremstyle{remark}
\newtheorem{remark}[theorem]{Remark}

\numberwithin{equation}{section}

\begin{document}

\begin{frontmatter}
\title{MAP-based Problem-Agnostic Diffusion Model for Inverse Problems}
\author[label1]{Pingping Tao\fnref{fntext1}}
\fntext[fntext1]{These authors contributed to the work equally and should be regarded as co-first authors.}
\affiliation[label1]{organization={Library, Shandong University},
	addressline={No. 180 West Culture Road},
	city={Weihai},
	postcode={264209},
	state={Shandong},
	country={China}}
\ead{pingping.tao@sdu.edu.cn}
\author[label2]{Haixia Liu\fnref{fntext1}\corref{cortext}}
\cortext[cortext]{Corresponding author.}
\affiliation[label2]{organization={School of Mathematics and Statistics \& Hubei Key Laboratory of Engineering Modeling and Scientific Computing \& Institute of Interdisciplinary Research for Mathematics and Applied Science, Huazhong University of Science and Technology},
	addressline={No. 1037 Luoyu road},
	city={Wuhan},
	postcode={430074},
	state={Hubei},
	country={China}}
\ead{liuhaixia@hust.edu.cn}
\author[label3]{Jing Su}
\affiliation[label3]{organization={Department of Basic Sciences, Dalian University of Science and Technology},
	addressline={No. 999-26 Harbor Road},
	city={Dalian},
	postcode={116052},
	state={Liaoning},
	country={China}}
 \ead{xujing@dlust.edu.cn}
\begin{abstract}
Diffusion models have indeed shown great promise in
solving inverse problems in image processing. In this paper, we propose a novel, problem-agnostic diffusion model called the maximum a posteriori (MAP)-based guided term estimation
method for inverse problems. To leverage unconditionally pretrained diffusion models to address conditional generation tasks, we divide the conditional score function into two terms according to Bayes’ rule: an unconditional score function (approximated by a pretrained score network) and a guided term, which is estimated using a novel MAP-based method that incorporates a Gaussian-type prior of natural images. This innovation allows us to better capture the intrinsic properties of the data, leading to improved performance. 
Numerical results demonstrate that our method preserves contents more effectively compared to state-of-the-art methods—for example, maintaining the structure of glasses in super-resolution tasks and producing more coherent results in the neighborhood of masked regions during inpainting. Our numerical implementation is available at  \url{https://github.com/liuhaixias1/MAP-DIFFUSION-IP}.
\end{abstract}
\end{frontmatter}

\section{Introduction}
\label{sec:Introduction}
\begin{figure}[htp]
	\centering	\includegraphics[width=1.0\textwidth]{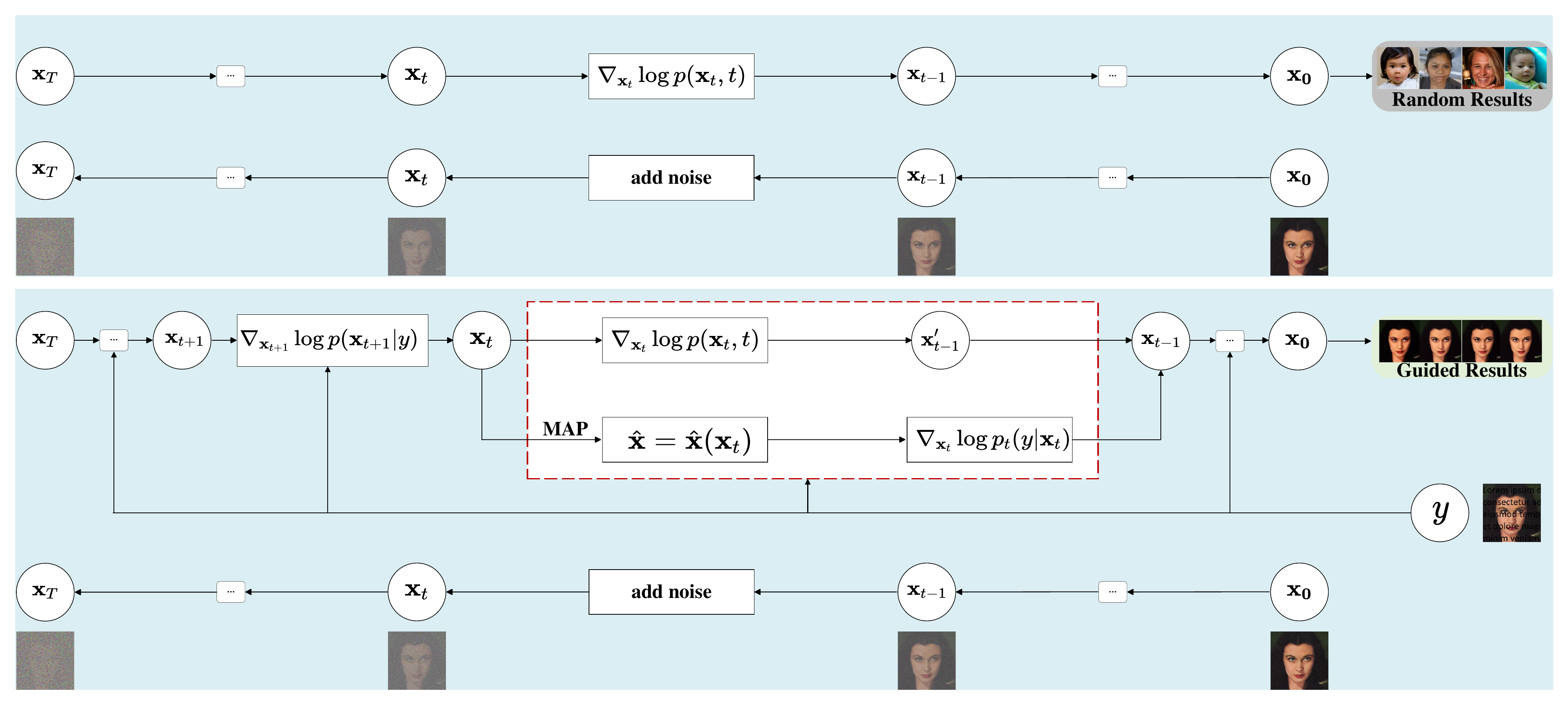}\vspace{-4mm}
    \caption{Illustrations of unconditional diffusion model (Top), and conditional diffusion model (Bottom), respectively. }
    \label{fig:Figure1}
\end{figure}
Diffusion models have demonstrated their power as both generative models and unsupervised inverse problem solvers, as evidenced by recent research \cite{choi2021ilvr,song2020score,song2021solving,chung2022improving}. They learn the true data distribution $p(x)$ of clean, high-quality images (or signals) from a large dataset. This learned distribution acts as an extremely strong, implicit prior. Instead of using simple, hand-crafted priors (like sparsity or smoothness) in inverse problems, the diffusion prior is rich and captures complex, natural structures (edges, textures, anatomical features). The solution is guided to lie on the manifold of realistic data. Owing to their capacity to effectively model complex data distributions and their ability to be trained without relying on specific problem formulations, these models hold great promise and versatility for future research and practical applications across a wide range of domains. 
Compared to Generative Adversarial Networks (GANs), another popular class of generative models, diffusion models are less prone to mode-collapse and training instabilities.
Additionally, diffusion models are more interpretable and provide a natural trade-off between sample quality and diversity.

The diffusion model comprises two diffusion processes. The first is the forward diffusion, also known as the noising process, which drives any data distribution to a tractable distribution by adding noise to the data. The second is the backward diffusion, also known as the denoising process, which sequentially removes noise from noisy data to generate realistic samples.  Diffusion models can be categorized into unconditional and conditional diffusion models based on the absence or presence of conditions. Both unconditional and conditional diffusion models share the same forward process, but their backward processes differ. For completeness, we illustrate  unconditional diffusion model (Top) and conditional diffusion model (Bottom) in Figure \ref{fig:Figure1}, respectively. For a comprehensive overview of diffusion models, we refer readers to the following survey papers  \cite{yang2023diffusion,croitoru2023diffusion,rombach2022high,luo2022understanding,ma2025efficient,fuest2024diffusion,chen2024overview,cao2024survey,kazerouni2022diffusion,chen2025comprehensive,li2023diffusion,moser2024diffusion,shen2025efficient}.

The core of diffusion models is the score function, which represents the gradient of the log-density of the $t$-th latent image distribution. When solving inverse problems, a conditional score function $\nabla_{\mathbf{x}_t}\log  p(\mathbf{x}_t|y)$ is applied, where $y$ is the given input and $\mathbf{x}_t$ is the $t$-th latent image.

There are two main approaches to solving inverse problems using diffusion models. The first is to train a problem-specific, conditional diffusion model for a particular inverse problem \cite{saharia2022image, saharia2022palette, whang2022deblurring}. 
The second approach involves problem-agnostic diffusion models, which leverage unconditionally pretrained diffusion models to tackle conditional generation tasks. That is, the conditional score function is derived from an unconditional score function combined with a given measurement. This plug-and-play technique enables the diffusion model to be applied to a wide range of inverse problems without requiring problem-specific training.

Several existing works are based on problem-agnostic diffusion models, including Denoising Diffusion Restoration Models (DDRM) \cite{kawar2022denoising}, Diffusion Posterior Sampling (DPS) \cite{chung2022diffusion}, Pseudoinverse-Guided Diffusion Models ($\Pi$GDM) \cite{song2022pseudoinverse}, Diffusion Model-Based Posterior Sampling (DMPS) \cite{meng2022diffusion}, Manifold Constrained Gradient (MCG) \cite{chung2022improving}, and others.  For a detailed comparison, we refer readers to Section \ref{sec:rw}. However, these methods primarily rely on probabilistic properties rather than leveraging the inherent structural characteristics of images. To improve the performance of problem-agnostic diffusion models, this work proposes a novel maximum a posteriori (MAP)-based guided term estimation method for inverse problems. Our approach is grounded in the assumption that the space of clean natural images is inherently smooth. We introduce a MAP estimate of the true image conditioned on the $t$-th latent image and substitute this estimation into the expression of the inverse problem, which  allows us to derive an approximation of the guided term.

Our key contributions are as follows:

\textbf{Our proposed method is a training-free diffusion model for solving inverse problems.} The approach leverages unconditionally pretrained diffusion models to address conditional generation tasks. By applying Bayes' rule, we decompose the problem-specific score function into two components: an unconditional score function (approximated by a pretrained score network) and a guided term, which is estimated using a novel MAP-based method that incorporates a Gaussian-type prior of natural images. 

\textbf{We propose a novel MAP-based method to estimate the guided term.} Most existing methods to estimate the true image are based on probabilistic properties.  Our approach builds on the
assumption that the space of clean natural images is inherently smooth, which is then used to compute the guided term by combining the given measurement with an explicit measurement model. This innovation enables us to better capture the intrinsic properties of the data, resulting in significantly improved performance.

\textbf{The plug-and-play nature of our approach enables its application to a wide range of inverse problems without requiring problem-specific training.} Our approach alternates between unconditional generation and the adjustment of the generated results (guided term). As a result,  only the model operator used in the guided term needs to be changed for different inverse problems.

\textbf{We extensively evaluate our method on several inverse problems, including super-resolution,  denoising, and inpainting.} The results demonstrate that our approach achieves performance comparable to state-of-the-art methods, such as DDRM, DPS, $\Pi$GDM, DMPS and MCG. Notably, our method preserves contents more effectively—for example, maintaining the structure of glasses in super-resolution tasks and producing more coherent results in the neighborhood of masked regions during inpainting.

The remainder of this paper is structured as follows: Section \ref{sec:rw} reviews related works on conditional diffusion models. Section \ref{sec:overview} overviews the score-based diffusion models, which is followed by  our proposed method, called MAP-based problem-agnostic diffusion model for inverse problems, in Section \ref{sec:map}. Numerical implementations are discussed in Section \ref{sec:numerical}. Finally, we conclude with our findings and highlight the limitations of this work in Section \ref{sec:conclusion}.
\section{Related Works}
\label{sec:rw}
Diffusion models have indeed shown great promise in solving inverse problems in image processing, including blind inverse problems \cite{kadkhodaie2021stochastic,chung2023parallel}, compressed sensing \cite{jalal2021robust,bora2017compressed} and plug-and-play priors for inverse problems \cite{graikos2022diffusion}. The goal of image inverse problems is to recover the original high-quality image from observed, often degraded or incomplete, measurement data. 

 Most methods for solving inverse problems with diffusion models can be divided into two main categories. The first involves training a problem-specific conditional diffusion model directly. While effective, this approach is limited to specific inverse problems and lacks generalizability. For instance, prior work has demonstrated success in super-resolution \cite{saharia2022image}, image-to-image translation \cite{saharia2022palette}, and deblurring  \cite{whang2022deblurring}. The second category leverages unconditional diffusion models in a plug-and-play manner, enabling their application to various tasks without retraining. In the following, we will focus on the details of the second category.

\textbf{Filtering-Based Methods:}
The Iterative Latent Variable Refinement (ILVR) method, proposed by Choi et al. \cite{choi2021ilvr}, guided the generative process in Denoising Diffusion Probabilistic Models (DDPM) \cite{ho2020denoising} using a reference image. However, its iterative nature could lead to error accumulation, causing the solution path to deviate from the original data manifold. Filtering Posterior Sampling (FPS) was introduced by Dou and Song \cite{dou2024diffusion}, who established a connection between Bayesian posterior sampling and Bayesian filtering in diffusion models by assuming the backward process followed a Markov chain. This method leveraged sequential Monte Carlo techniques to address filtering problems.
Frequency-Guided Posterior Sampling (FGPS), established by Thaker et al. \cite{thaker2025frequency}, leveraged the frequency structure of the reverse diffusion process. This method executed a time-varying low-pass filter in the frequency domain of the input data, progressively incorporating high-frequency information in a data-dependent manner to unify diverse frequency components throughout the restoration process.

\textbf{Methods Inspired by DDIM (Denoising Diffusion Implicit Models) \cite{song2020denoising}:} 
Denoising Diffusion Restoration Models (DDRM) was proposed by Kawar et al. \cite{kawar2022denoising} and constructed a non-Markovian process to enable flexible skip-step sampling, similar to DDIM, while maintaining conditioning for solving inverse problems. Unfortunately, the variational distribution $q(\bar{\bx}^{(i)}_t|\bx_{t+1},\bx_0,y)$ did not guided by the measurement $y$ if the $i$-th singular value of linear model operator was zero, where $\bar{\bx}^{(i)}_t$ was the $i$-th entry of $\bar{\bx}_t$.

\textbf{Methods Based on Tweedie's Formula:}
The Manifold Constrained Gradient (MCG) method, proposed by Chung et al. \cite{chung2022improving}, introduced correction terms inspired by manifold constraints to ensure the iterative process remained close to the original data manifold. Building on this, He et al. \cite{he2024manifold} proposed Manifold Preserving Guided Diffusion (MPGD), based on the hypothesis that \( p(\mathbf{x}_t) \) was concentrated on a \((d-1)\)-dimensional linear subspace manifold. This method formulated an optimization problem involving tangent spaces and established relationships between \(\mathbf{x}_{t-1}\) and \(\mathbf{x}_0, \mathbf{x}_t\). Both MCG and MPGD relied on the linear manifold assumption, which might be too restrictive for cases involving complex data. Chung et al. \cite{chung2022diffusion} also developed Diffusion Posterior Sampling (DPS), which estimated \(\hat{\mathbf{x}}_0(\mathbf{x}_t) := \mathbb{E}(\mathbf{x}_0|\mathbf{x}_t)\) using Tweedie's formula. 
In a different approach, Song et al. \cite{song2022pseudoinverse} introduced Pseudo-Inverse Guided Diffusion (\(\Pi\)GDM), assuming \( p_t(\mathbf{x}_0|\mathbf{x}_t) \) was approximately normal. This method used pseudo-inverse guidance to reverse the measurement model, improving approximation accuracy. In DPS and $\Pi$GDM, the gradient of the neural network \(\mathbf{S}_\theta(\mathbf{x}_t, t)\) was computed via backpropagation, which was slightly more computationally expensive per Neural Function Evaluation (NFE) compared to other methods. Finally, Meng and Kabashima \cite{meng2022diffusion} proposed Diffusion Model Posterior Sampling (DMPS), which operated under the assumption that \(p(\mathbf{x}_0|\mathbf{x}_t) \propto p(\mathbf{x}_t|\mathbf{x}_0)\). This method employed reconstruction guidance to effectively measure diffusion model guidance, with the assumption that \(p(\mathbf{x}_0)/p(\mathbf{x}_t)\) remained constant. The Diffusion Priors by Expectation Maximization (DP-EM) method, originated by Rozet et al. \cite{rozet2024learning}, combined expectation-maximization principles with diffusion priors. Its core mechanism involved moment matching through explicit variance matrix computation, utilizing Jacobian matrix-vector products. 
Yang et al. \cite{yang2024guidance} established Diffusion with Spherical Gaussian constraint (DSG), which retained guidance steps within the intermediate data manifold via optimization, thus facilitating the adoption of larger guidance steps.
Peng et al. \cite{peng2024improving} proposed Diffusion Models using Optimal Posterior Covariance (DM-OPC), which aimed to optimize the diffusion model's optimal diagonal posterior covariance matrix. The optimization was performed via maximum likelihood estimation within the DDPM framework.
Kim et al. \cite{kim2025flowdps} proposed Flow-Driven Posterior Sampling (FlowDPS), which extended the posterior sampling  from diffusion models to general affine flows by decomposing a single Euler step into a linear combination of pure noise estimates and employing the generalized Tweedie formula.
\section{Overview of Score-based Models}\label{sec:overview}
The forward SDE process of diffusion model can be formalized as the It\^o stochastic differential equation \cite{song2020score}
\begin{equation}\label{eq:forward}
    d\mathbf{x}=f(\mathbf{x},t)dt+g(\mathbf{x},t)d\omega,
\end{equation}
where $\omega$ is the standard Wiener process. When $f(\mathbf{x}_t,t)=-\frac{1}{2}\beta_t\mathbf{x}_t$ and $g(\mathbf{x}_t,t)=\sqrt{\beta_t}$, it corresponds to the Variance-Preserving Model (VP-SDE) \cite{song2020score}, where $\beta_t$ is a non-negative continuous function about $t$. Let $p_0(\mathbf{x}_0)$ be the data distribution and $p_t(\mathbf{x}_t)$ be the distribution obtained by adding Gaussian noise to $p_0(\mathbf{x}_0)$ with $p(\mathbf{x}_t,t|\mathbf{x}_0,0)=\mathcal{N}(\mathbf{x}_t|\sqrt{1-\zeta_t}\mathbf{x}_0,\zeta_t I)$, where $\zeta_t=1-e^{-\int^t_0\beta_tdt}$. 

 By the Anderson's theorem \cite{anderson1982reverse,song2020score}, the reverse SDE corresponding to the forward SDE process \eqref{eq:forward} is 
\begin{equation}\label{eq:backward}
    d\bx=\left\{
    \begin{array}{ll}
      [f(\bx,t)-g(\bx,t)\nabla_{\bx}\log p_t(\bx)]dt+g(\bx,t)d\omega,   & \hbox{unconditioned}, \\
      \left[f(\bx,t)-g(\bx,t)\nabla_{\bx}\log p_t(\bx| y)\right]dt+g(\bx,t)d\omega,   & \hbox{conditioned}.
    \end{array}
    \right.
\end{equation}
Note that the reverse SDE defines the generative procedure through the score function $\nabla_{\bx}\log p_t(\bx)$ or $\nabla_{\bx}\log p_t(\bx|y)$. Once the score is available, solutions can be obtained by solving the reverse SDE.

In diffusion-based generative models, one estimates the score function $\nabla_{\mathbf{x}_t}\log p(\mathbf{x}_t,t)$ by a neural network $\mathbf{S}_\theta(\mathbf{x}_t,t)$. We have the following representation
\[\frac{-\mathbf{S}_\theta(\mathbf{x}_t,t)}{\sqrt{\zeta_t}}\overset{model}{\approx}\nabla_{\mathbf{x}_t}\log p(\mathbf{x}_t,t)\underset{t\ne0}{\overset{almost\ equal}{\approx}}\nabla_{\mathbf{x}_t}\log p(\mathbf{x}_t,t|\mathbf{x}_0,0)=-\frac{\mathbf{x}_t-\sqrt{1-\zeta_t}\mathbf{x}_0}{\zeta_t},\]
where the almost equal equality is from \cite{tachibana2021quasi}. As score-based generative models, the score function of diffusion models is approximated with a neural network $\mathbf{S}_\theta(\mathbf{x}_t,t)$, trained with the denoising score matching objective \cite{vincent2011connection}. Throughout of the paper, we use the VP-SDE, which is equivalent to DDPM \cite{ho2020denoising}.
\section{MAP-based Guided Term Estimation}
\label{sec:map}
Suppose we have measurements $y \in R^m$ of some image $\mathbf{x_0} \in R^n$,   the linear inverse problem can be expressed as follows:
\begin{eqnarray}\label{eq:pro}
	y =  H\mathbf{x_0}+z,
\end{eqnarray}
where $H \in R^{m\times n}$ is a known measurement matrix, and $z \sim  \mathcal N(0, \sigma_y^2I)$ is a Gaussian noise with a mean 0 and a standard deviation $\sigma_y$.
We aim to solve the inverse problem and recover $\mathbf{x_0} \in R^n$ from the measurements $y$.

In the following, we focus on a conditional diffusion model for the inverse problem described above, where solutions can be obtained by reverse SDE or ODE if the problem-specific scores $\{\nabla_{\mathbf{x}_t} \log p(\mathbf{x}_t|y)\}_{t=T,\cdots,1}$ are available. 
To compute the problem-specific scores, one approach is to train a diffusion model specifically for the problem using paired samples $(\mathbf{x}_0,y)$. However, this method requires retraining the model for each new problem, which can be computationally expensive. An alternative approach is to decompose the problem-specific score into two terms. By applying Bayes' rule, we can express as 
$p(\mathbf{x}_t|y)= p(\mathbf{x}_t)\cdot p(y|\mathbf{x}_t)/p(y)$. Consequently, the problem-specific score $\nabla_{\mathbf{x}_t} \log p(\mathbf{x}_t|y)$ can be broken down into:
\begin{eqnarray}\label{eq:bay_equation}
	\nabla_{\mathbf{x}_t} \log p(\mathbf{x}_t|y) =\nabla_{\mathbf{x}_t} \log p(\mathbf{x}_t) + \nabla_{\mathbf{x}_t} \log p(y|\mathbf{x}_t).
\end{eqnarray}
The first term, $\nabla_{\mathbf{x}_t} \log p(\mathbf{x}_t)$, can be approximated using a pretrained score network $\mathbf{S}_{\theta}(\mathbf{x}_t, t)$, which was trained via the denoising score matching objective \cite{vincent2011connection}. As a result, the main challenge in estimating the problem-specific score reduces to computing the second term, $\nabla_{\mathbf{x}_t} \log p(y|\mathbf{x}_t)$, which is referred to as the {\it guided term}.

In the diffusion model, we need to estimate the problem-specific score function $\nabla_{\mathbf{x}_t}\log p(\mathbf{x}_t|y)$ for each $t=T,\cdots,1$. From Equation \eqref{eq:bay_equation}, our task in this paper is to estimate the guided term $\nabla_{\mathbf{x}_t}\log p(y|\mathbf{x}_t)$ for each $t$.
To begin, we start by estimating $\mathbf{x}_0$ and represent $\mathbf{x}_0$ as a function of $\mathbf{x}_t$ in Equation \eqref{eq:solution}. 
Then, we substitute the estimated $\mathbf{x}_0$ into Equation \eqref{eq:pro} and represent $y$ as a function of $\mathbf{x}_t$. Furthermore, in Subsection \ref{subsec:estimate_guide}, we obtain the conditional distribution of $p(y|\mathbf{x}_t)$.
\subsection{Estimation of True Image}\label{subsec:estimate_image}
In this subsection, we focus on estimating the true image conditioned on a $t$-th latent image $\mathbf{x}_t$. Building on the assumption that the space of clean natural images is inherently smooth, we introduce a maximum a posteriori (MAP) estimate given $\mathbf{x}_t$. 

Let $\widetilde{\mathbf{x}}$ represent a potential image in the natural image space and $\mathbf{x}$ denote any arbitrary image. To quantify the differences between $\widetilde{\mathbf{x}}$ and $\mathbf{x}$, we employ the following utility function \cite{arjomand2017deep}:
\begin{eqnarray*}
	G(\widetilde {\mathbf{x}},\mathbf{x})= g_{\sigma}(\widetilde {\mathbf{x}}-\mathbf{x}) p(\mathbf{x})/p(\widetilde {\mathbf{x}}),
\end{eqnarray*}
where $g_{\sigma}$ is the Gaussian function with a mean of $0$ and a standard deviation of $\sigma$. 

 In this paper, our goal is to estimate the true solution by considering all possible candidates in the natural image space conditioned on the $t$-th latent image $\mathbf{x}_t$. 
We take all possible $\widetilde{\mathbf{x}}$ conditioned on $\mathbf{x}_t$, the conditional expectation is as follows:
\begin{equation}\label{eq:bay}
	\begin{split}
		E_{\widetilde {\mathbf{x}}|\mathbf{x}_t} [G(\widetilde {\mathbf{x}},\mathbf{x})]=&\int G(\widetilde {\mathbf{x}},\mathbf{x})p(\widetilde {\mathbf{x}}|\mathbf{x}_t )d\widetilde {\mathbf{x}}
		=\frac{1}{p(\mathbf{x}_t)}\int G(\widetilde {\mathbf{x}},\mathbf{x})p(\mathbf{x}_t | \widetilde {\mathbf{x}})p(\widetilde {\mathbf{x}})d\widetilde {\mathbf{x}},
	\end{split}
\end{equation}
where the second equality is from Bayes's formula, $\widetilde {\mathbf{x}}$ belongs to the natural image space, $\mathbf{x}$ is an arbitrary image, and $\mathbf{x}_t$ is a noisy image generated by diffusion model. 
Then we consider the following optimization problem:
 \[\max_{\mathbf{x}}E_{\widetilde {\mathbf{x}}|\mathbf{x}_t} [G(\widetilde {\mathbf{x}},\mathbf{x})].\]
Unfortunately, directly optimizing this objective is computationally challenging. To address this, we employ the Minorization-Maximization (MM) algorithm. This involves estimating a lower bound of $E_{\widetilde {\mathbf{x}}|\mathbf{x}_t} [G(\widetilde {\mathbf{x}},\mathbf{x})]$ and then maximizing that lower bound. 
By inserting our utility function into Equation \eqref{eq:bay},we obtain
\begin{equation}\label{eq: exp_utility}
	\begin{split}
		E_{\widetilde {\mathbf{x}}|\mathbf{x}_t} [G(\widetilde {\mathbf{x}},\mathbf{x})]
		=\frac{1}{p(\mathbf{x}_t)}\int g_{\sigma}(\widetilde {\mathbf{x}}-\mathbf{x})p(\mathbf{x}_t | \widetilde {\mathbf{x}})p(\mathbf{x})d\widetilde {\mathbf{x}} 
		=\frac{1}{p(\mathbf{x}_t)}\int g_{\sigma}(r)p(\mathbf{x}_t | \mathbf{x}+r)p(\mathbf{x})dr,
	\end{split}
\end{equation}
where we introduce a substitution $r=\widetilde {\mathbf{x}}-\mathbf{x}$.

To establish the lower bound, we leverage Jensen's inequality, taking advantage of the concavity property of the logarithmic function. We have:
\begin{align}\label{eq:log_utility}
	\log E_{\widetilde {\mathbf{x}}|\mathbf{x}_t} [G(\widetilde {\mathbf{x}},\mathbf{x})]
	=& \log \int g_{\sigma}(r)p(\mathbf{x}_t | \mathbf{x}+r)p(\mathbf{x})dr-\log p(\mathbf{x}_t) \notag\\
	\ge& \int g_{\sigma}(r)\log[p(\mathbf{x}_t | \mathbf{x}+r)p(\mathbf{x})]dr-\log p(\mathbf{x}_t) \notag\\
	=& \int g_{\sigma}(r)\log p(\mathbf{x}_t | \mathbf{x}+r)dr + \log p(\mathbf{x})-\log p(\mathbf{x}_t),
\end{align}

The advantage of this lower bound is that the objective can be converted to two simple forms. Next, we consider the log-likelihood
\[\log p(\mathbf{x}_t | \mathbf{x}+r)=-\frac{\|\mathbf{x}_t-\sqrt{1-\zeta_t}(\mathbf{x}+r)\|^2}{2\zeta_t}+const1,\]
which implies
\begin{eqnarray}\label{eq:data_term}
	\int g_{\sigma}(r)\log p(\mathbf{x}_t | \mathbf{x}+r)dr&= &
    \int \frac{1}{Z}\exp{\left(-\frac{\|r\|^2}{2\sigma^2}\right)}\frac{-\|\mathbf{x}_t-\sqrt{1-\zeta_t}(\mathbf{x}+r)\|^2}{2\zeta_t}dr+const1 
     \notag\\&= &
    \int \frac{1}{Z}\exp{\left(-\frac{\|r\|^2}{2\sigma^2}\right)}\frac{-\|\mathbf{x}_t-\sqrt{1-\zeta_t}\mathbf{x}\|^2}{2\zeta_t}dr+const 
     \notag\\
    &= &\frac{-\|\mathbf{x}_t-\sqrt{1-\zeta_t}\mathbf{x}\|^2}{2\zeta_t}+const,
\end{eqnarray}
where $Z$ is a constant such that $\int \frac{1}{Z}\exp{\left(-\frac{\|r\|^2}{2\sigma^2}\right)}dr=1$ and $r\sim\mathcal{N}(0,\sigma^2I)$.  
Combining \eqref{eq:data_term}, we take the derivative of the right-hand side of Equation \eqref{eq:log_utility} about $\mathbf{x}$ and obtain
\begin{eqnarray}\label{eq:deviration}
	\frac{-\sqrt{1-\zeta_t}(\sqrt{1-\zeta_t}\mathbf{x}-\mathbf{x}_t)}{\zeta_t}+\nabla_{\mathbf{x}} \log p(\mathbf{x})=0.
\end{eqnarray}
We have $\mathbf{x} = \frac{\mathbf{x}_t}{\sqrt{1-\zeta_t}}+\frac{\zeta_t}{1-\zeta_t}\nabla_{\mathbf{x}} \log p(\mathbf{x})$ from \eqref{eq:deviration}. 
The following lemma provides an estimate of $\mathbf{x}_0$ based on $\mathbf{x}_t$ and the neural network $\mathbf{S}_\theta(\mathbf{x}_t,t)$.
\begin{lemma}\label{theo:estimation_x}
	Let $\mathbf{x}_t$ be the $t$-th latent image in the backward process of diffusion model and the neural network $\mathbf{S}_{\theta}(\mathbf{x}_t, t)$ be an approximation of score function $\nabla_{\mathbf{x}_t}\log p(\mathbf{x}_t)$. Define $\bar{\alpha}_t=1-\zeta_t$, then the estimation of $\mathbf{x}_0$ can be represented as
	\begin{equation}\label{eq:solution}
		\begin{split}
            \hat{\mathbf{x}}  = \frac {\left(\sqrt{\overline{\alpha _{t}}}+\frac{q_1 t\beta _t}{2}+q_2\right)\mathbf{x}_t-\left(\sqrt{1-\overline{\alpha _{t}}}+\frac{q_1 t\beta _t}{2\sqrt{1-\overline{\alpha _{t}}}}\right)\mathbf{S}_{\theta}(\mathbf{x}_{t},t)}{(\overline{\alpha _{t}} +q_2)},
		\end{split}
	\end{equation}
	where  $q_1$ and  $q_2$ satisfy $\mathbf{S}_{\theta}(\hat{\mathbf{x}},0)=\mathbf{S}_{\theta}(\mathbf{x}_{t},t)+q_1(0-t)\partial_{t}\mathbf{S}_{\theta}(\mathbf{x}_{t},t)+q_2\nabla_{\mathbf{x}_{t}}\mathbf{S}_{\theta}(\mathbf{x}_{t},t)^\top(\mathbf{\hat x}-\mathbf{x}_{t})$.
    
\end{lemma}
\begin{proof}
Let's compute the spatial derivative $\nabla_{\mathbf{x}_{t}}
	\mathbf{S}_{\theta}(\mathbf{x}_{t},t)$ and time derivative $\partial_{t}\mathbf{S}_{\theta}(\mathbf{x}_{t},t)$ first. 
Note that 
\[\frac{-\mathbf{S}_\theta(\mathbf{x}_t,t)}{\sqrt{\zeta_t}}\overset{model}{\approx}\nabla_{\mathbf{x}_t}\log p(\mathbf{x}_t,t)\underset{t\ne 0}{\overset{almost\ equal}{\approx}}\nabla_{\mathbf{x}_t}\log p(\mathbf{x}_t,t|\mathbf{x}_0,0)=-\frac{\mathbf{x}_t-\sqrt{1-\zeta_t}\mathbf{x}_0}{\zeta_t}.\]
Therefore,
\begin{eqnarray}\label{eq:express_s_xt}
	\nabla_{\mathbf{x}_{t}} \mathbf{S}_{\theta}(\mathbf{x}_{t},t) = \frac{1}{\sqrt{\zeta_{t}}} I.
\end{eqnarray}
Next, let us compute $\partial_{t}\mathbf{S}_{\theta}(\mathbf{x}_{t},t)$. Note that $\mathbb E[\mathbf{x}_0|\mathbf{x}_t]=\frac{1}{\sqrt{1-\zeta_t}}\left(\mathbf{x}_t-\sqrt{\zeta_t}\mathbf{S}_\theta(\mathbf{x}_t,t)\right)$, then 
\begin{eqnarray}\label{eq:express_s_theta}
    \mathbf{S}_{\theta}(\mathbf{x}_{t},t) =\frac{\mathbf{x}_{t}-\sqrt{1-\zeta_{t}}\mathbb E[\mathbf{x}_0|\mathbf{x}_t]}{\sqrt{\zeta_{t}}}.
\end{eqnarray}
Since $\zeta_t=1-e^{-\int^t_0\beta_tdt}$, We have the derivative of $\zeta_t$ about $t$ is 
\begin{eqnarray}\label{eq:dot_zeta}
	\dot{\zeta}_t=(1-\zeta_t)\beta(t).
\end{eqnarray}
Note that we consider $x_t,t$ as two variables,  $\mathbb{E}[x_0|x_t]$ is only related to $x_t$, but not $t$, we have $\partial_t \mathbb{E}[x_0|x_t]=0$. Next, we write $\mathbf{S}_{\theta}(\hat{\mathbf{x}},0)$ as
\begin{equation}\label{eq:equality}
    \mathbf{S}_{\theta}(\hat{\mathbf{x}},0)=\mathbf{S}_{\theta}(\mathbf{x}_{t},t)+q_1(0-t)\partial_{t}\mathbf{S}_{\theta}(\mathbf{x}_{t},t)+q_2\nabla_{\mathbf{x}_{t}}\mathbf{S}_{\theta}(\mathbf{x}_{t},t)^\top(\mathbf{\hat x}-\mathbf{x}_{t}),
\end{equation}
where $q_1,q_2$ may be related to $\hat{\mathbf{x}}$, $\mathbf{x}_t$ and $t$. 
Then we get from \eqref{eq:express_s_theta}, \eqref{eq:dot_zeta} and \eqref{eq:equality}
\begin{align}\label{eq:express_s_t}
	&\partial_{t}\mathbf{S}_{\theta}(\mathbf{x}_{t},t) =\partial_{t}\frac{\mathbf{x}_{t}-\sqrt{1-\zeta_{t}}\mathbb E[\mathbf{x}_0|\mathbf{x}_t]}{\sqrt{\zeta_{t}}}  \notag\\
	\approx&\frac{1}{\sqrt{\zeta_{t}}}\left(\frac{1}{2}\dot{\zeta}_{t}(1-\zeta_{t})^{-1/2}\mathbb E[\mathbf{x}_0|\mathbf{x}_t]\right)+(\mathbf{x}_{t}-\sqrt{1-\zeta_{t}}\mathbb E[\mathbf{x}_0|\mathbf{x}_t])\left(-\frac{1}{2}\dot{\zeta}_{t}\zeta_{t}^{-3/2}\right) \notag\\
	=&\frac{\dot{\zeta}_{t}}{2\zeta_{t}^{3/2}}\left(\frac{\zeta_{t}}{\sqrt{1-\zeta_{t}}\mathbb E[\mathbf{x}_0|\mathbf{x}_t]}-(\mathbf{x}_{t}-\sqrt{1-\zeta_{t}}\mathbb E[\mathbf{x}_0|\mathbf{x}_t])\right)
	=\frac{\dot{\zeta}_{t}}{2\zeta_{t}^{3/2}}\left(-\mathbf{x}_{t}+\frac{1}{\sqrt{1-\zeta_{t}}}\mathbb E[\mathbf{x}_0|\mathbf{x}_t]\right) \notag\\
	=&\frac{\dot{\zeta}_{t}}{2\zeta_{t}^{3/2}}\left(-\mathbf{x}_{t}+\frac{1}{\sqrt{1-\zeta_{t}}}\frac{1}{\sqrt{1-\zeta_{t}}}\left(\mathbf{x}_{t}-\sqrt{\zeta_{t}}\mathbf{S}_{\theta}(\mathbf{x}_{t},t)\right)\right) \notag\\
        =&\frac{\dot{\zeta}_{t}}{2\zeta_{t}^{3/2}}\left(\left(-1+\frac{1}{1-\zeta_{t}}\right)\mathbf{x}_{t}-\frac{\sqrt{\zeta_{t}}}{1-\zeta_{t}}\mathbf{S}_{\theta}(\mathbf{x}_{t},t)\right) \notag\\
	=&\frac{1}{2\zeta_{t}^{3/2}}\frac{\dot{\zeta}_{t}}{1-\zeta_{t}}\left(\zeta_{t}\mathbf{x}_{t}-\sqrt{\zeta_{t}}\mathbf{S}_{\theta}(\mathbf{x}_{t},t)\right) 
       =\frac{1}{2\zeta_{t}^{3/2}}\beta(t)\left(\zeta_{t}\mathbf{x}_{t}-\sqrt{\zeta_{t}}\mathbf{S}_{\theta}(\mathbf{x}_{t},t)\right)\notag\\
       =&\frac{\beta(t)}{2\sqrt{\zeta_{t}}}\left(\mathbf{x}_{t}-\frac{\mathbf{S}_{\theta}(\mathbf{x}_{t},t)}{\sqrt{\zeta_{t}}}\right).
\end{align}
	Since $\bar{\alpha}_t=1-\zeta_t$, using the results in Subsections \eqref{eq:express_s_xt} and \eqref{eq:express_s_t}, we have
	\begin{align}\label{eq:estimat_taylor}
		\hat{\mathbf{x}} &= \frac{\mathbf{x}_t}{\sqrt{1-\zeta_t}}+\frac{\zeta_t}{1-\zeta_t}\nabla_{\mathbf{\hat x}} \log p(\mathbf{\hat x}) \approx\frac{\mathbf{x}_{t}}{\sqrt{\overline{\alpha _{t}} }}-\frac{\sqrt{1-\overline{\alpha _{t}}}}{\overline{\alpha _{t}}}\mathbf{S}_{\theta}(\mathbf{\hat x},0) \notag\\
		&=\frac{\mathbf{x}_t}{\sqrt{\overline{\alpha _{t}}}}-\frac{\sqrt{1-\overline{\alpha _{t}}}}{\overline{\alpha _{t}}}\mathbf{S}_{\theta}(\mathbf{x}_{t},t)+ q_1\left(t\mathbf{x}_t\frac{\beta _t}{2\overline{\alpha _{t}}}-t\frac{\beta _t \mathbf{S}_{\theta}(\mathbf{x}_{t},t) }{2\overline{\alpha _{t}}\sqrt{1-\overline{\alpha _{t}}}}\right)-q_2\frac{\mathbf{\hat x}-\mathbf{x}_t}{\overline{\alpha _{t}}}\notag\\
        &=\frac{\mathbf{x}_t}{\sqrt{\overline{\alpha _{t}}}}-\frac{\sqrt{1-\overline{\alpha _{t}}}}{\overline{\alpha _{t}}}\mathbf{S}_{\theta}(\mathbf{x}_{t},t)+q_1 t\mathbf{x}_t\frac{\beta _t}{2\overline{\alpha _{t}}}-q_1 t\frac{\beta _t \mathbf{S}_{\theta}(\mathbf{x}_{t},t) }{2\overline{\alpha _{t}}\sqrt{1-\overline{\alpha _{t}}}}-q_2\frac{\mathbf{\hat x}-\mathbf{x}_t}{\overline{\alpha _{t}}}\notag\\
		&=\frac{\mathbf{x}_t}{\sqrt{\overline{\alpha _{t}}}}-\frac{\sqrt{1-\overline{\alpha _{t}}}}{\overline{\alpha _{t}}}\mathbf{S}_{\theta}(\mathbf{x}_{t},t)+\frac{q_1 t\beta _t}{2\overline{\alpha _{t}}}\mathbf{x}_t-\frac{q_1 t\beta _t \mathbf{S}_{\theta}(\mathbf{x}_{t},t) }{2\overline{\alpha _{t}}\sqrt{1-\overline{\alpha _{t}}}}-\frac{q_2}{\overline{\alpha_{t}}}(\mathbf{\hat x}-\mathbf{x}_t),
	\end{align}
 where the second equality is from the condition $\mathbf{S}_{\theta}(\hat{\mathbf{x}},0)=\mathbf{S}_{\theta}(\mathbf{x}_{t},t)+q_1(0-t)\partial_{t}\mathbf{S}_{\theta}(\mathbf{x}_{t},t)+q_2\nabla_{\mathbf{x}_{t}}\mathbf{S}_{\theta}(\mathbf{x}_{t},t)^\top(\mathbf{\hat x}-\mathbf{x}_{t})$. The final result of $\hat{\mathbf{x}}$ is:
	\begin{eqnarray*}
		\hat{\mathbf{x}} & = &\frac {\left(\sqrt{\overline{\alpha _{t}}}+\frac{q_1 t\beta _t}{2}+q_2\right)\mathbf{x}_t-\left(\sqrt{1-\overline{\alpha _{t}}}+\frac{q_1 t\beta _t}{2\sqrt{1-\overline{\alpha _{t}}}}\right)\mathbf{S}_{\theta}(\mathbf{x}_{t},t)}{(\overline{\alpha _{t}} +q_2)}.
	\end{eqnarray*}
\end{proof}
\begin{remark}
 From Lagrangian mean value theorem, there exist $\xi_t$, $\xi_{\bx}$ such that $\mathbf{S}_{\theta}(\hat{\mathbf{x}},0)=\mathbf{S}_{\theta}(\mathbf{x}_{t},t)+(0-t)\partial_{t}\mathbf{S}_{\theta}(\xi_{\bx},\xi_t)+\nabla_{\mathbf{x}_{t}}\mathbf{S}_{\theta}(\xi_{\bx},\xi_t)^\top(\mathbf{\hat x}-\mathbf{x}_{t})$. We set $q_1=\partial_{t}\mathbf{S}_{\theta}(\xi_{\bx},\xi_t)/\partial_{t}\mathbf{S}_{\theta}(\mathbf{x}_{t},t)$ and $q_2=\nabla_{\mathbf{x}_{t}}\mathbf{S}_{\theta}(\xi_{\bx},\xi_t)^\top(\mathbf{\hat x}-\mathbf{x}_{t})/\nabla_{\mathbf{x}_{t}}\mathbf{S}_{\theta}(\mathbf{x}_{t},t)^\top(\mathbf{\hat x}-\mathbf{x}_{t})$, and the expression $\mathbf{S}_{\theta}(\hat{\mathbf{x}},0)=\mathbf{S}_{\theta}(\mathbf{x}_{t},t)+q_1(0-t)\partial_{t}\mathbf{S}_{\theta}(\mathbf{x}_{t},t)+q_2\nabla_{\mathbf{x}_{t}}\mathbf{S}_{\theta}(\mathbf{x}_{t},t)^\top(\mathbf{\hat x}-\mathbf{x}_{t})$ is valid. 
The parameters $q_1$ and $q_2$ may vary depending on the specific $\hat{\bx}$ being predicted. In practical applications, we treat them as adjustable parameters and tune them case-by-case.
\end{remark}
\subsection{Estimation of Guided Term}\label{subsec:estimate_guide}
Next, we estimate the conditional probability density function $p(y|\mathbf{x}_t)$. We substitute the estimated value $\hat {\mathbf{x}}$ in Lemma \ref{theo:estimation_x} into Equation \eqref{eq:pro} and combine it with the distribution of $z$. Then the conditional distribution of $y$ conditioned on $\mathbf{x}_t$ can be approximated by a normal distribution with mean $\mu=H\hat{\mathbf{x}}$ and covariance matrix $\Sigma=\sigma_y^2I$.
The approximation of  corresponding guided term is
	\begin{eqnarray}\label{eq:log_proba}
		\nabla_{\mathbf{x}_t} \log p(y|\mathbf{x}_t) \approx \frac{1}{\sigma_y^2} \left(H\frac{\partial \hat{\mathbf{x}}}{\partial \mathbf{x}_t}\right)^\top(y-H\hat{\mathbf{x}}),
	\end{eqnarray}
	where $\hat {\mathbf{x}}$ is defined in \eqref{eq:solution}, $y$ is the measurement defined in \eqref{eq:pro}, and $\mathbf{x}_t$ is the $t$-th latent image in the backward process of diffusion model. A detailed derivation of \eqref{eq:log_proba} is provided in Appendix A.

When implementing automatic differentiation in Python, we first detach the residual $y-H\hat{\mathbf{x}}$ (denoted as $\tilde{y}$) from the computational graph. Backpropagation is then applied through the scalar $\tilde{y}^\top H\hat{\mathbf{x}}$. As shown in \citep{song2022pseudoinverse}, the time cost of automatic differentiation is approximately two to three times the time cost of the forward procedure.

\begin{algorithm}[ht]
	\caption{MAP-based Problem-Agnostic Method}
	\label{alg:ialm}
	\begin{algorithmic}
		\STATE {\bfseries Input:} an observation $y$, model operator $H$, $\{\tilde{\sigma}_{t}\}_{t=1}^{T}$, learning rate $\eta$
		\STATE Initialize $x_T \sim  \mathcal{N}(0, I)$
		\FOR{$t=T,...,1$}
		\STATE $z_t \sim  \mathcal{N}(0, I)$
		\STATE $\mathbf{x}_{t-1}^{\prime}=\frac{1}{\sqrt{\alpha_t}}(\mathbf{x}_t-\frac{1-\alpha_t}{\sqrt{1-\bar{\alpha}_t}}\mathbf{S}_\theta(\mathbf{x}_t,t))+\tilde{\sigma}_tz_t$
		\STATE compute $\nabla_{\mathbf{x}_{t}}\log p(y\mid \mathbf{x}_{t})$ as Equation \eqref{eq:log_proba}
		\STATE $\mathbf{x}_{t-1}=\mathbf{x}_{t-1}^{\prime}+\eta\nabla_{\mathbf{x}_{t}}\log p(y\mid \mathbf{x}_{t})$
		\ENDFOR
		\STATE {\bfseries Output:} $\mathbf{x}_0$
	\end{algorithmic}
\end{algorithm} 
By integrating the prior score $\mathbf{S}_{\theta}(\mathbf{x}_t, t)$ derived from a pre-trained diffusion model with the guided term outlined in Equation \eqref{eq:log_proba}, we can execute posterior sampling in a manner analogous to the reverse diffusion process of the diffusion model.
The resulting algorithm is presented in Algorithm \ref{alg:ialm} and the corresponding flowchart appears at the bottom of Figure \ref{fig:Figure1}.

\section{Numerical Experiments}\label{sec:numerical}
The main tasks in the experiments consist of super-resolution (SR), denoising, and inpainting. Before proceeding further, we present the implementation details of the experiments, which is then followed by the numerical results for super-resolution, denoising, inpainting and the runtime.
\subsection{Experimental Setup}
\textbf{Experimental Procedure:} 
For the SR task, the input images are the downscaled versions of the ground truth high-resolution images, typically using a bicubic downsampling technique. This downscaling process simulates the loss of detail that is commonly experienced in real-world imaging scenarios. 

In the denoising task, the images are corrupted with Gaussian noise at a higher level of intensity (with a standard deviation of $\sigma=0.5$) to test the models' ability to remove noise and restore the original image quality. The models are evaluated on their performance to recover a clean image while preserving details.

For the inpainting task, portions of the images are masked out, either by a box shape or a text image, to simulate missing or damaged regions. The models are trained to inpaint these missing areas with plausible content that matches the surrounding context. For the SR and inpainting tasks, Gaussian noise is also added with a mean of zero and a standard deviation of $\sigma=0.05$. 

\textbf{Datasets Quality and Pre-trained Models:} We use the pretrained models from the Denoising Diffusion Probabilistic Models (DDPM), which were originally trained on the FFHQ 256$\times$256 dataset\footnote{The checkpoint can be downloaded from \url{https://drive.google.com/drive/folders/1jElnRoFv7b31fG0v6pTSQkelbSX3xGZh}}. These pretrained models are directly used for testing without further fine-tuning for specific tasks. To evaluate the performance of our method, we conduct experiments on FFHQ 256x256-1k (the in-distribution validation set) and CelebA-HQ 256x256-1k (out-of-distribution (OOD) validation set)\footnote{FFHQ 256x256-1k can be downloaded from \url{https://paperswithcode.com/sota/image-generation-on-ffhq-256-x-256} and CelebA-HQ 256x256-1k can be downloaded from \url{https://huggingface.co/datasets/korexyz/celeba-hq-256x256}}.

\textbf{Contrastive Methods and Evaluation Metrics:} We conduct comparisons with five methods: DDRM, DPS, $\Pi$GDM, DMPS, and MCG. Except for DDRM, all methods are based on the DDPM framework but differ in their approaches to posterior sampling. For DDRM, we utilize the publicly available code provided by the authors. To ensure a fair and objective comparison, we employ the same evaluation metrics across all diffusion-based methods.

Model performance is evaluated using four standard metrics covering distortion, perceptual quality, and generation fidelity: (1) Peak Signal-to-Noise Ratio (PSNR), which measures the signal-to-noise ratio; (2) Structural Similarity Index Measure (SSIM), which models perceptual changes in structural information, luminance, and contrast; (3) Learned Perceptual Image Patch Similarity (LPIPS), which assesses similarity in the feature space of a pre-trained network; and (4) Fr\'echet Inception Distance (FID), which compares the feature distributions of generated and real images to evaluate quality and diversity. 
All metrics assess the fidelity of the reconstructed images to the ground truth, as well as their visual quality, but they may have different preferences or trade-offs.

\subsection{Super-Resolution}
\begin{table}[htb]
	\centering
	\resizebox{0.9\textwidth}{!}{
		\begin{tabular}{llllllllllll}
			\hline
			\multicolumn{1}{c}{}    & \multicolumn{1}{c}{\textbf{}} &  & \multicolumn{4}{c}{\textbf{SR($\times$4)}}  &           & \multicolumn{4}{c}{\textbf{DENOISE}} \\ \cline{4-7} \cline{9-12} 
			\textbf{Dataset}        & \textbf{Method}               &  & PSNR↑          & LPIPS↓      & SSIM↑ & FID↓      & \textbf{} & PSNR↑        & LPIPS↓      & SSIM↑ & FID↓     \\ \hline
			\multirow{6}{*}{FFHQ}   & ours                          &  & \textbf{30.63} & \textbf{0.2347}   &\textbf{0.90} & \textbf{30.34}  &      & \textbf{30.24}   & \textbf{0.2344}   & \textbf{0.87}   & \textbf{40.24}  \\
			& DDRM                          &  & 29.25          & 0.3087   & 0.83 & 66.17        &           & 27.87            & 0.3048   & 0.85 & 58.95         \\
			& DPS                           &  & 26.68          & 0.2717     & 0.87  & 40.49       &           & 29.16            & 0.2574    & 0.83 & 58.96         \\
			& $\Pi$GDM                          &  & 28.69          & 0.2408    & 0.83 & 38.61        &           & 24.64            & 0.2688    & 0.84 & 42.65         \\
			& DMPS                          &  & 27.23          & 0.2533    & 0.85 & 39.29        &           & 28.69            & 0.2419   & 0.80 & 47.27           \\ 
            & MCG                          &  & 29.02          & 0.3069     & 0.87  & 60.12      &           & 28.41            & 0.2456     &0.78	&51.80        \\ \hline
			\multirow{6}{*}{CelebA-HQ} & ours                          &  & \textbf{31.85} & \textbf{0.2355} & \textbf{0.90}  & \textbf{31.93}  &           & \textbf{31.48}   & \textbf{0.2243}  & \textbf{0.90}   & \textbf{35.47} \\
			& DDRM                          &  & 30.12          & 0.2614     & 0.77 & 63.06       &           & 30.46            & 0.2332   & 0.87 & 55.75           \\
			& DPS                           &  & 24.62          & 0.3071     & 0.86 & 61.97      &           & 29.63            & 0.2969    & 0.84 & 44.08         \\
			& $\Pi$GDM                          &  & 30.57          & 0.2509      & 0.87 & 44.43     &           & 23.97            & 0.2510    & 0.82 & 63.99         \\
			& DMPS                          &  & 26.70          & 0.2603      & 0.89 & 49.88         &           & 28.96            & 0.3060     & 0.86 & 52.21         \\ 
            & MCG                          &  & 28.79          & 0.2424     & 0.87          & 47.99     &           & 28.78           & 0.2624     &0.84	&45.68       \\ \hline
		\end{tabular}
	}
       \caption{Quantitative comparisons (PSNR (dB), LPIPS, SSIM and FID) of different methods for the SR and Denoising tasks on the FFHQ 256$\times$256-1k validation dataset and the CelebA-HQ 256$\times$256-1k validation dataset, respectively. The pre-trained model used in our proposed method, as well as in DPS, \(\Pi\)GDM, DMPS, and MCG, is trained on the FFHQ dataset. For DDRM, we utilize the original code provided by the authors.}
       \label{table1}
\end{table}

\begin{figure}[ht]
	\centering
	\begin{subfigure}[Input]{
			\begin{minipage}{0.115\textwidth}
				\centering
				\includegraphics[width=\linewidth]{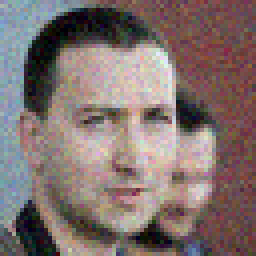}
				\includegraphics[width=\linewidth]{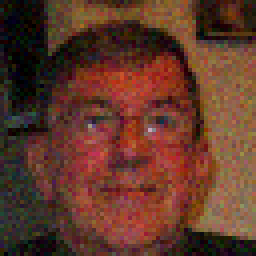}
				\includegraphics[width=\linewidth]{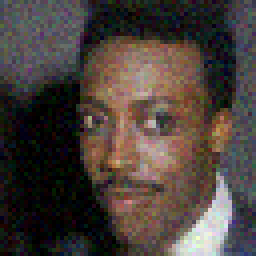}
		\end{minipage}}
	\end{subfigure}\hspace{-3mm}
	\begin{subfigure}[GT]{
			\begin{minipage}{0.115\textwidth}
				\centering
				\includegraphics[width=\linewidth]{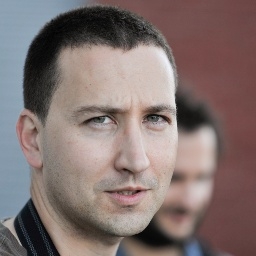}
				\includegraphics[width=\linewidth]{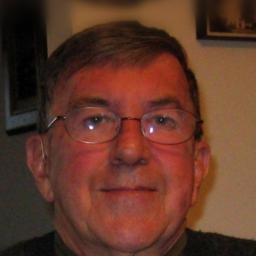}
				\includegraphics[width=\linewidth]{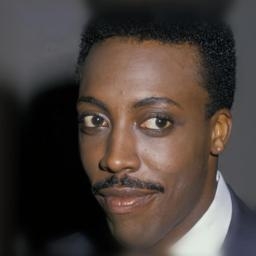}
		\end{minipage}}
	\end{subfigure}\hspace{-3mm}
	\begin{subfigure}[Our]{
			\begin{minipage}{0.115\textwidth}
				\centering
				\includegraphics[width=\linewidth]{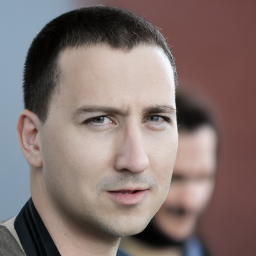}
				\includegraphics[width=\linewidth]{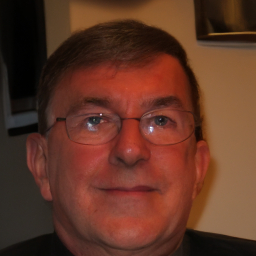}
				\includegraphics[width=\linewidth]{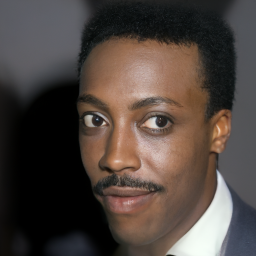}
		\end{minipage}}
	\end{subfigure}\hspace{-3mm}
	\begin{subfigure}[DDRM]{
			\begin{minipage}{0.115\textwidth}
				\centering
				\includegraphics[width=\linewidth]{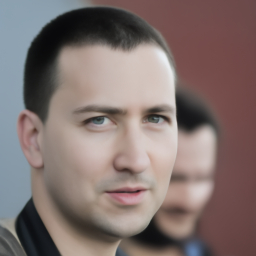}
				\includegraphics[width=\linewidth]{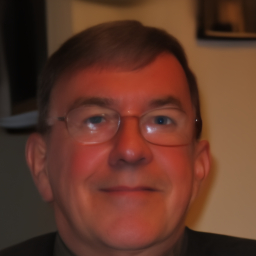}
				\includegraphics[width=\linewidth]{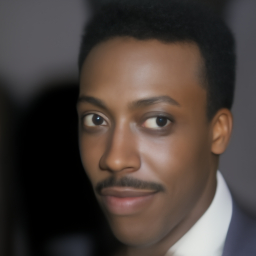}   
		\end{minipage}}
	\end{subfigure}\hspace{-3mm}
	\begin{subfigure}[DPS]{
			\begin{minipage}{0.115\textwidth}
				\centering
				\includegraphics[width=\linewidth]{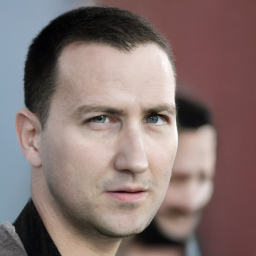}
				\includegraphics[width=\linewidth]{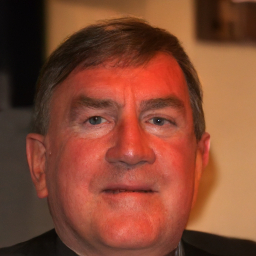}
				\includegraphics[width=\linewidth]{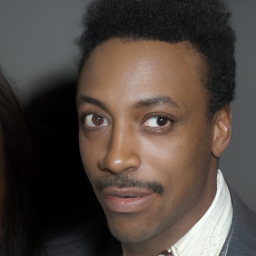}  
		\end{minipage}}
	\end{subfigure}\hspace{-3mm}
	\begin{subfigure}[$\Pi$GDM]{
			\begin{minipage}{0.115\textwidth}
				\centering
				\includegraphics[width=\linewidth]{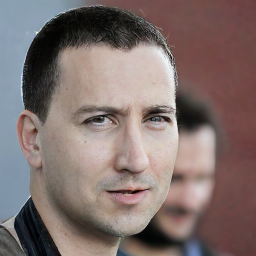}
				\includegraphics[width=\linewidth]{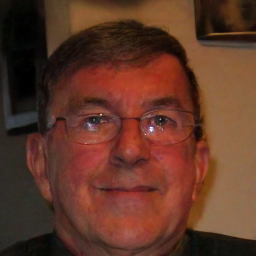}
				\includegraphics[width=\linewidth]{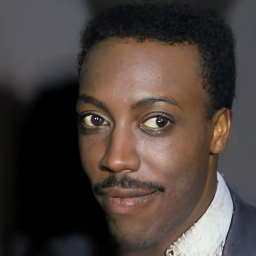} 
		\end{minipage}}
	\end{subfigure}\hspace{-3mm}
	\begin{subfigure}[DMPS]{
			\begin{minipage}{0.115\textwidth}
				\centering
				\includegraphics[width=\linewidth]{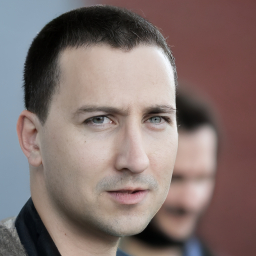}
				\includegraphics[width=\linewidth]{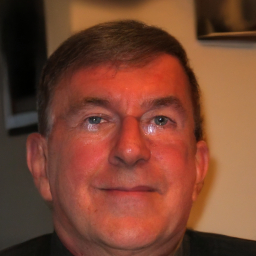}
				\includegraphics[width=\linewidth]{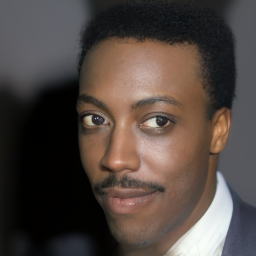}
		\end{minipage}}
	\end{subfigure}\hspace{-3mm}
	\begin{subfigure}[MCG]{
			\begin{minipage}{0.115\textwidth}
				\centering
				\includegraphics[width=\linewidth]{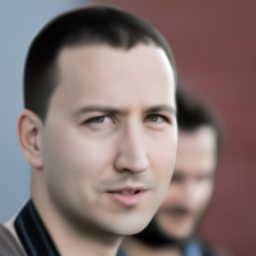}
				\includegraphics[width=\linewidth]{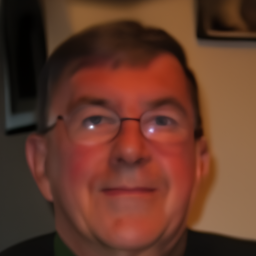}
				\includegraphics[width=\linewidth]{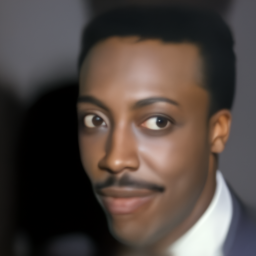}
		\end{minipage}}
	\end{subfigure}\vspace{-3mm}
    \caption{The results for super-resolution. The first column is the input image, the second one is  the Ground Truth (denoted as GT) and the third to eighth columns are our proposed method, DDRM, DPS, $\Pi$GDM, DMPS and MCG, respectively.}
	\label{fig:SR}
\end{figure}

For the super-resolution task, the original high-resolution images undergo bicubic downsampling to create low-resolution versions. These low-resolution images serve as the input measurements to the super-resolution model, which then generates the $4\times$ super-resolved output image.

For our proposed method, we fix the parameters $q_1 = 2$, $q_2 = 10$, and set $\eta = 200$. It is clearly demonstrated in Table \ref{table1} for SR that our model outperforms other state-of-the-art diffusion models by a large margin in both in-distribution (FFHQ) and out-of-distribution (CelebA-HQ) experiments. Our method achieves the highest PSNR (30.63 and 31.85), the lowest LPIPS (0.2347 and 0.2355), the highest SSIM (0.90), and the lowest FID (30.34 and 31.93), highlighting its outstanding quality and impressive performance. 

In Figure \ref{fig:SR}, we show some examples of super-resolution on two datasets and compare them with other models. From these illustrations in Figure \ref{fig:SR}, we can see that the images generated by DDRM are too smooth and lose a lot of details. The other models also handle the special features very unnaturally, and the generated images are far from the truth and reality, particularly evident in the portrayal of the eyes. Specifically, the depiction of the eyes in the generated images lacks realism and fails to capture the intricate details that make them lifelike. Additionally, the models struggle to accurately represent eyeglasses, resulting in subpar visual representation. However, our model overcomes these challenges and achieves better results in these aspects, providing more realistic and detailed images. 
It also demonstrates adaptability of the model to CelebA-HQ dataset, although the pretrained models are originally trained on FFHQ 256x256 dataset.  

\begin{figure}[ht!]
	\begin{center}
			\begin{minipage}[b]{0.49\linewidth}
            \includegraphics[width=1.05\textwidth]{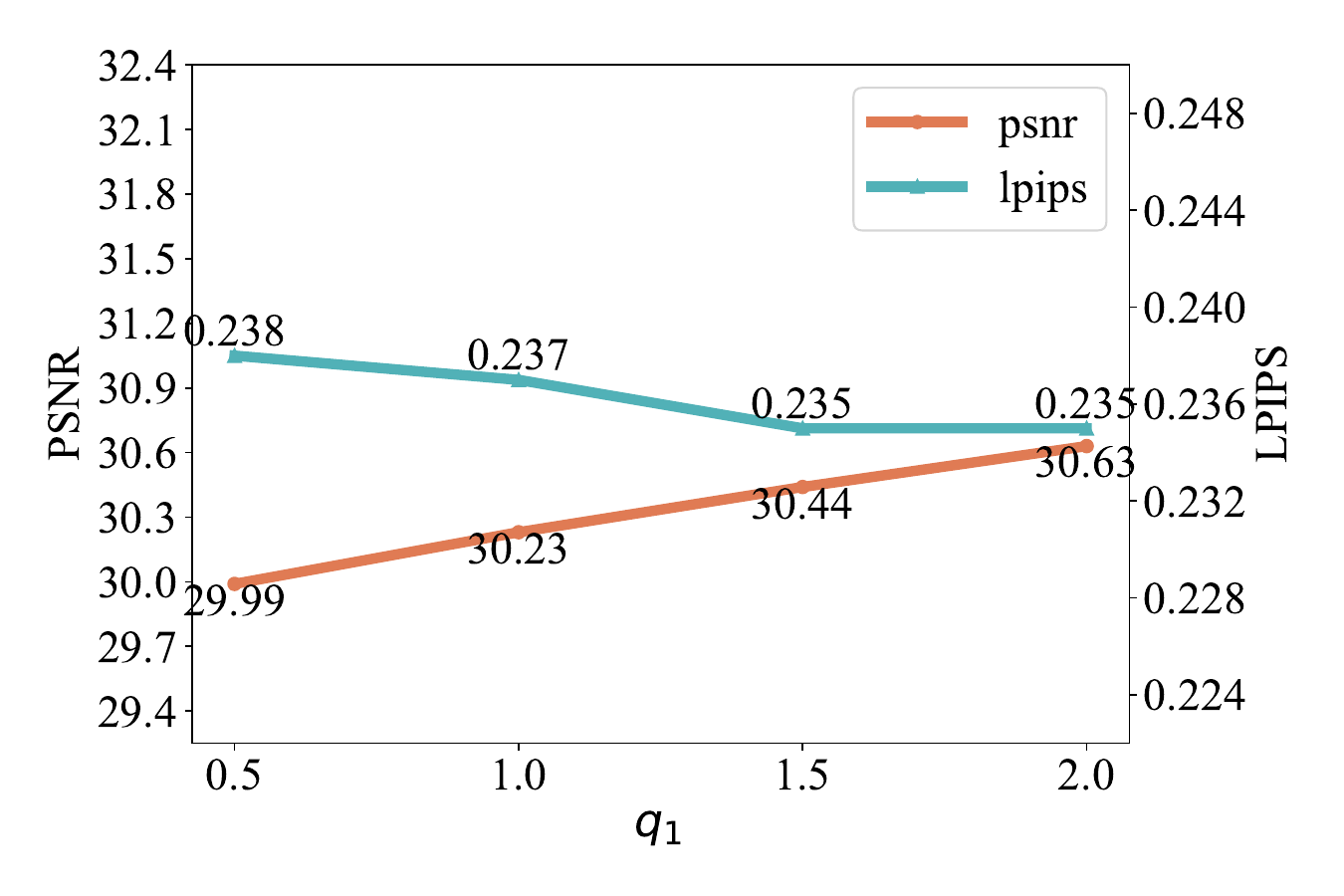}
            \includegraphics[width=1.05\textwidth]{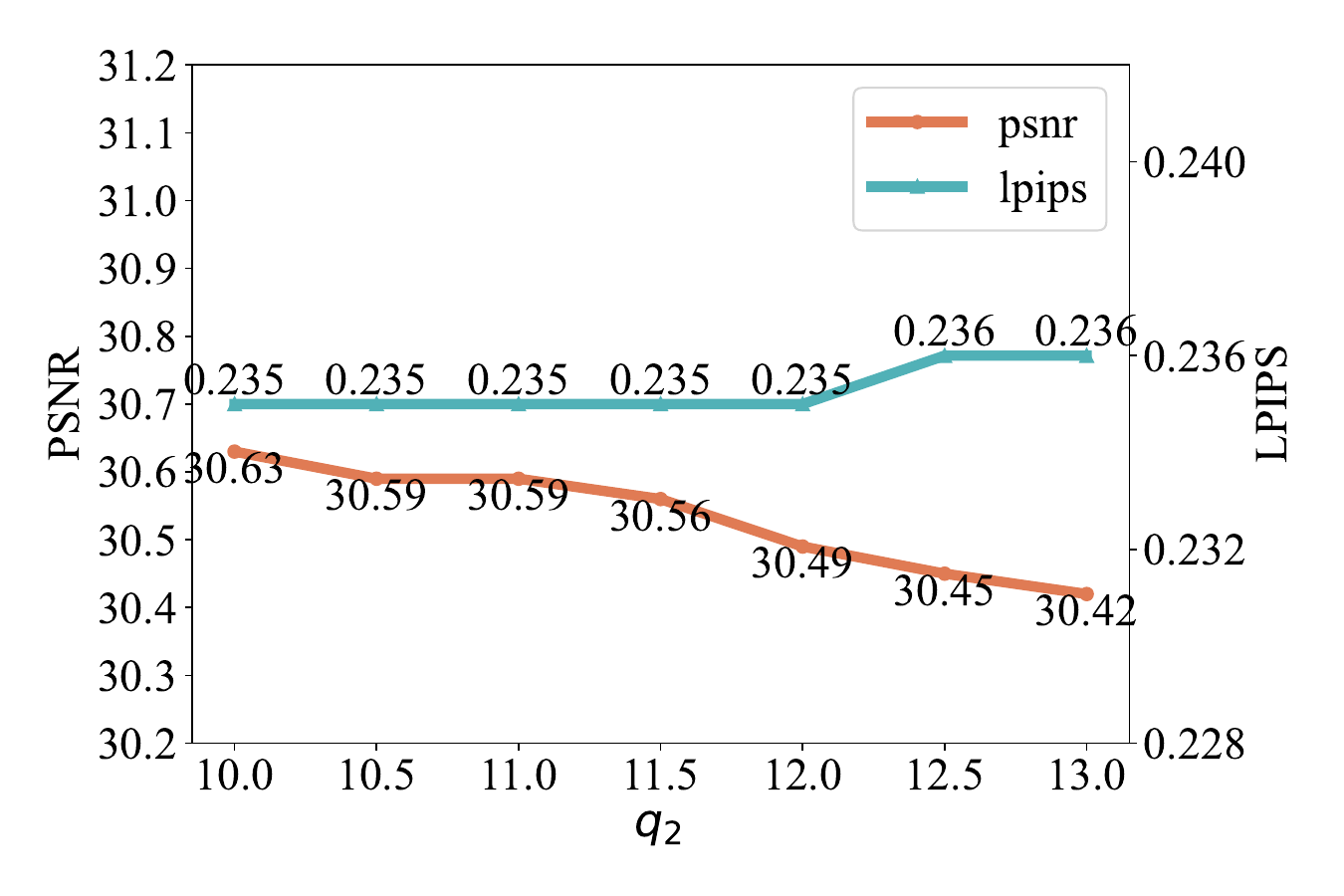}    \includegraphics[width=1.05\textwidth]{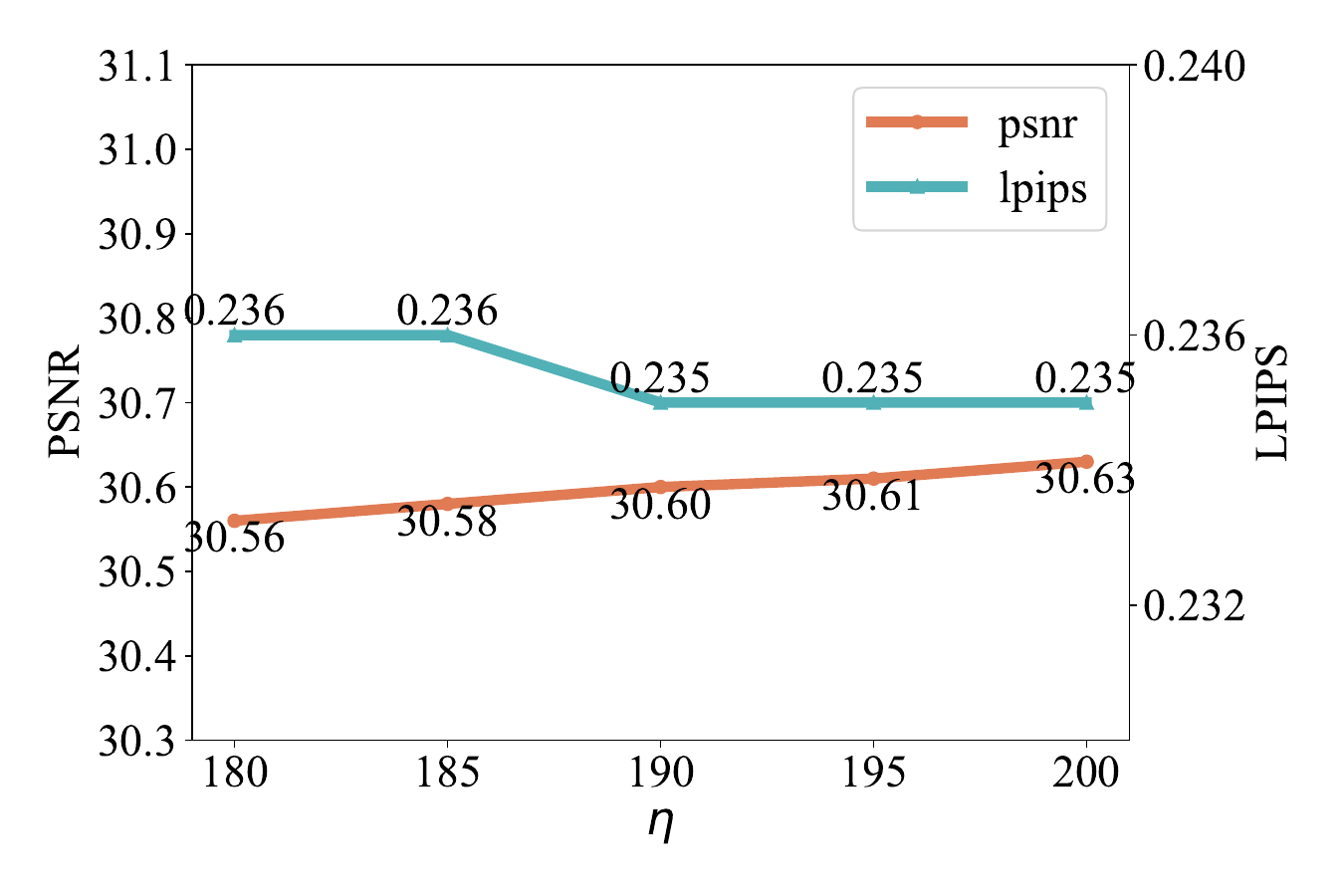} 
			\end{minipage}
            \begin{minipage}[b]{0.49\linewidth}	
            \includegraphics[width=1.05\textwidth]{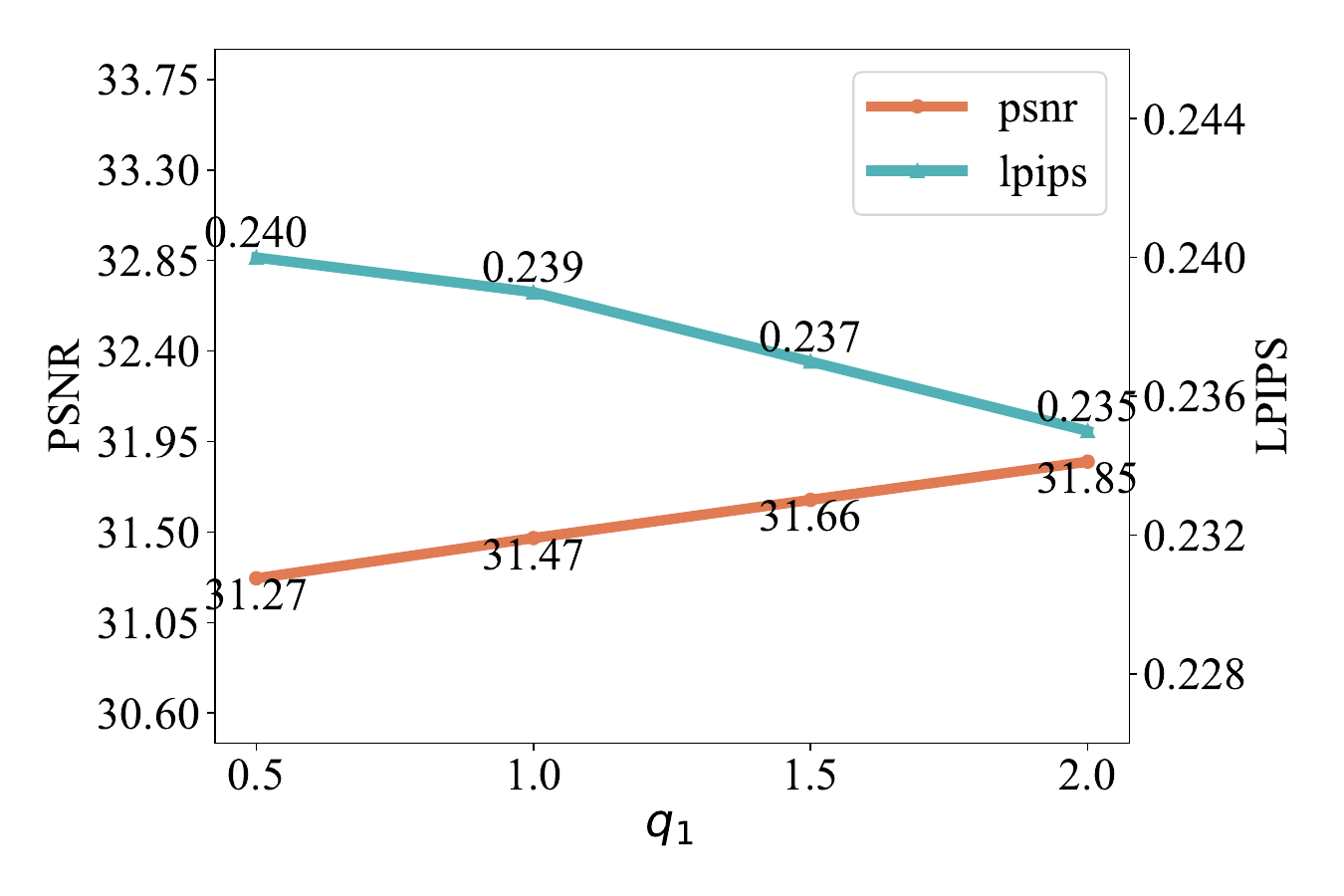}  
            \includegraphics[width=1.05\textwidth]{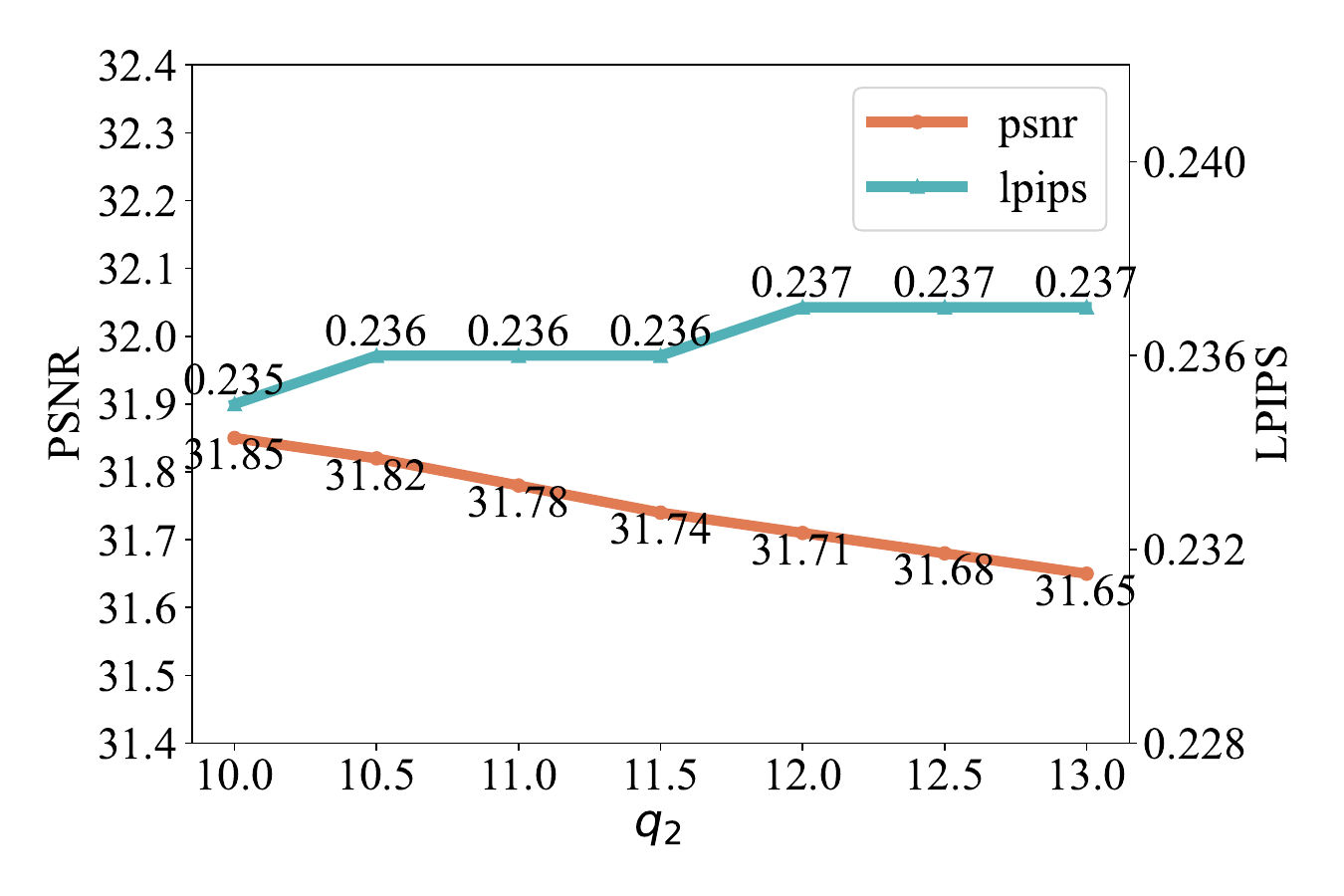} 
            \includegraphics[width=1.05\textwidth]{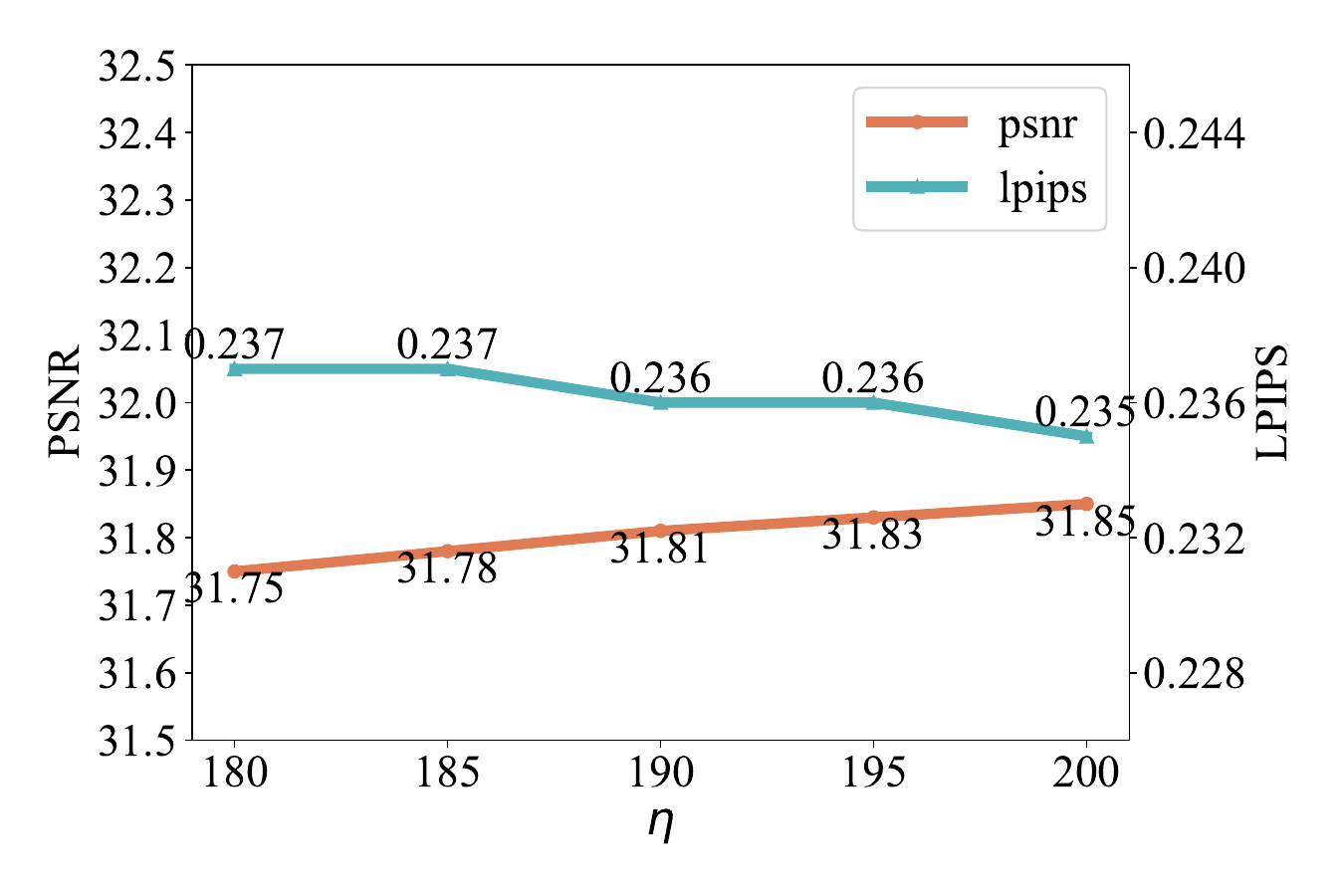}
		\end{minipage}
        \caption{The robustness analysis results for SR on  FFHQ 256×256-1k (first row) and CelebA-HQ 256×256-1k (second row) validation sets with different parameters. Rows 1-3 are the plots of PSNR and LPIPS values versus the changes of parameters $q_1$, $q_2$, and $\eta$, with the other two fixed. }
		\label{fig:Super}
	\end{center}
\end{figure}
In order to select the most suitable parameters of our proposed method, extensive experiments are conducted on the validation set for all tasks. The primary objective of these experiments is to identify the optimal parameter combinations. Numerical results for the super-resolution tasks are presented in Figure \ref{fig:Super}. 
Interestingly, our finding reveals that the changes in these parameters had minimal impact on the PSNR and LPIPS values, indicating the robustness of our model. This observation underscores the ability of our model to consistently deliver high-quality results across different parameter settings. Such robustness is a highly desirable characteristic as it ensures the model's performance remains stable and reliable even when faced with variations in inputs or parameter values.

We further evaluate our method on $2\times$ and $8\times$ super-resolution tasks. The corresponding experimental results are detailed in  \ref{subsec:additional_sr}.
\subsection{Denoising}
\begin{figure}[htp]
	\centering
	\begin{subfigure}[Input]{
			\begin{minipage}{0.115\textwidth}
				\centering
				\includegraphics[width=\linewidth]{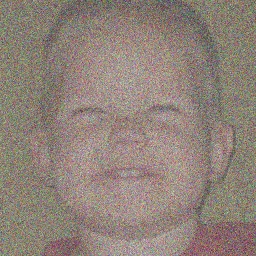}
				\includegraphics[width=\linewidth]{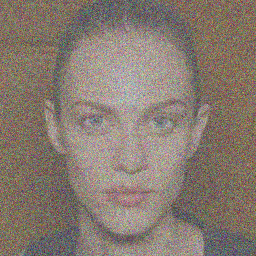}    
		\end{minipage}}
	\end{subfigure}\hspace{-3mm}
	\begin{subfigure}[GT]{
			\begin{minipage}{0.115\textwidth}
				\centering
				\includegraphics[width=\linewidth]{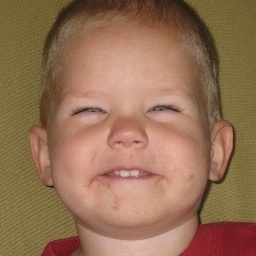}
				\includegraphics[width=\linewidth]{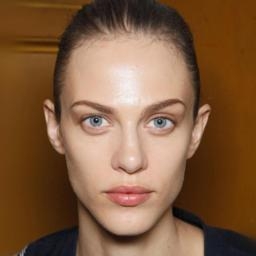}
		\end{minipage}}
	\end{subfigure}\hspace{-3mm}
	\begin{subfigure}[Our]{
			\begin{minipage}{0.115\textwidth}
				\centering
				\includegraphics[width=\linewidth]{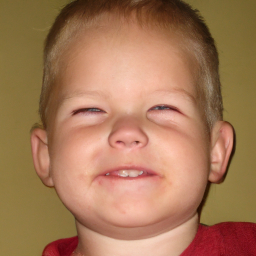}
				\includegraphics[width=\linewidth]{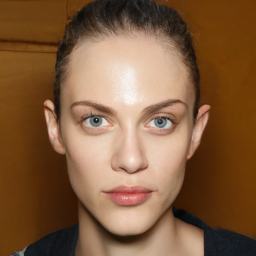}
		\end{minipage}}
	\end{subfigure}\hspace{-3mm}
	\begin{subfigure}[DDRM]{
			\begin{minipage}{0.115\textwidth}
				\centering
				\includegraphics[width=\linewidth]{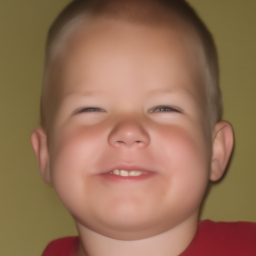}
				\includegraphics[width=\linewidth]{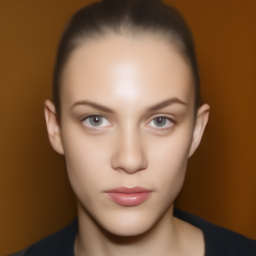}
		\end{minipage} }
	\end{subfigure}\hspace{-4mm}
	\begin{subfigure}[DPS]{
			\begin{minipage}{0.115\textwidth}
				\centering
				\includegraphics[width=\linewidth]{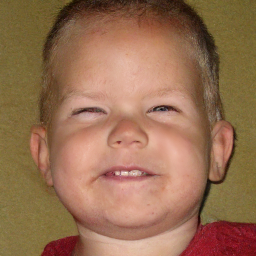}
				\includegraphics[width=\linewidth]{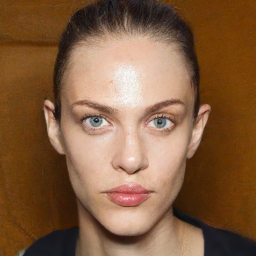}
		\end{minipage}}
	\end{subfigure}\hspace{-3mm}
	\begin{subfigure}[$\Pi$GDM]{
			\begin{minipage}{0.115\textwidth}
				\centering
				\includegraphics[width=\linewidth]{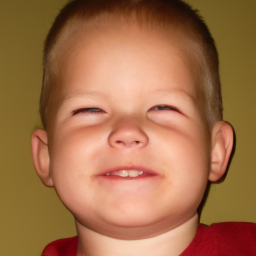}
				\includegraphics[width=\linewidth]{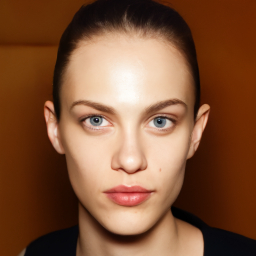}
		\end{minipage}}
	\end{subfigure}\hspace{-3mm}
	\begin{subfigure}[DMPS]{
			\begin{minipage}{0.115\textwidth}
				\centering
				\includegraphics[width=\linewidth]{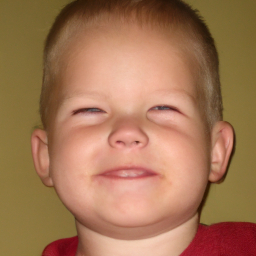}
				\includegraphics[width=\linewidth]{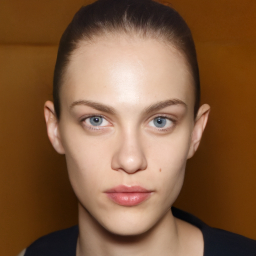}
		\end{minipage}}
	\end{subfigure}
    \hspace{-4mm}
	\begin{subfigure}[MCG]{
			\begin{minipage}{0.115\textwidth}
				\centering
				\includegraphics[width=\linewidth]{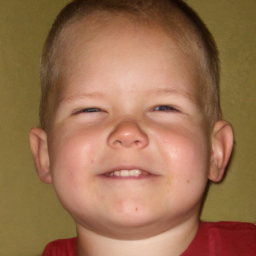}
				\includegraphics[width=\linewidth]{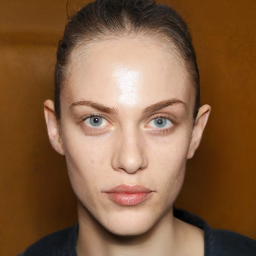}
		\end{minipage}}
	\end{subfigure}\vspace{-4mm}
    \caption{The results for denoising.  
 All measurements are with Gaussian noise $\sigma=0.5$, where GT stands for Ground Truth.}
	\label{fig:Denoise}
\end{figure}

In denoising tasks, Gaussian noise is typically added to the image to simulate image degradation in the real world. The intensity of the noise determines the degree to which the noise affects the image, with higher values indicating higher noise and worse image quality degradation. To evaluate the model's performance in denoising tasks, we deliberately introduce high-intensity ($\sigma$=0.5) Gaussian noise to corrupt the images. The goal is to test the model's ability to remove noise while preserving the original image's clarity and fine details. We set the parameters $q_1=12,q_2=22,\eta=2.2$ and $q_1=10,q_2=22,\eta=2.2$ for the FFHQ 256$\times$256-1k and the CelebA-HQ 256$\times$256-1k validation sets, respectively. To assess the effectiveness and performance of the model in denoising tasks, we adopt four evaluation metrics, PSNR, LPIPS, SSIM and FID. Similar to the case of super-resolution, our model shows an impressive performance in the denoising task (see the right part in Table \ref{table1}).

In Figure \ref{fig:Denoise}, we show some examples of denoising on two datasets and compare them with other models. Analyzing the examples provided, it becomes evident that the images generated by DDRM and $\Pi$GDM exhibit an overly smooth appearance, resulting in a loss of fine details. Additionally, $\Pi$GDM tends to produce images that appear overly vibrant with higher color saturation. On the other hand, DPS generates images that are extremely sharp and may still retain some noise, resulting in an exaggerated emphasis on details and potential imperfections in the generated images. Moreover, there are instances where DPS and MCG introduce additional details on the faces for the lady, resembling noise artifacts. The performance of the little boy by MCG is not satisfactory, particularly in the depiction of the eyes. Moving on to DMPS, specific examples further illustrate its strengths and weaknesses. For instance, in the first image, the teeth details of the person are missing, and in the second image, an imperfection such as a mole appears on the right side of the person's nose, which does not exist in the original image. Overall, our model outperforms these methods in terms of preserving fine details and achieving more realistic results. We also conduct numerical experiments to identify the optimal parameter combinations for the denoising task. Numerical results are deferred to Figure \ref{fig:Denoise1}. Additional experiments across various noise levels ($\sigma\in\{0.05, 0.1, 0.2\}$) are presented in  \ref{subsec:additinal_denoising}.
\subsection{Inpainting}

\begin{table}[htbp]
	\centering
	\resizebox{1\textwidth}{!}{
		\begin{tabular}{lllllllllllllllll}
			\hline
			\multicolumn{1}{c}{}    & \multicolumn{1}{c}{} &  & \multicolumn{4}{c}{\textbf{BOX}} &  & \multicolumn{4}{c}{\textbf{LOLCAT}} &  & \multicolumn{4}{c}{\textbf{LOREM}} \\ \cline{4-7} \cline{9-12} \cline{14-17}
			\textbf{DATASET}        & \textbf{METHOD}                   &  & PSNR↑          & LPIPS↓     &SSIM↑ &FID↓     &  & PSNR↑            & LPIPS↓  &SSIM↑ &FID↓         &  & PSNR↑           & LPIPS↓     &SSIM↑ &FID↓      \\ \hline
			\multirow{6}{*}{FFHQ}   & ours                              &  & \textbf{30.06} & 0.2768    &\textbf{0.87}	&\textbf{45.75}
      &  & \textbf{31.24}   & 0.2587    &\textbf{0.88}	&\textbf{44.12}
       &  & \textbf{31.31}           & \textbf{0.2097}  & \textbf{0.89}	& \textbf{36.97}
\\
			& DDRM                              &  & 29.88          & 0.2373    &0.86	&55.01
      &  & 30.50            & 0.2458    &  0.82	& 46.21
      &  & 31.27  & 0.2222     &0.85	&48.84
      \\
			& DPS                               &  & 23.78          & 0.2455   &0.81	&55.35
       &  & 30.46            & \textbf{0.2367}  &0.82	&44.32
 &  & 28.57           & 0.2124       &0.85	&43.26
    \\
			& $\Pi$GDM                              &  & 20.46          & \textbf{0.2312}
            & 0.79	&73.22
&  & 17.55            & 0.3416    &0.71 &118.38
       &  & 22.69           & 0.3088     & 0.77 &67.74
     \\
			& DMPS                              &  & 24.37          & 0.2338  &0.84	&64.04
        &  & 22.97            & 0.2429     &0.77	&64.35
      &  & 27.65           & 0.2132       &0.85	&45.71
    \\
            & MCG                             &  & 26.91          & 0.2411    &0.85	&56.59
      &  & 28.53            & 0.2484     & 0.86	&45.12
     &  & 28.75          & 0.2176        &0.87 &	42.84
   \\\hline
			\multirow{6}{*}{CelebA-HQ} & ours                              &  & \textbf{31.10} & \textbf{0.2071} & \textbf{0.90} 	& \textbf{37.41}
&  & \textbf{33.42}   & \textbf{0.2400}    & \textbf{0.89}  	& \textbf{38.62}
     &  & \textbf{32.93}  & \textbf{0.2077}  & \textbf{0.90}   & \textbf{37.46}	
\\
			& DDRM                              &  & 29.88          & 0.2373   & 0.83	& 43.94
       &  & 31.23            & 0.2403  &0.85	&43.49
&  & 32.07           & 0.2085 & 0.88	&41.88
\\
			& DPS                               &  & 23.20          & 0.2215   & 0.82 & 	54.13
       &  & 29.63            & 0.2421      &0.81	&40.54
     &  & 29.06           & 0.2185    &0.83	&45.86
       \\
			& $\Pi$GDM                              &  & 21.09          & 0.2287    &0.80	&78.73
      &  & 15.59            & 0.4110      & 0.73	 &123.83
     &  & 21.59           & 0.3369    & 0.76	&83.98
      \\
			& DMPS                              &  & 23.19          & 0.2106   &0.81	&61.86
       &  & 18.97            & 0.3031     & 0.71	& 97.33
      &  & 28.20           & 0.2187     & 0.82	&42.35
     \\
            &  MCG                             &  &  28.18          &  0.2280   &0.89	&52.66
       &  &  27.80            &  0.2603       &  0.85	& 52.11
   &  &  28.84          &  0.2328       & 0.85 &	57.43
   \\\hline
		\end{tabular}
	}
    \caption{Quantitative comparison (PSNR (dB), LPIPS, SSIM and FID) of different methods for different inpainting tasks on the FFHQ 256$\times$256-1k validation dataset and the CelebA-HQ 256$\times$256-1k validation dataset. The pre-trained model used in our proposed method, as well as in DPS, \(\Pi\)GDM, DMPS, and MCG, is trained on the FFHQ dataset. For DDRM, we utilize the original code provided by the authors.}
	\label{table2}
\end{table}

\begin{figure}[htbp] 
	\centering
	\begin{subfigure}[Input]{
			\centering
			\begin{minipage}{0.115\linewidth}
				\includegraphics[width=\linewidth]{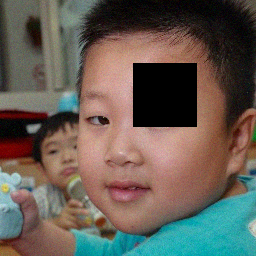}
				\includegraphics[width=\linewidth]{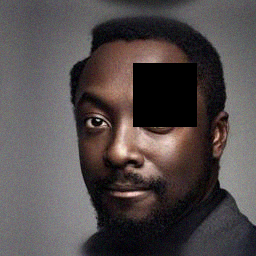}
		\end{minipage}}
	\end{subfigure}\hspace{-3mm}
	\begin{subfigure}[GT]{
			\centering
			\begin{minipage}{0.115\linewidth}
				\includegraphics[width=\linewidth]{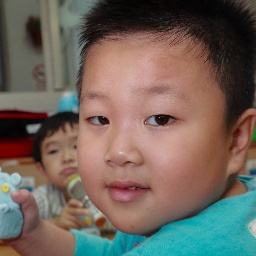}
				\includegraphics[width=\linewidth]{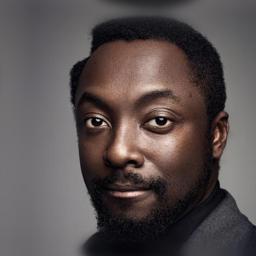}
		\end{minipage}}
	\end{subfigure}\hspace{-3mm}
	\begin{subfigure}[Our]{
			\begin{minipage}{0.115\linewidth}
				\includegraphics[width=\linewidth]{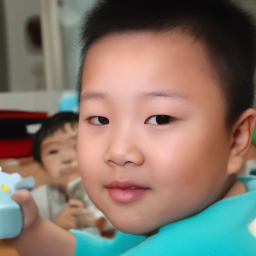}
				\includegraphics[width=\linewidth]{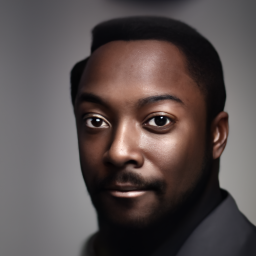}
		\end{minipage}}
	\end{subfigure}\hspace{-3mm}
	\begin{subfigure}[DDRM]{
			\centering
			\begin{minipage}{0.115\linewidth}
				\includegraphics[width=\linewidth]{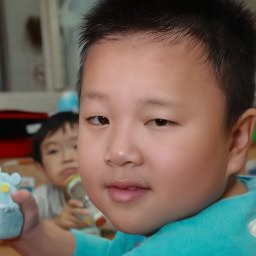}
				\includegraphics[width=\linewidth]{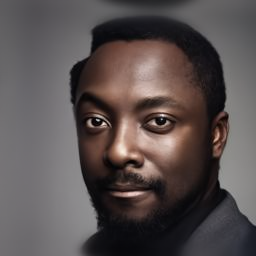}
		\end{minipage}}
	\end{subfigure}\hspace{-3mm}
	\begin{subfigure}[DPS]{
			\begin{minipage}{0.115\linewidth}
				\centering
				\includegraphics[width=\linewidth]{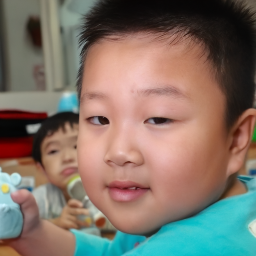}
				\includegraphics[width=\linewidth]{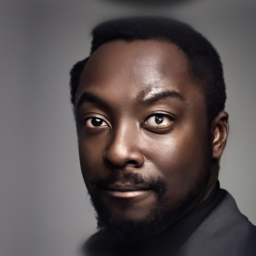}
		\end{minipage}}
	\end{subfigure}\hspace{-3mm}
	\begin{subfigure}[$\Pi$GDM]{
			\centering
			\begin{minipage}{0.115\linewidth}
				\includegraphics[width=\linewidth]{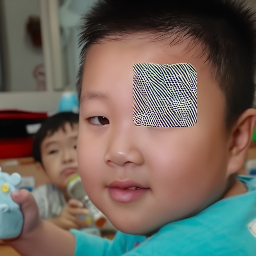}
				\includegraphics[width=\linewidth]{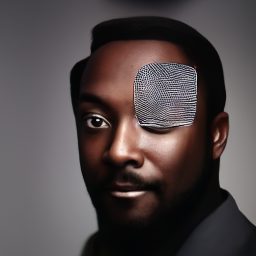}
		\end{minipage}}
	\end{subfigure}\hspace{-3mm}
	\begin{subfigure}[DMPS]{
			\centering
			\begin{minipage}{0.115\linewidth}
				\includegraphics[width=\linewidth]{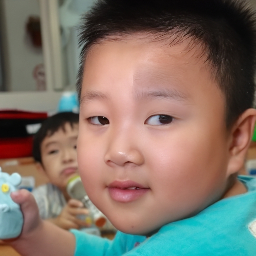}
				\includegraphics[width=\linewidth]{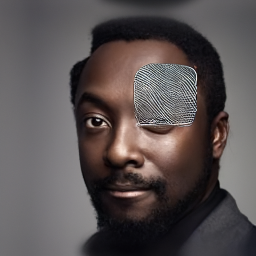}
		\end{minipage}}
	\end{subfigure}\hspace{-3mm}
	\begin{subfigure}[MCG]{
			\centering
			\begin{minipage}{0.115\linewidth}
				\includegraphics[width=\linewidth]{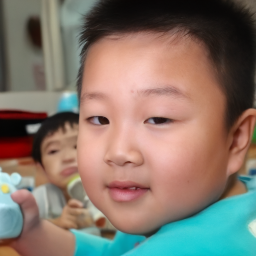}
				\includegraphics[width=\linewidth]{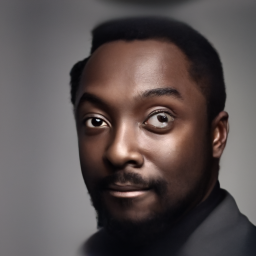}
		\end{minipage}}
	\end{subfigure}\vspace{-5mm}
	\caption{The results for Inpainting (Box).} 
	\label{fig:Box_1}
\end{figure}
We evaluate the proposed approach against baselines diffusion models on three types of image 
inpainting tasks: Inpainting (Box), Inpainting (Lolcat), and Inpainting (Lorem). We use the parameters $q_1 = 12$, $q_2 = 23$, and $\eta = 3$ for Inpainting (Lolcat), Inpainting (Lorem) and the parameters $q_1 = 10$, $q_2 = 24$, and $\eta = 4$ for Inpainting (Box). Table \ref{table2} presents the quantitative results of our model and other methods on the image inpainting tasks. 

\begin{figure}[htbp]
	\centering
	\begin{subfigure}[Input]{
			\centering
			\begin{minipage}{0.115\linewidth}
				\includegraphics[width=\linewidth]{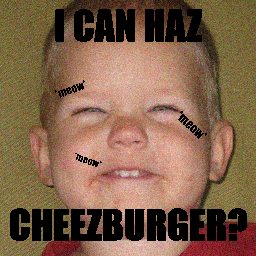}
				\includegraphics[width=\linewidth]{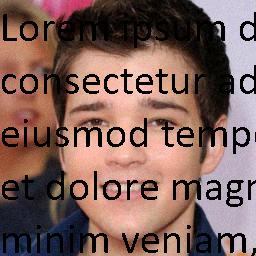}
		\end{minipage}}
	\end{subfigure}\hspace{-3mm}
	\begin{subfigure}[GT]{
			\centering
			\begin{minipage}{0.115\linewidth}
				\includegraphics[width=\linewidth]{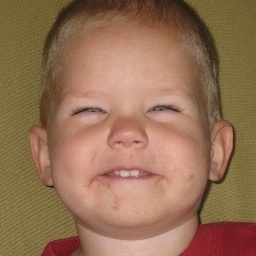}
				\includegraphics[width=\linewidth]{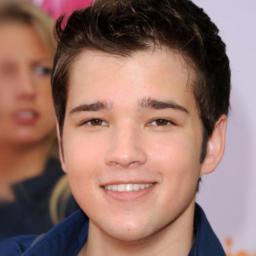}
		\end{minipage}}
	\end{subfigure}\hspace{-3mm}
	\begin{subfigure}[Our]{
			\begin{minipage}{0.115\linewidth}
				\includegraphics[width=\linewidth]{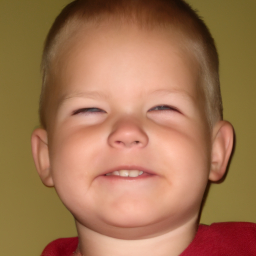}
				\includegraphics[width=\linewidth]{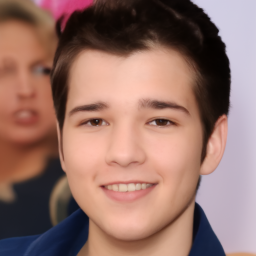}
		\end{minipage}}
	\end{subfigure}\hspace{-3mm}
	\begin{subfigure}[DDRM]{
			\centering
			\begin{minipage}{0.115\linewidth}
				\includegraphics[width=\linewidth]{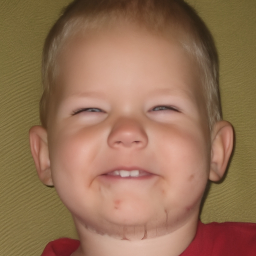}
				\includegraphics[width=\linewidth]{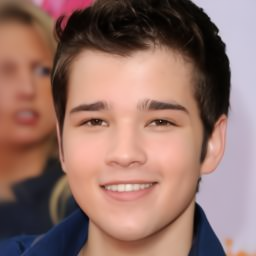}
		\end{minipage}}
	\end{subfigure}\hspace{-3mm}
	\begin{subfigure}[DPS]{
			\begin{minipage}{0.115\linewidth}
				\centering
				\includegraphics[width=\linewidth]{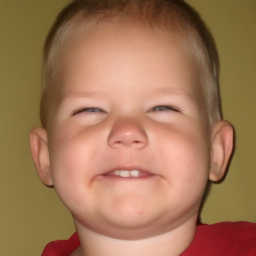}
				\includegraphics[width=\linewidth]{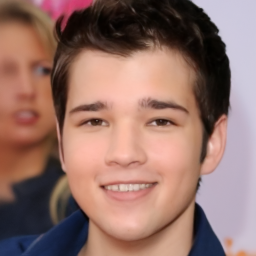}
		\end{minipage}}
	\end{subfigure}\hspace{-3mm}
	\begin{subfigure}[$\Pi$GDM]{
			\centering
			\begin{minipage}{0.115\linewidth}
				\includegraphics[width=\linewidth]{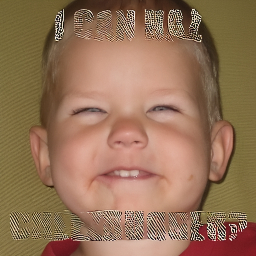}
				\includegraphics[width=\linewidth]{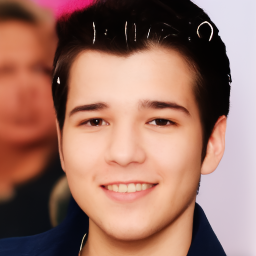}
		\end{minipage}}
	\end{subfigure}\hspace{-3mm}
	\begin{subfigure}[DMPS]{
			\centering
			\begin{minipage}{0.115\linewidth}
				\includegraphics[width=\linewidth]{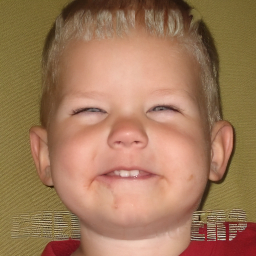}
				\includegraphics[width=\linewidth]{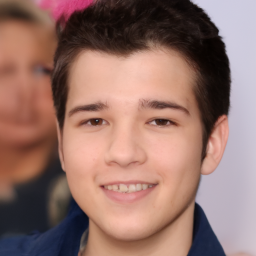}
		\end{minipage}}
	\end{subfigure}\hspace{-3mm}
	\begin{subfigure}[MCG]{
			\centering
			\begin{minipage}{0.115\linewidth}
				\includegraphics[width=\linewidth]{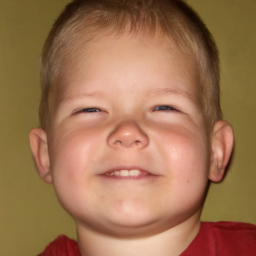}
				\includegraphics[width=\linewidth]{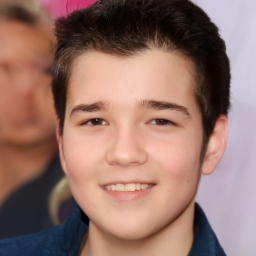}
		\end{minipage}}
	\end{subfigure}\vspace{-4mm}
    \caption{The results for Inpainting (Lolcat) and Inpainting (Lorem), which are presented in the first and second rows, respectively.}
	\label{fig:Inpainting}
\end{figure}

In this set of experiments, our method achieves the best performance in PSNR, SSIM and FID across almost all tasks, while exhibiting relatively lower performance in LPIPS for some tasks (see Table \ref{table2}). However, on average, our model still demonstrates excellent quality compared to other models, as illustrated in Figure \ref{fig:Box_1} and Figure \ref{fig:Inpainting}, which showcase qualitative examples of the inpainted images generated by different methods.

From the examples presented, it is evident that DDRM handles the special features very unnaturally. The generated images exhibit highly unnatural depiction of the person's eyes, and in some cases, there are noticeable traces of text shapes appearing on the person's chin, where was originally covered by text. As a result, the generated images produced by DDRM are quite far from real images and do not reflect reality accurately. $\Pi$GDM is only able to remove lighter masks and completely fail to remove masks that covered a significant amount of relevant information. DMPS performs somewhat better than $\Pi$GDM, as it is able to partially remove some masks, but still left noticeable traces in the generated images. DPS and MCG are unable to accurately depict images with sharp edges, as seen in the first image of the Inpainting (Box) task where the edge of the person's forehead, covered by the mask, appears twisted. In the Inpainting (Lolcat) and Inpainting (Lorem) tasks, the forehead and eyes in the first image, as well as the woman in the background of the second image, exhibit lower performance compared to the real image in the results produced by MCG.  Given these observations, it is clear that our model achieves better performance overall in the inpainting task compared to the other models. For the inpainting tasks, numerical results to identify the optimal parameter combinations are deferred to Figures \ref{fig:Box}, \ref{fig:Lolcat}, and \ref{fig:Lorem}.

\subsection{Runtime}
Since the runtime for diffusion models dominates the total runtime (as other computations are negligible), we use the number of Neural Function Evaluations (NFEs) as a criterion to estimate the runtime for different algorithms. Table \ref{table:nfes} reports the NFEs used in each algorithm, including the results for our proposed method with  unconditional sampling via DDPM and DDIM.
 \begin{table}[htbp]
    \centering
    \begin{tabular}{cccccccc}
        \hline
        Method & Our (DDPM) & Our (DDIM) & DDRM & DPS & DMPS & $\Pi$GDM & MCG
        \\ \hline
         NFEs & 1000 &20 & 20 & 1000 & 1000 & 1000 & 1000
         \\ \hline
    \end{tabular}\vspace{-2mm}
        \caption{The NFEs used in each algorithm.}
    \label{table:nfes}
\end{table}

In the numerical experiments provided in this paper, we use a pretrained unconditional score for all methods except for DDRM, which constructed a non-Markovian process to enable flexible skip-step sampling, similar to DDIM. 

In addition, the gradient of the neural network $\mathbf{S}_\theta(\bx_t,t)$ is required for computations in DPS, MCG, $\Pi$GDM, and our proposed method, which incurs a slightly higher computational cost. While DMPS does not require autogradients, it relies on a strong assumption that \(p(\mathbf{x}_0)/p(\mathbf{x}_t)\) remains constant. Runtime comparisons for the different algorithms are provided in Table \ref{table:runtime}. Notably, our method with DDIM requires only one-fifth the time of DDRM when using 20 NFEs.
 \begin{table}[htbp]
    \centering
    \begin{tabular}{cccccccc}
        \hline
        Method & Our(DDPM) & Our (DDIM) & DDRM & DPS & DMPS & $\Pi$GDM & MCG 
        \\ \hline
         Time & 79.176 &1.643  & 8.168 & 82.332 & 33.018 & 81.640 & 82.080 
         \\ \hline
    \end{tabular} \vspace{-2mm}
        \caption{The runtime (seconds) in each algorithm.}
    \label{table:runtime}
\end{table}

\section{Conclusion}\label{sec:conclusion}
In this paper, we propose a novel, problem-agnostic diffusion model called the MAP-based Guided Term Estimation method for inverse problems. First, we divide the conditional score function into two terms according to Bayes' rule: an unconditional score function (approximated by a pretrained score network) and a guided term, which is estimated using a novel MAP-based method that incorporates a Gaussian-type prior of natural images. This method enables us to better capture the intrinsic properties of the data, resulting in significantly improved performance. Numerical experiments validate the efficacy of our proposed method on a variety of linear inverse problems, such as super-resolution, inpainting, and denoising. Through extensive evaluations, we have demonstrated that our method achieves comparable performance to state-of-the-art diffusion models, including DDRM, DPS, $\Pi$GDM, DMPS and MCG.

\textbf{Limitations \& Future Works:}  
While this paper makes valuable contributions, there are some limitations that suggest future research directions. 
(1) Our approach relies on the assumption that the space of clean natural images is inherently smooth, which may result in the loss of certain features. 
(2) The numerical experiments in this work focus solely on linear inverse problems and do not extend to nonlinear cases.
(3) Our approach leverages unconditionally pretrained diffusion models to address conditional generation tasks (i.e., inverse problems). In this paper, we conduct experiments using pretrained diffusion models on the FFHQ 256×256 dataset. If score functions are not available, they must be trained in advance. 
(4) While the present works focus on inpainting with known measurement matrices, our diffusion-based framework readily extends to learning irregular inpainting masks through an iterative mask-inpainting procedure.
\section*{Acknowledgement}
The work of H.X. Liu was supported in part by NSFC 11901220, Interdisciplinary Research Program of HUST 2024JCYJ005, National Key Research and Development Program of China 2023YFC3804500.

\bibliographystyle{elsarticle-num}
\bibliography{smi_12_template}

\newpage
\appendix
\section{Derivation of \texorpdfstring{\eqref{eq:log_proba}}{3.1}}
\begin{proof} Note that $\hat {\mathbf{x}}$ is the estimation of $\mathbf{x}_0$, then $y =  H\mathbf{x}_0+z \approx H\hat{\mathbf{x}}+z$, 
where $z \sim  \mathcal N(0, \sigma_y^2I)$ is a Gaussian noise with a mean 0 and a standard deviation $\sigma_y$. Therefore $p(y|\mathbf{x}_t)$ obeys approximately $\mathcal N(H\hat{\mathbf{x}},\sigma^2_yI)$ and 
\begin{eqnarray*}
p(y|\mathbf{x}_t) \approx  \frac{1}{\left(\sqrt{2\pi\sigma_y^2}\right)^m}\exp{-\left(\frac{\|y-H\hat{\mathbf{x}}\|^2}{2\sigma^2_y}\right)},
\end{eqnarray*}
where $m$ is the dimension of $y$. Thus, we have the following approximation:
\begin{eqnarray*}
    \nabla_{\mathbf{x}_t} \log p(y|\mathbf{x}_t) \approx \frac{1}{\sigma_y^2} \left(H\frac{\partial \hat{\mathbf{x}}}{\partial \mathbf{x}_t}\right)^\top(y-H\hat{\mathbf{x}}).
\end{eqnarray*}
\end{proof}
\section{Numerical Comparisons}
\subsection{Super-Resolution}\label{subsec:additional_sr}

\begin{table}[H]
	\centering
	\resizebox{0.9\textwidth}{!}{
		\begin{tabular}{llllllllllll}
			\hline
			\multicolumn{1}{c}{}    & \multicolumn{1}{c}{\textbf{}} &  & \multicolumn{4}{c}{\textbf{SR($\times$2)}}  &     &      & \multicolumn{2}{c}{\textbf{SR($\times$8)}} \\ \cline{4-7} \cline{9-12} 
			\textbf{Dataset}        & \textbf{Method}               &  & PSNR↑          & LPIPS↓  & SSIM↑ & FID↓        & \textbf{} & PSNR↑            & LPIPS↓   & SSIM↑ & FID↓     \\ \hline
			\multirow{5}{*}{FFHQ}   & ours                          &  & \textbf{33.74} & \textbf{0.1964}  & \textbf{0.92}  & \textbf{31.03} &  &  \textbf{27.55} & \textbf{0.2734}  &\textbf{0.82} & \textbf{57.11} \\  	
			& DDRM                          &  & 30.58          & 0.2559 & 0.91 & 62.55         &           & 26.42         & 0.3173 & 0.75 & 77.27       \\ 
			& DPS                           &  & 28.34         & 0.2554    & 0.88  & 38.32      &           & 23.61         & 0.3023    & 0.78  & 71.89     \\
  			& $\Pi$GDM                          &  & 30.05          & 0.2309   & 0.90 & 43.33      &           & 24.03        & 0.2989   & 0.78 & 76.33     \\ 
            & DMPS                          &  & 29.88          & 0.2371    & 0.91 & 50.35     &           & 24.38         & 0.2837    & 0.73 & 65.49        \\ 
            & MCG                          &  & 29.48          & 0.2401   & 0.89   &40.29   &           & 25.49          & 0.2796    & 0.74  & 65.09      \\ \hline
			\multirow{5}{*}{CelebA-HQ} & ours     
          &  & \textbf{36.54} & \textbf{0.1620}  & \textbf{0.93} & \textbf{35.17} 	&  & \textbf{28.29} & \textbf{0.2884}  & \textbf{0.81}  & \textbf{57.88}  \\  
			& DDRM                          &  & 34.83          & 0.1974     & 0.91 & 51.62      &          & 27.74          & 0.2941     & 0.77 & 68.36        \\ 
            & DPS                           &  & 29.67          & 0.2637    & 0.87 & 69.54      &           & 25.26         & 0.3507    & 0.72 & 71.13       \\
 			& $\Pi$GDM                          &  & 31.95          & 0.1884     & 0.90 & 43.73    &           & 25.44         & 0.3119     & 0.70 & 78.36        \\ 
			& DMPS                          &  & 31.60          & 0.1934   & 0.90 & 44.22       &           & 25.42         & 0.2991   & 0.74 & 67.57        \\ 
            & MCG                          &  & 27.91          & 0.2451  & 0.89  & 50.49  &           & 24.59         & 0.3036      & 0.70          & 79.94      \\ \hline 
		\end{tabular}
	}
    \caption{Quantitative comparisons (PSNR (dB), LPIPS, SSIM and FID) of $2\times$ and $8\times$ super-resolution tasks on the FFHQ 256$\times$256-1k validation dataset and the CelebA-HQ 256$\times$256-1k validation dataset, respectively. The pre-trained model used in our proposed method, as well as in DPS, \(\Pi\)GDM, DMPS, and MCG, is trained on the FFHQ dataset. For DDRM, we utilize the original code provided by the authors.}
	\label{table1-3}
\end{table}
\begin{figure}[H]
	\centering
	\begin{subfigure}[Input]{
			\begin{minipage}{0.115\textwidth}
				\centering
				\includegraphics[width=\linewidth]{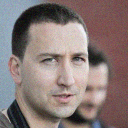}
				\includegraphics[width=\linewidth]{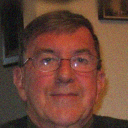}
		\end{minipage}}
	\end{subfigure}\hspace{-3mm}
	\begin{subfigure}[GT]{
			\begin{minipage}{0.115\textwidth}
				\centering
				\includegraphics[width=\linewidth]{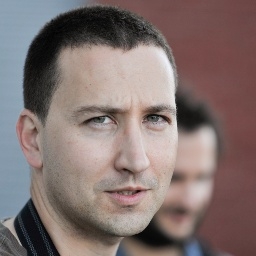}
				\includegraphics[width=\linewidth]{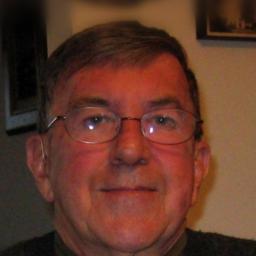}
		\end{minipage}}
	\end{subfigure}\hspace{-3mm}
	\begin{subfigure}[Our]{
			\begin{minipage}{0.115\textwidth}
				\centering
				\includegraphics[width=\linewidth]{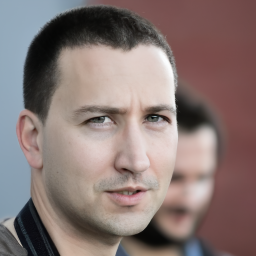}
				\includegraphics[width=\linewidth]{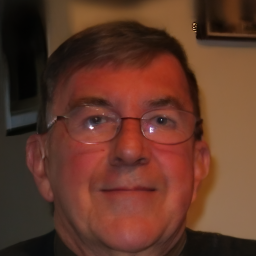}
		\end{minipage}}
	\end{subfigure}\hspace{-3mm}
	\begin{subfigure}[DDRM]{
			\begin{minipage}{0.115\textwidth}
				\centering
				\includegraphics[width=\linewidth]{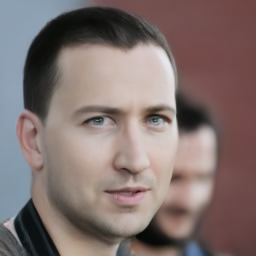}
				\includegraphics[width=\linewidth]{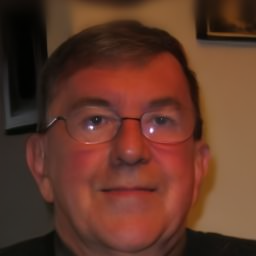} 
		\end{minipage}}
	\end{subfigure}\hspace{-3mm}
	\begin{subfigure}[DPS]{
			\begin{minipage}{0.115\textwidth}
				\centering
				\includegraphics[width=\linewidth]{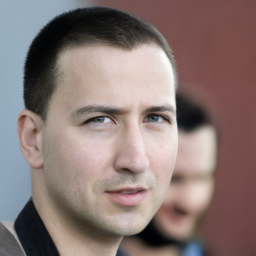}
				\includegraphics[width=\linewidth]{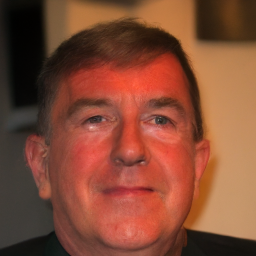}
		\end{minipage}}
	\end{subfigure}\hspace{-3mm}
	\begin{subfigure}[$\Pi$GDM]{
			\begin{minipage}{0.115\textwidth}
				\centering
				\includegraphics[width=\linewidth]{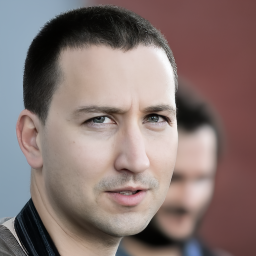}
				\includegraphics[width=\linewidth]{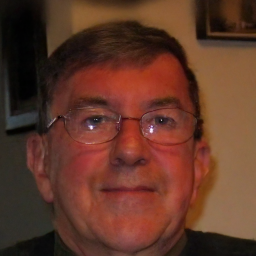}
		\end{minipage}}
	\end{subfigure}\hspace{-3mm}
	\begin{subfigure}[DMPS]{
			\begin{minipage}{0.115\textwidth}
				\centering
				\includegraphics[width=\linewidth]{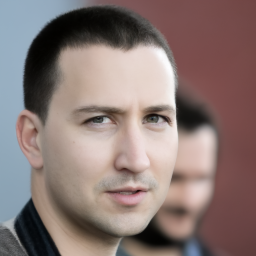}
				\includegraphics[width=\linewidth]{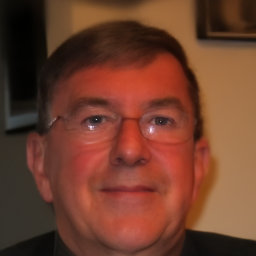}
		\end{minipage}}
	\end{subfigure}\hspace{-3mm}
	\begin{subfigure}[MCG]{
			\begin{minipage}{0.115\textwidth}
				\centering
				\includegraphics[width=\linewidth]{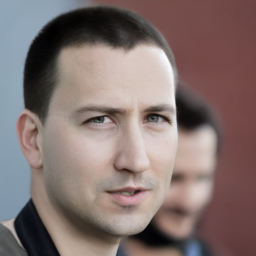}
				\includegraphics[width=\linewidth]{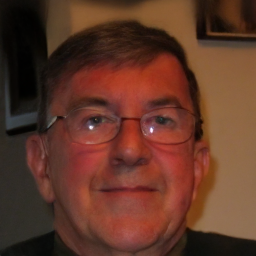}
		\end{minipage}}
	\end{subfigure}\vspace{-4mm}
    \caption{The results for super-resolution $(2\times)$. The first column is the input image, the second one is  the Ground Truth (denoted as GT) and the third to eighth columns are our proposed method, DDRM, DPS, $\Pi$GDM, DMPS and MCG, respectively.}
	\label{fig:SR1}
\end{figure}

\begin{figure}[H]
	\centering
	\begin{subfigure}[Input]{
			\begin{minipage}{0.11\textwidth}
				\centering
				\includegraphics[width=\linewidth]{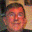}
		\end{minipage}}
	\end{subfigure}\hspace{-2mm}
	\begin{subfigure}[GT]{
			\begin{minipage}{0.11\textwidth}
				\centering
				\includegraphics[width=\linewidth]{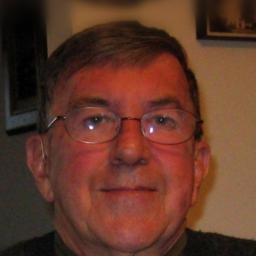}
		\end{minipage}}
	\end{subfigure}\hspace{-2mm}
	\begin{subfigure}[Our]{
			\begin{minipage}{0.11\textwidth}
				\centering
				\includegraphics[width=\linewidth]{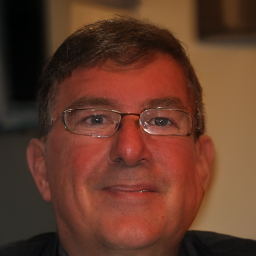}
		\end{minipage}}
	\end{subfigure}\hspace{-2mm}
	\begin{subfigure}[DDRM]{
			\begin{minipage}{0.11\textwidth}
				\centering
				\includegraphics[width=\linewidth]{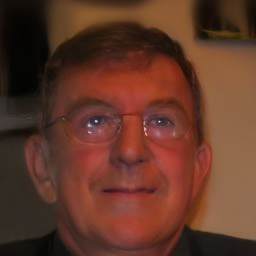}
		\end{minipage}}
	\end{subfigure}\hspace{-2mm}
	\begin{subfigure}[DPS]{
			\begin{minipage}{0.11\textwidth}
				\centering
				\includegraphics[width=\linewidth]{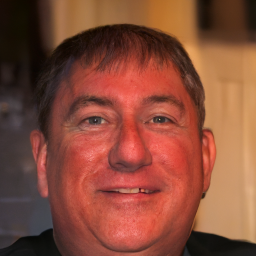}
		\end{minipage}}
	\end{subfigure}\hspace{-2mm}
	\begin{subfigure}[$\Pi$GDM]{
			\begin{minipage}{0.11\textwidth}
				\centering
				\includegraphics[width=\linewidth]{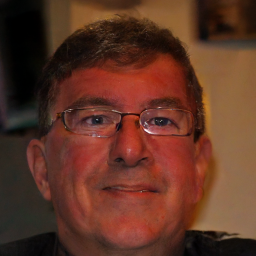}
		\end{minipage}}
	\end{subfigure}\hspace{-2mm}
	\begin{subfigure}[DMPS]{
			\begin{minipage}{0.11\textwidth}
				\centering
				\includegraphics[width=\linewidth]{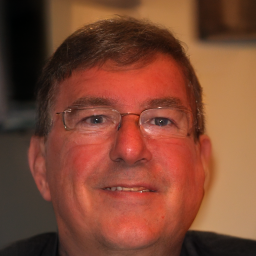}
		\end{minipage}}
	\end{subfigure}\hspace{-2mm}
	\begin{subfigure}[MCG]{
			\begin{minipage}{0.11\textwidth}
				\centering
				\includegraphics[width=\linewidth]{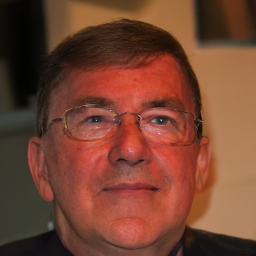}
		\end{minipage}}
	\end{subfigure}\vspace{-4mm}
    \caption{The results for super-resolution $(8\times)$. The first column is the input image, the second one is  the Ground Truth (denoted as GT) and the third to eighth columns are our proposed method, DDRM, DPS, $\Pi$GDM, DMPS and MCG, respectively.}
	\label{fig:SR2}
\end{figure}

Table \ref{table1-3} shows that our model significantly outperforms current state-of-the-art diffusion methods on the FFHQ and CelebA-HQ datasets. This quantitative advantage of the $2\times$ super-resolution task is reflected in the visual quality presented in Figure \ref{fig:SR1}. Existing methods tend to yield either oversmoothed results that lack detail or unnatural reconstructions with artifacts in complex regions like eyes and glasses. Our method overcomes these limitations, generating images with enhanced detail richness and superior visual fidelity. Specifically, it better preserves facial features and eye texture in the first example and achieves a more accurate structural reconstruction of glasses in the second row.

The 8$\times$ super-resolution task presents a substantially greater challenge, as the extreme upsampling ratio exacerbates information loss and reconstruction complexity, particularly for high-frequency and structural details. As a result, all evaluated methods, as demonstrated in Figure \ref{fig:SR2}, failed to deliver satisfactory performance.
\subsection{Denoising}\label{subsec:additinal_denoising}

\begin{table}[!htp]
	\centering
	\resizebox{1\textwidth}{!}{
		\begin{tabular}{lllllllllllllllll}
			\hline
			\multicolumn{1}{c}{}    & \multicolumn{1}{c}{\textbf{}} &  & \multicolumn{4}{c}{\textbf{DENOISE (0.05)}}  &           & \multicolumn{4}{c}{\textbf{DENOISE (0.1)}}  &       & \multicolumn{4}{c}{\textbf{DENOISE (0.2)}} \\ \cline{4-7} \cline{9-12} \cline{14-17}
			\textbf{Dataset}        & \textbf{Method}               &  & PSNR↑          & LPIPS↓  & SSIM↑ & FID↓        &  & PSNR↑            & LPIPS↓   & SSIM↑ & FID↓ & & PSNR↑ & LPIPS↓ & SSIM↑ & FID↓    \\ \hline
            
			\multirow{6}{*}{FFHQ}   & ours     
                     &  & \textbf{35.67} & \textbf{0.1721} & \textbf{0.92} & \textbf{33.69} &           & \textbf{35.37}   & \textbf{0.2074}  & \textbf{0.92}   & \textbf{36.02} &	& \textbf{32.67} & \textbf{0.2145} & \textbf{0.90} & \textbf{37.53} \\ 
			& DDRM                          &  & 34.13          & 0.2337 & 0.91 & 39.85         &           & 33.45           & 0.2482     & 0.88 & 46.05   &  & 30.22          & 0.2622 & 0.85 & 54.75      \\ 
			& DPS                           &  & 30.21         & 0.2182    & 0.90  & 42.34      &           & 29.66            & 0.2359      & 0.87 & 47.04    &  & 29.22          & 0.2478    & 0.85  & 51.03    \\ 
			& $\Pi$GDM                          &  & 33.37          & 0.2171   & 0.90 & 33.93       &           & 31.27            & 0.2387       & 0.87 & 36.70     &  & 27.25          & 0.2504   & 0.84 & 38.90\\ 	

			& DMPS                          &  & 31.04         & 0.2183   & 0.88 & 34.87      &           & 30.46            & 0.2371    & 0.87 & 36.89     &  & 29.69          & 0.2415    & 0.82 & 43.26      \\ 
            & MCG                          &  & 32.43          & 0.1919  & 0.90          & 37.20        &           & 31.66            & 0.2091      & 0.84            & 40.54	 &  & 29.52          & 0.2353  & 0.82         & 46.08    
      \\ \hline
			\multirow{6}{*}{CelebA-HQ} & ours   
                      &  & \textbf{38.19} & \textbf{0.1522}  & \textbf{0.93} & \textbf{33.14}&           & \textbf{37.69}   & \textbf{0.1593}  & \textbf{0.93}   & \textbf{34.85}	&  & \textbf{34.61} & \textbf{0.1770}   & \textbf{0.91} & \textbf{35.11} \\ 
			& DDRM                          &  & 35.98          & 0.1648     & 0.92 & 37.75      &           & 35.21            & 0.1663    & 0.91 & 47.18     &  & 31.22          & 0.2165     & 0.89 & 53.55        \\ 
			& DPS                           &  & 30.57          & 0.2017    & 0.92 & 37.53      &           & 30.44            & 0.2151     & 0.91 & 42.57  &  & 29.93         & 0.2510   & 0.87 & 43.67    \\ 
            & $\Pi$GDM                          &  & 31.27          & 0.1665     & 0.91 & 33.79     &           & 31.01            & 0.1681   & 0.90 & 37.57     &  & 29.76          & 0.1884     & 0.86 & 43.71      \\ 
			& DMPS                          &  & 31.95          & 0.1762   & 0.92 & 37.11       &           & 30.62           & 0.1771    & 0.91 & 45.37  &  & 29.73          & 0.2051   & 0.89 & 48.16      \\
            & MCG                          &  & 33.72         & 0.1694  & 0.92         & 38.34       &           & 32.28           & 0.1852   & 0.91           & 41.88	 &  & 29.17		& 0.2224   & 0.86  & 42.30
         \\ \hline
		\end{tabular}
	}
    \caption{Quantitative comparisons (PSNR (dB), LPIPS, SSIM and FID) for denoising tasks with noise levels $\sigma\in\{0.05,0.1,0.2\}$ on the FFHQ 256$\times$256-1k validation dataset and the CelebA-HQ 256$\times$256-1k validation dataset, respectively. The pre-trained model used in our proposed method, as well as in DPS, \(\Pi\)GDM, DMPS, and MCG, is trained on the FFHQ dataset. For DDRM, we utilize the original code provided by the authors.}
	\label{table1-1}
\end{table}
To systematically evaluate performance across denoising methods, we conducted comparative experiments at multiple noise levels ($\sigma\in\{0.05, 0.1, 0.2\}$), using four quantitative metrics: PSNR, SSIM, LPIPS, and FID. Detailed results are presented in Table \ref{table1-1}.

\begin{figure}[htbp]
	\centering
	\begin{subfigure}[Input]{
			\begin{minipage}{0.115\textwidth}
				\centering
				\includegraphics[width=\linewidth]{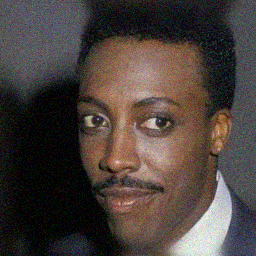}
				\includegraphics[width=\linewidth]{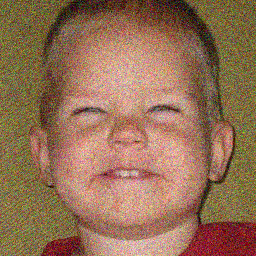} 
                \includegraphics[width=\linewidth]{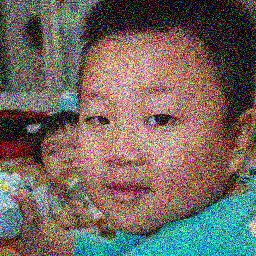} 
		\end{minipage}}
	\end{subfigure}\hspace{-3mm}
	\begin{subfigure}[GT]{
			\begin{minipage}{0.115\textwidth}
				\centering
				\includegraphics[width=\linewidth]{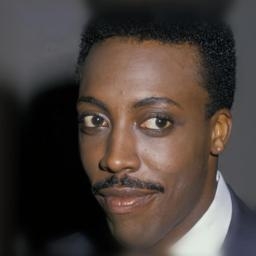}
				\includegraphics[width=\linewidth]{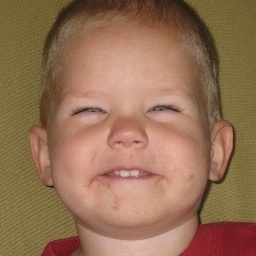}
                \includegraphics[width=\linewidth]{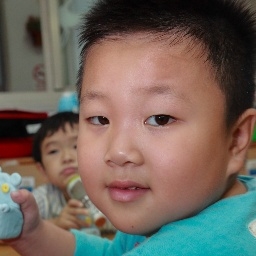}
		\end{minipage}}
	\end{subfigure}\hspace{-3mm}
	\begin{subfigure}[Our]{
			\begin{minipage}{0.115\textwidth}
				\centering
				\includegraphics[width=\linewidth]{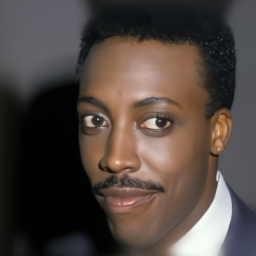}
				\includegraphics[width=\linewidth]{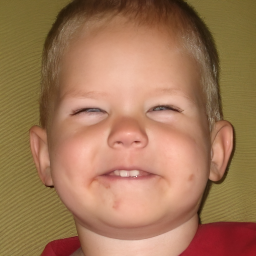}
                \includegraphics[width=\linewidth]{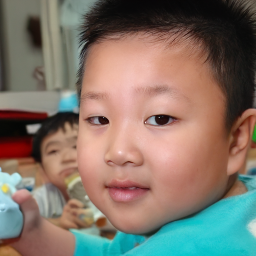}
		\end{minipage}}
	\end{subfigure}\hspace{-3mm}
	\begin{subfigure}[DDRM]{
			\begin{minipage}{0.115\textwidth}
				\centering
				\includegraphics[width=\linewidth]{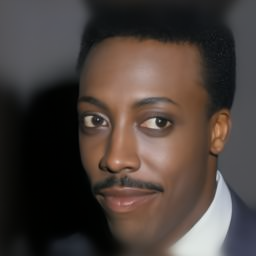}
				\includegraphics[width=\linewidth]{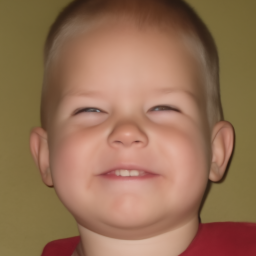}
                \includegraphics[width=\linewidth]{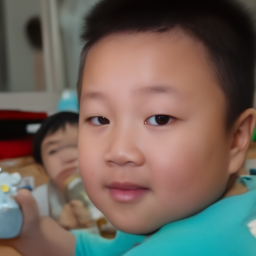}
		\end{minipage}}
	\end{subfigure}\hspace{-3mm}
	\begin{subfigure}[DPS]{
			\begin{minipage}{0.115\textwidth}
				\centering
				\includegraphics[width=\linewidth]{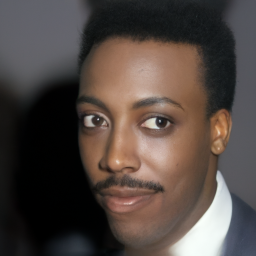}
				\includegraphics[width=\linewidth]{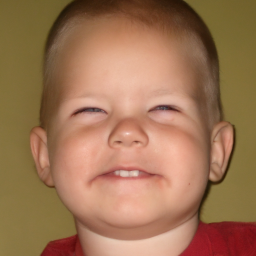}
                \includegraphics[width=\linewidth]{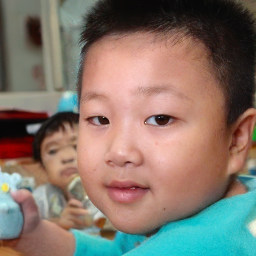}
		\end{minipage}}
	\end{subfigure}\hspace{-3mm}
	\begin{subfigure}[$\Pi$GDM]{
			\begin{minipage}{0.115\textwidth}
				\centering
				\includegraphics[width=\linewidth]{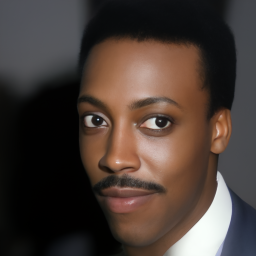}
				\includegraphics[width=\linewidth]{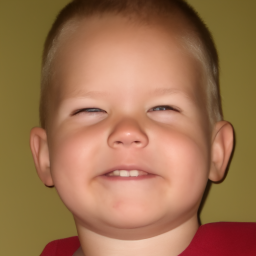}
                \includegraphics[width=\linewidth]{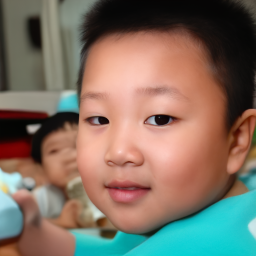}
		\end{minipage}}
	\end{subfigure}\hspace{-3mm}
	\begin{subfigure}[DMPS]{
			\begin{minipage}{0.115\textwidth}
				\centering
				\includegraphics[width=\linewidth]{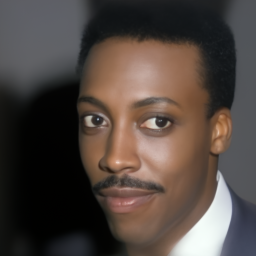}
				\includegraphics[width=\linewidth]{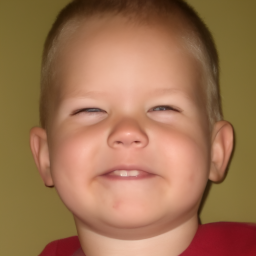}
                \includegraphics[width=\linewidth]{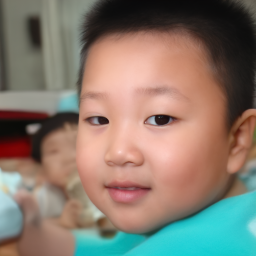}
		\end{minipage}}
	\end{subfigure}\hspace{-3mm}
	\begin{subfigure}[MCG]{
			\begin{minipage}{0.115\textwidth}
				\centering
				\includegraphics[width=\linewidth]{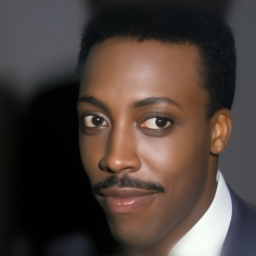}
				\includegraphics[width=\linewidth]{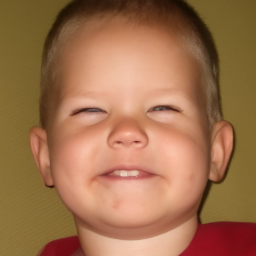}
                \includegraphics[width=\linewidth]{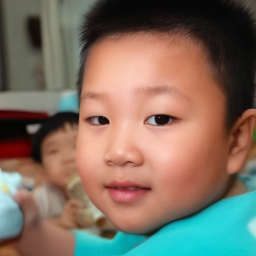}
		\end{minipage}}
	\end{subfigure}\vspace{-4mm}
    \caption{The results for denoising.  The measurements are subjected to Gaussian noise contamination with standard deviations of $\sigma=0.05$, $0.1$, and $0.2$ for the data in the first, second, and third rows, respectively. GT stands for Ground Truth.}
	\label{fig:Denoise2}
\end{figure}

The qualitative results in Figure \ref{fig:Denoise2} reveal common failure modes among existing methods. Most exhibit significant over-smoothing, leading to a loss of detail, while others suffer from excessive color saturation and distortion. The DPS method introduces noise artifacts due to over-sharpening, as seen in the unnatural facial details of the third image.

In contrast, our method achieves a superior balance, as reflected in its strong quantitative metrics. Qualitatively, it excels in key areas: the first image demonstrates precise restoration of the auricle contour and micro-textures; the second one preserves high-frequency details around the mouth and eyes; and the third one clearly recovers the child's face in the background, showing an optimal trade-off between noise suppression and detail fidelity.

\section{Robust Analysis}
\begin{figure}[!htp]
	\begin{center}
			\begin{minipage}[b]{0.32\linewidth}				\includegraphics[width=1.05\textwidth]{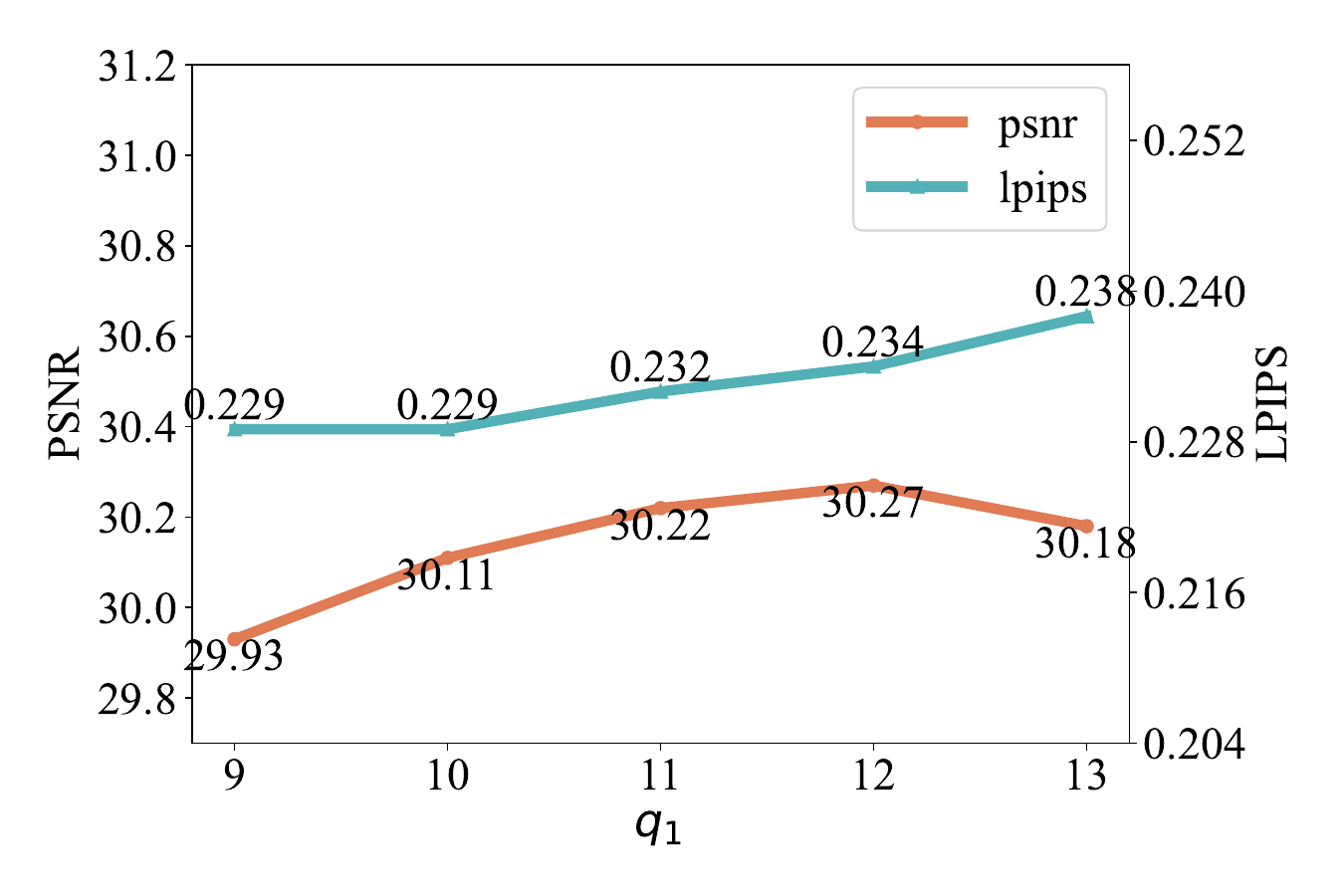}\vspace{4pt}
				\includegraphics[width=1.05\textwidth]{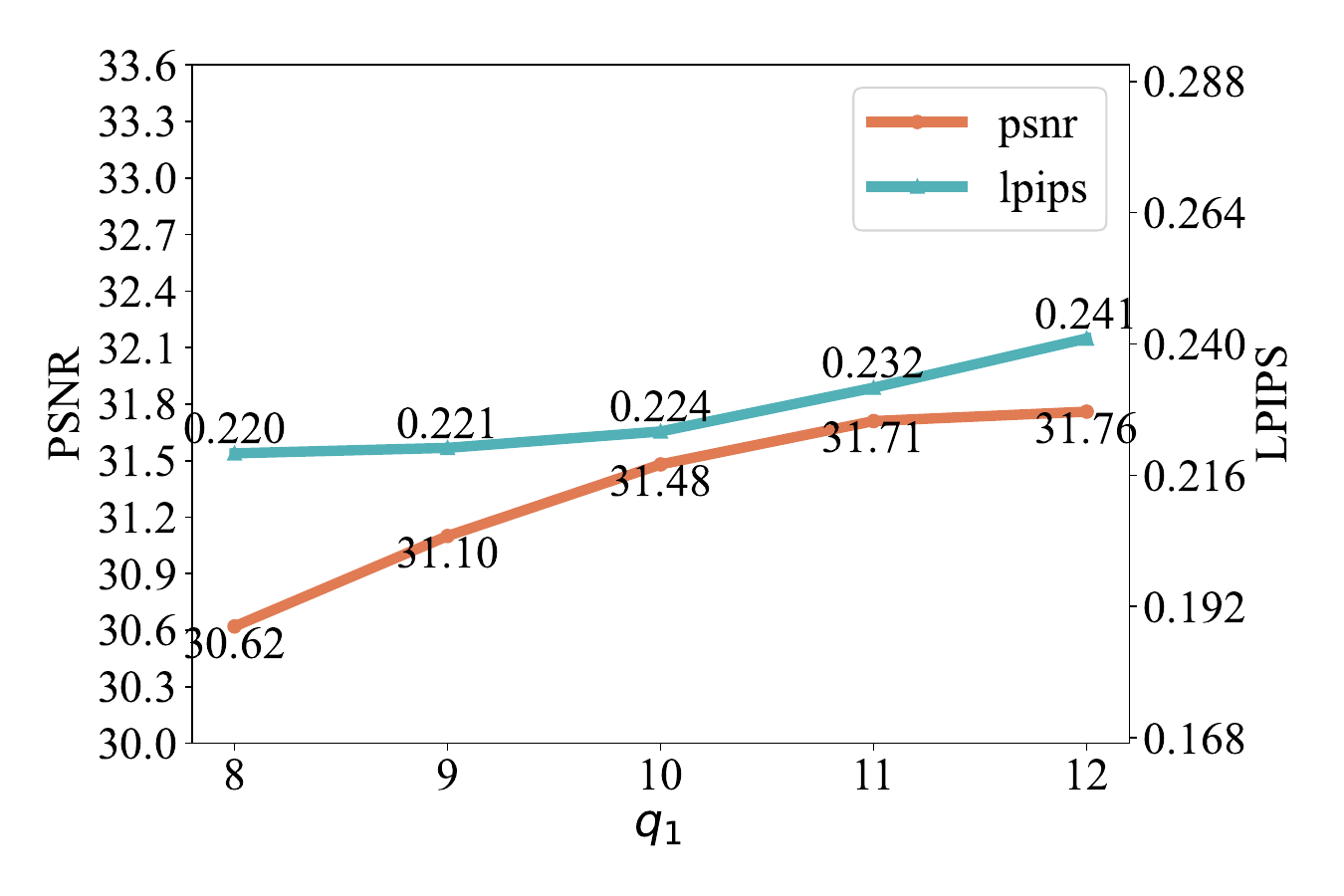}
		\end{minipage}
			\begin{minipage}[b]{0.32\linewidth}
				\includegraphics[width=1.05\textwidth]{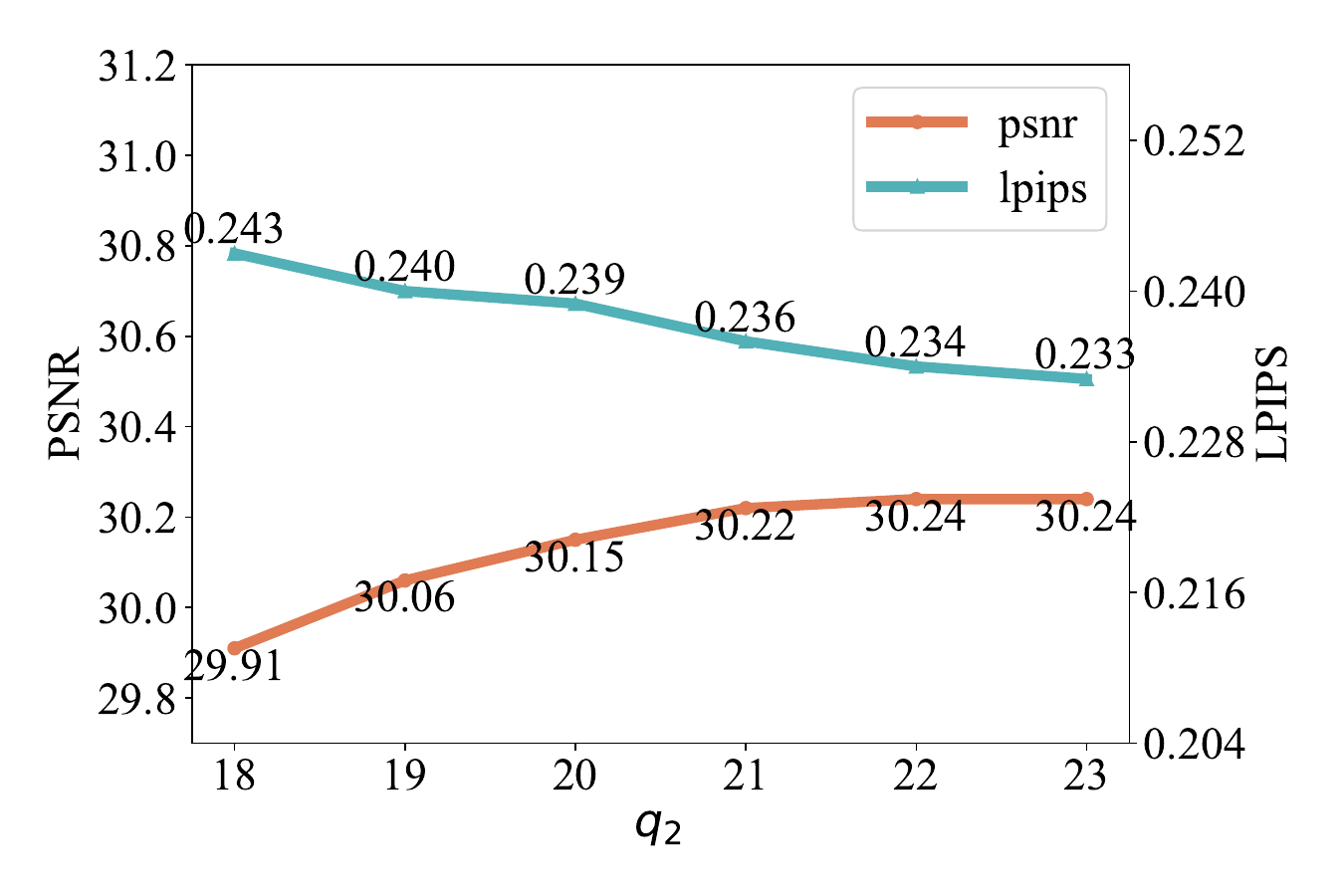}\vspace{4pt}
				\includegraphics[width=1.05\textwidth]{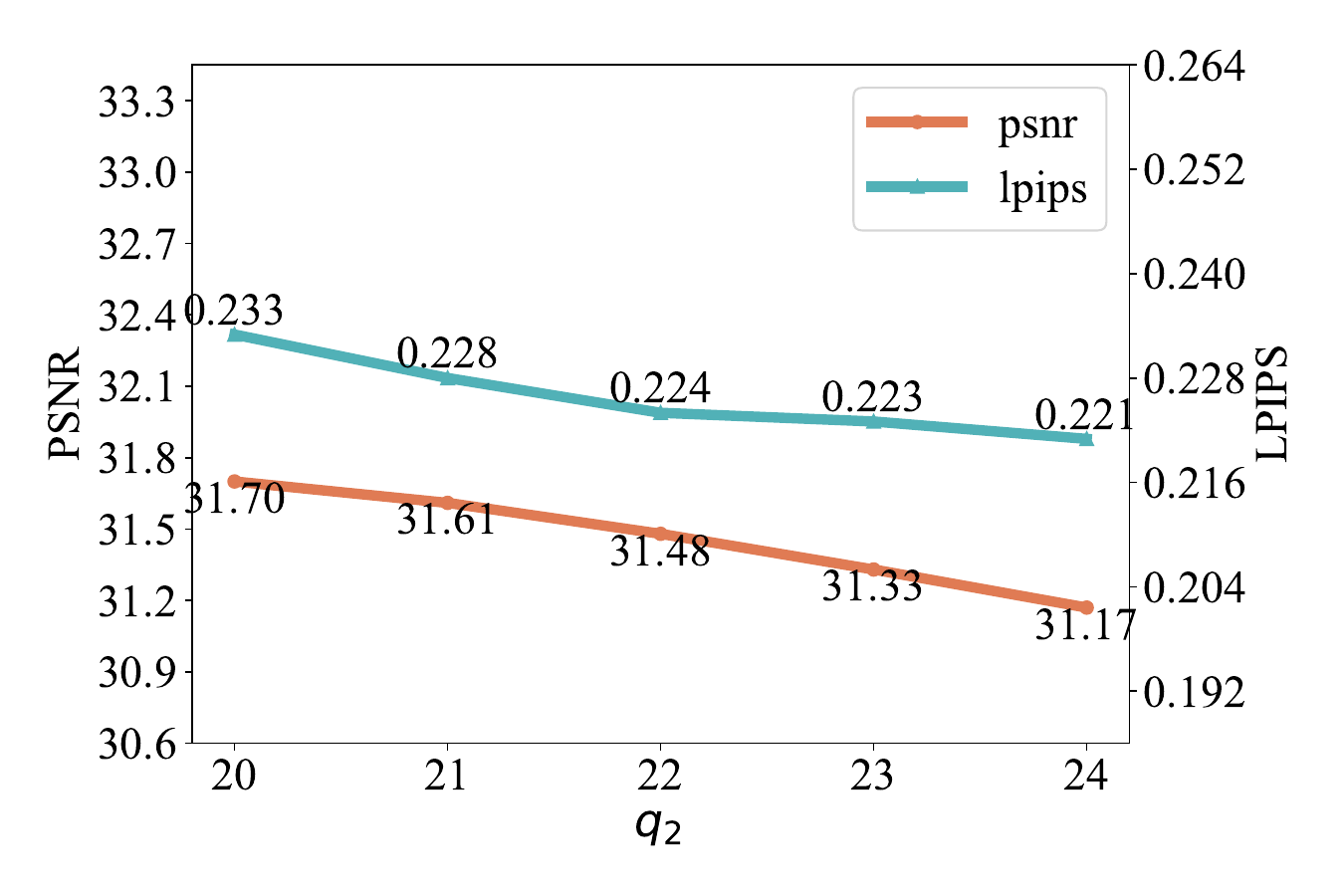}
		\end{minipage}
			\begin{minipage}[b]{0.32\linewidth}
				\includegraphics[width=1.05\textwidth]{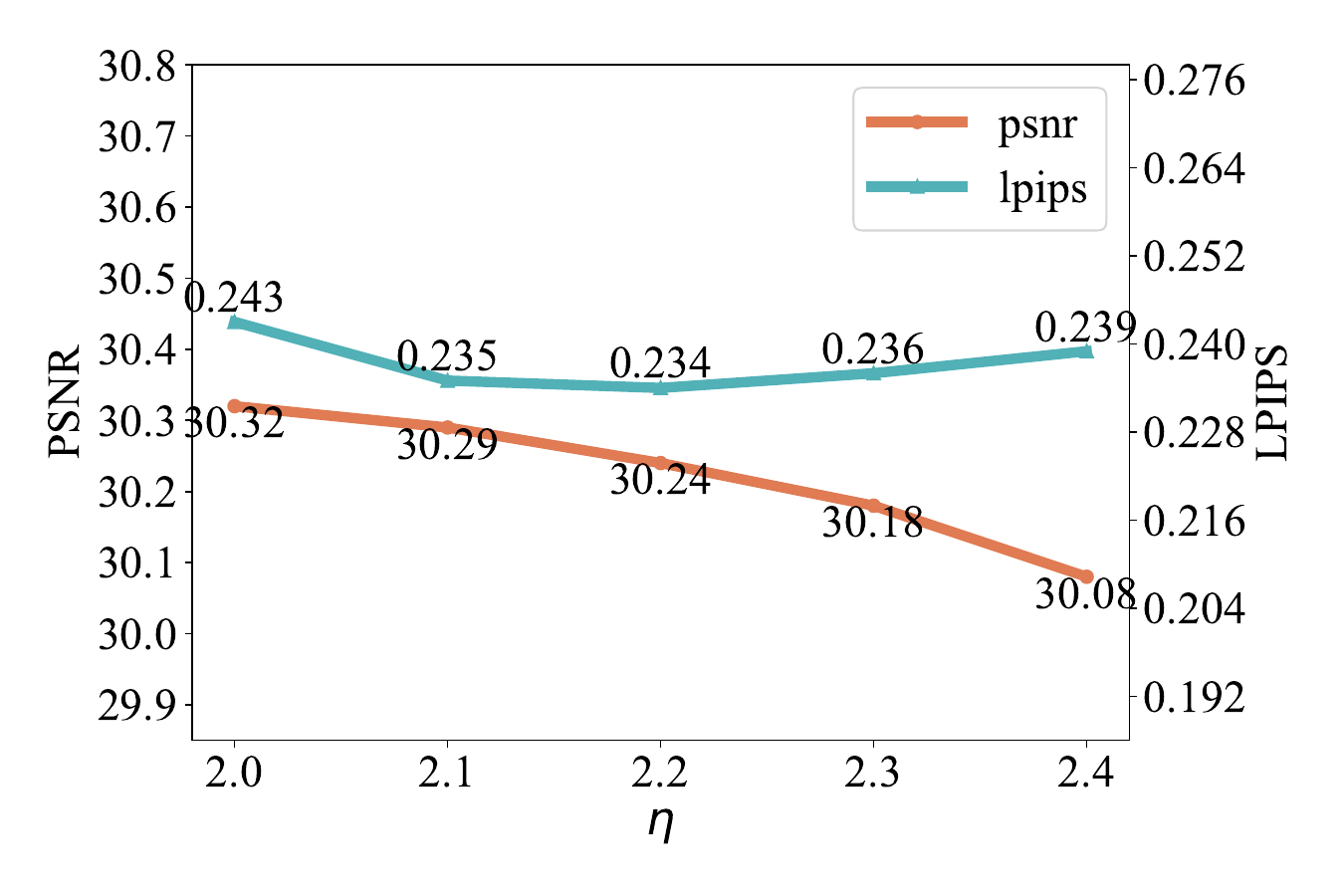}\vspace{4pt}
				\includegraphics[width=1.05\textwidth]{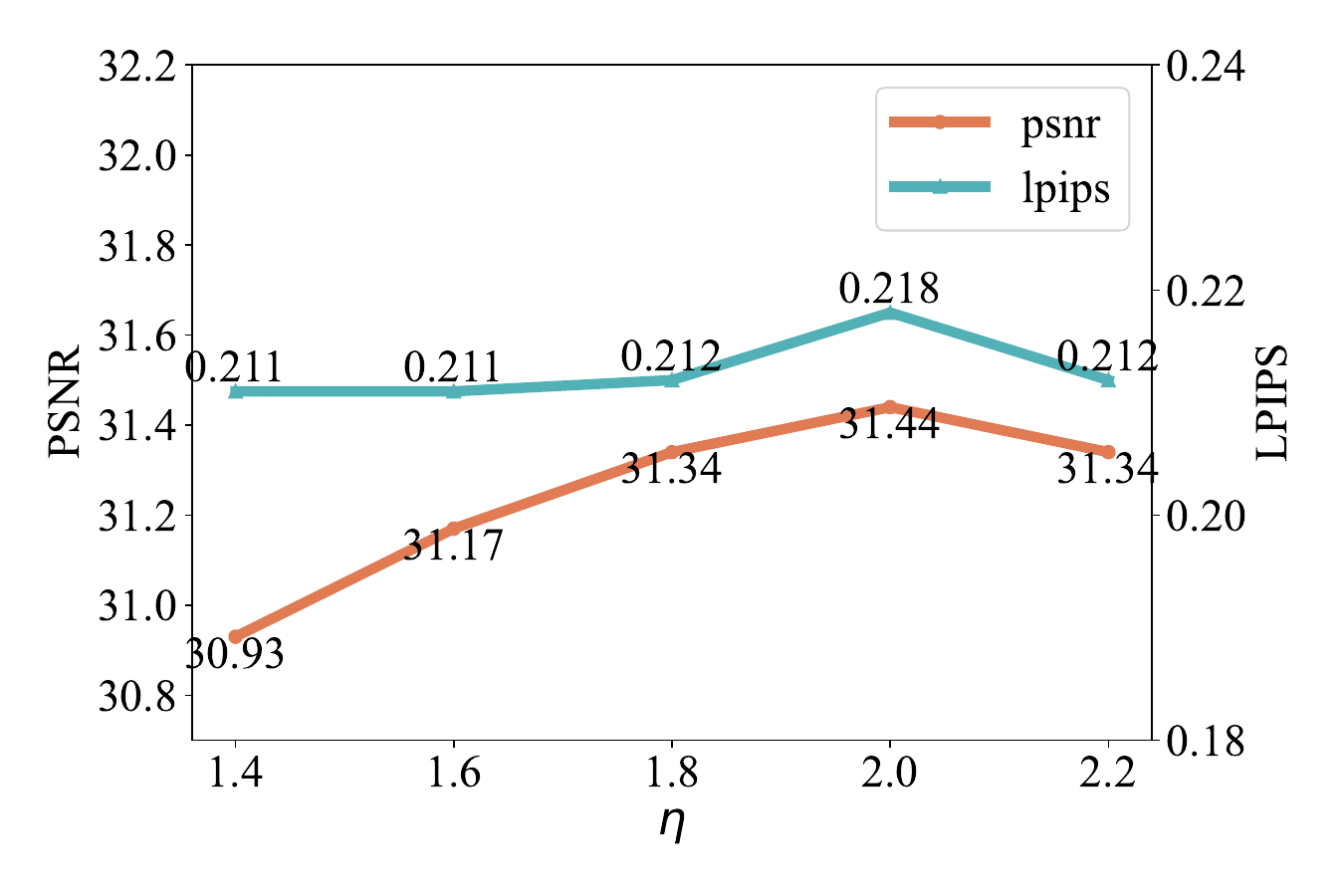}
		\end{minipage}
	\end{center}\vspace{-10mm}
    \caption{The robustness analysis results for denoising on the FFHQ 256×256-1k and the CelebA-HQ 256×256-1k validation sets with different parameters. Columns 1-3 are the plots of PSNR (blue) and LPIPS (orange) values versus the changes of parameters $q_1$, $q_2$, and $\eta$, with the other two fixed.}
		\label{fig:Denoise1}
\end{figure}

\begin{figure}[!htp]
	\begin{center}
			\begin{minipage}[b]{0.32\linewidth}
				\includegraphics[width=1.05\textwidth]{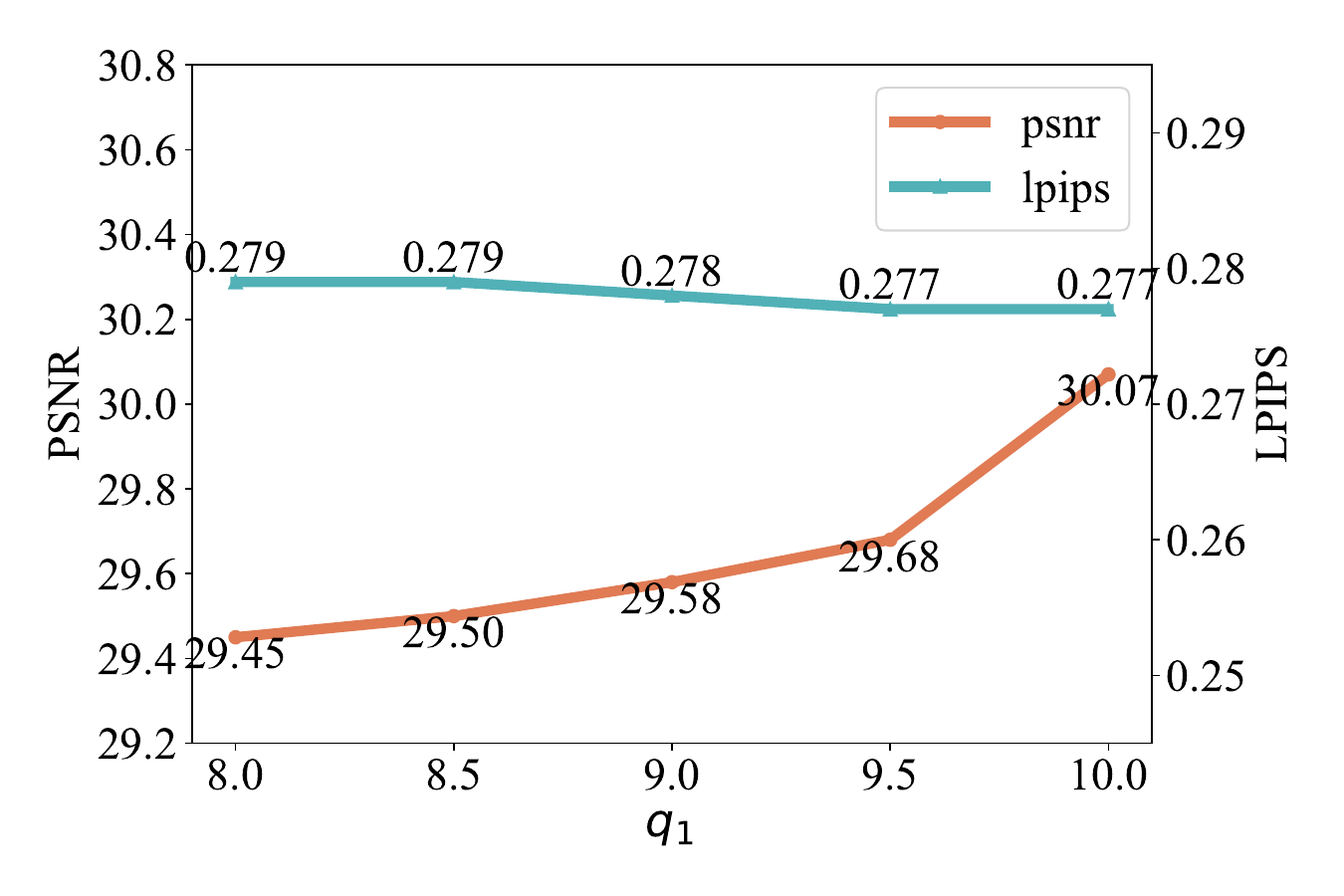}
				\includegraphics[width=1.05\textwidth]{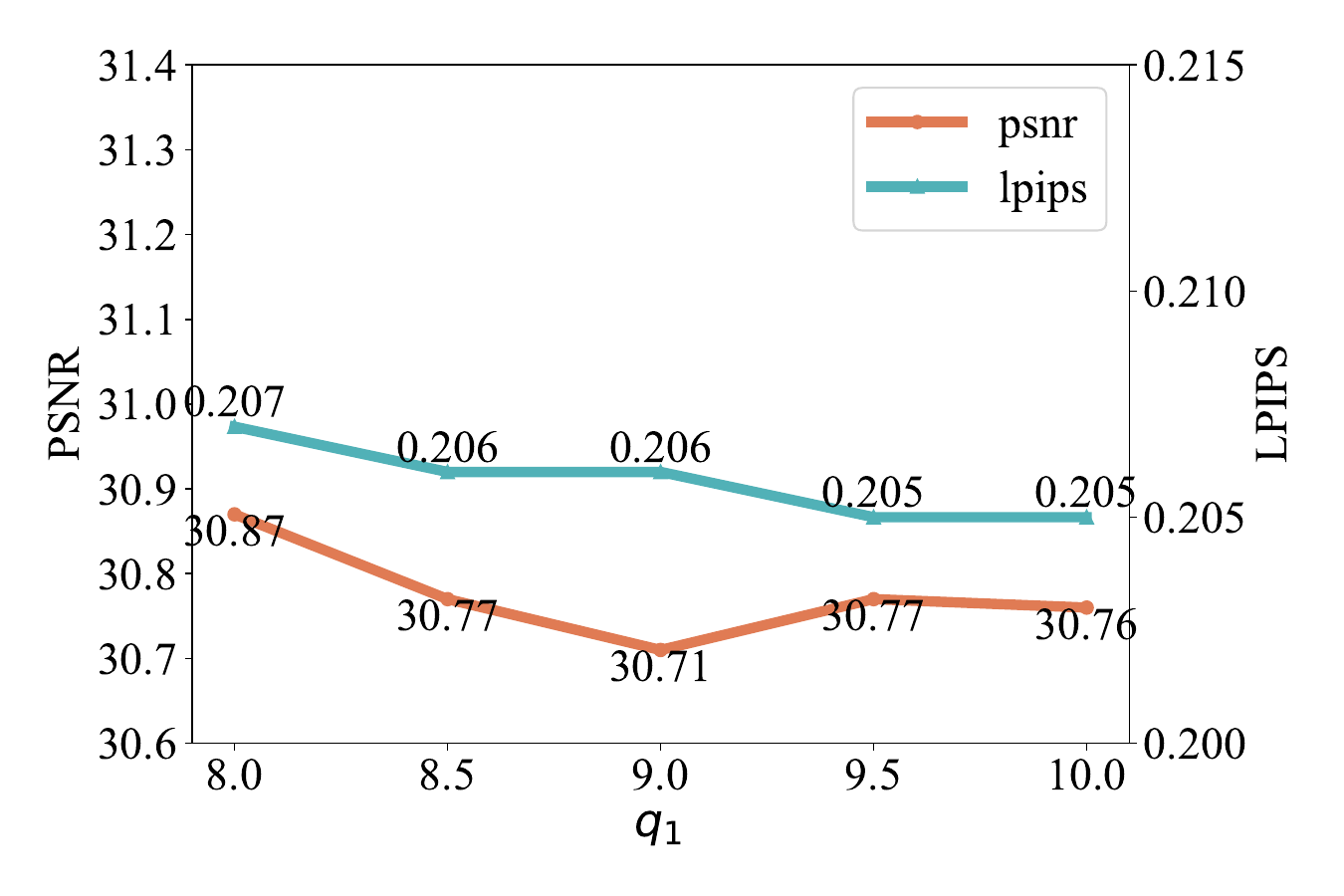}
		\end{minipage}
			\begin{minipage}[b]{0.32\linewidth}
				\includegraphics[width=1.05\textwidth]{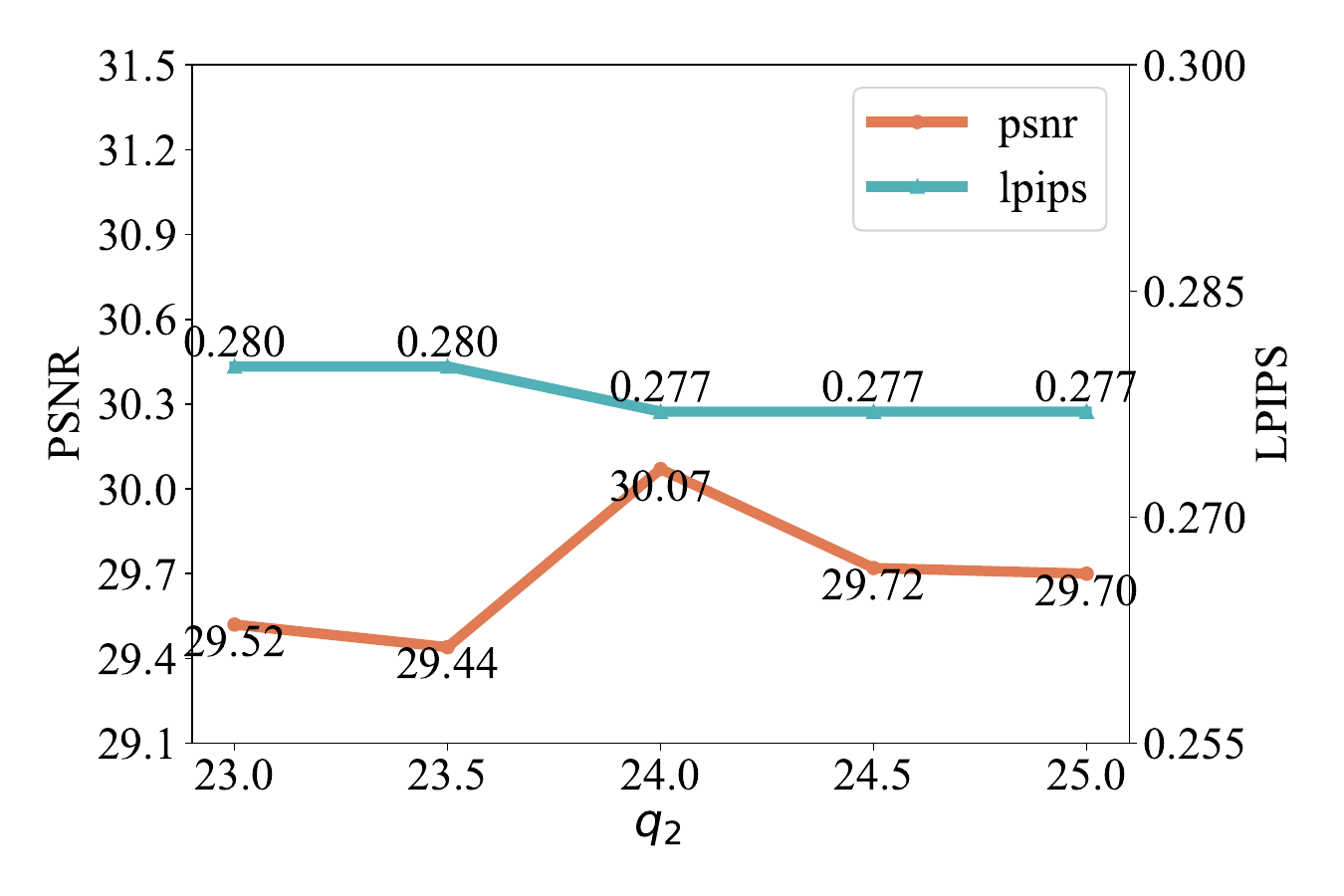}
				\includegraphics[width=1.05\textwidth]{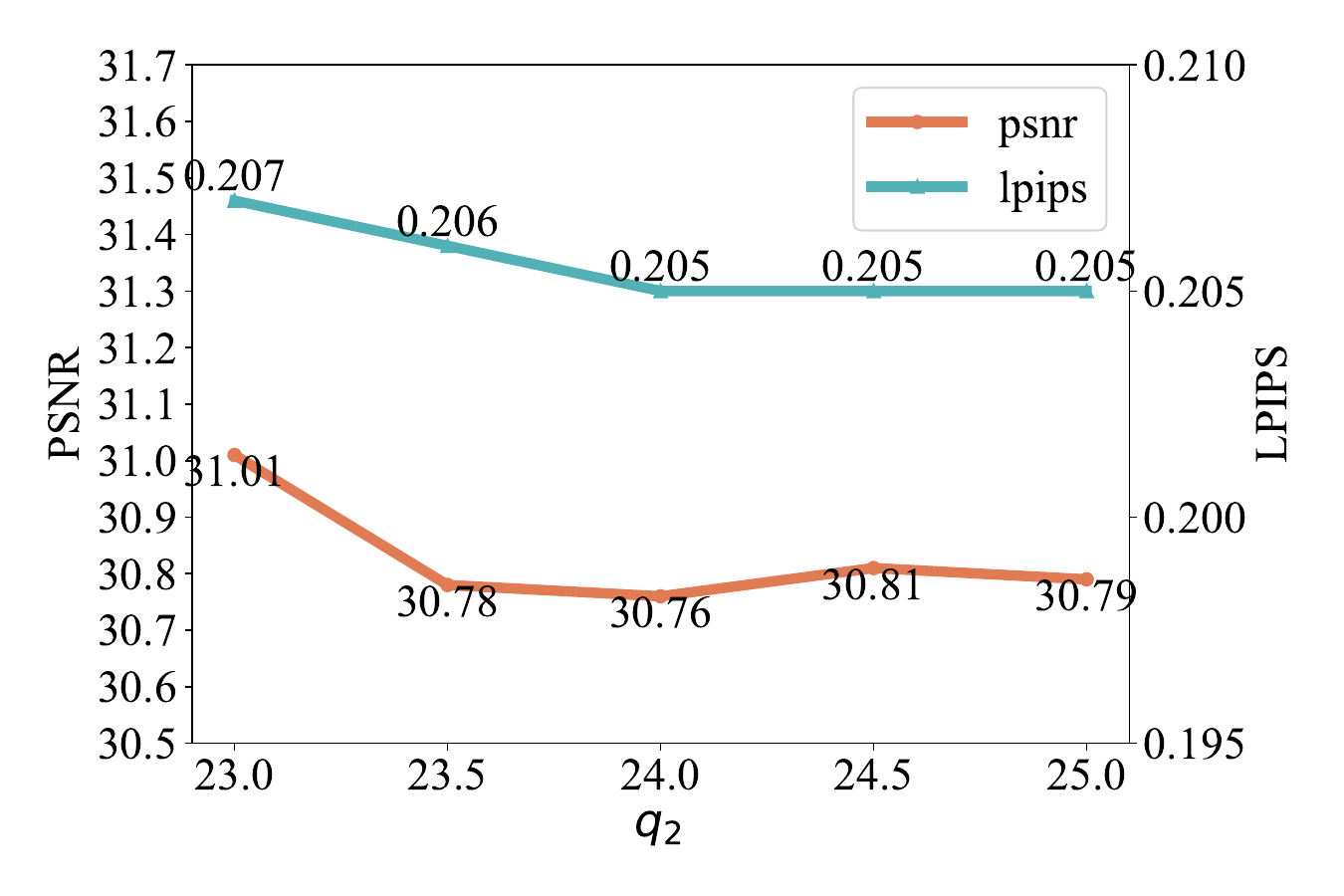}
		\end{minipage}
			\begin{minipage}[b]{0.32\linewidth}
				\includegraphics[width=1.05\textwidth]{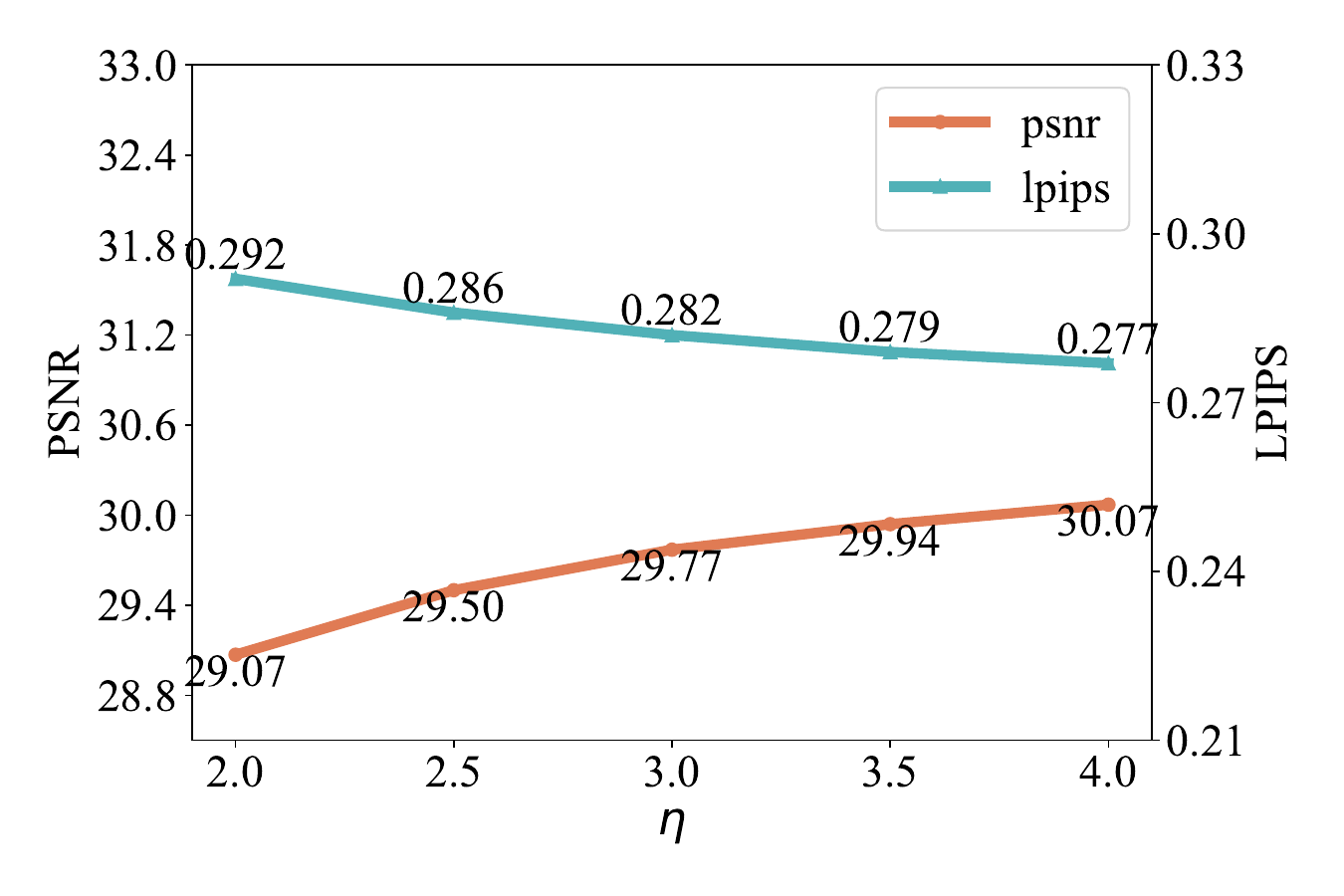}
				\includegraphics[width=1.05\textwidth]{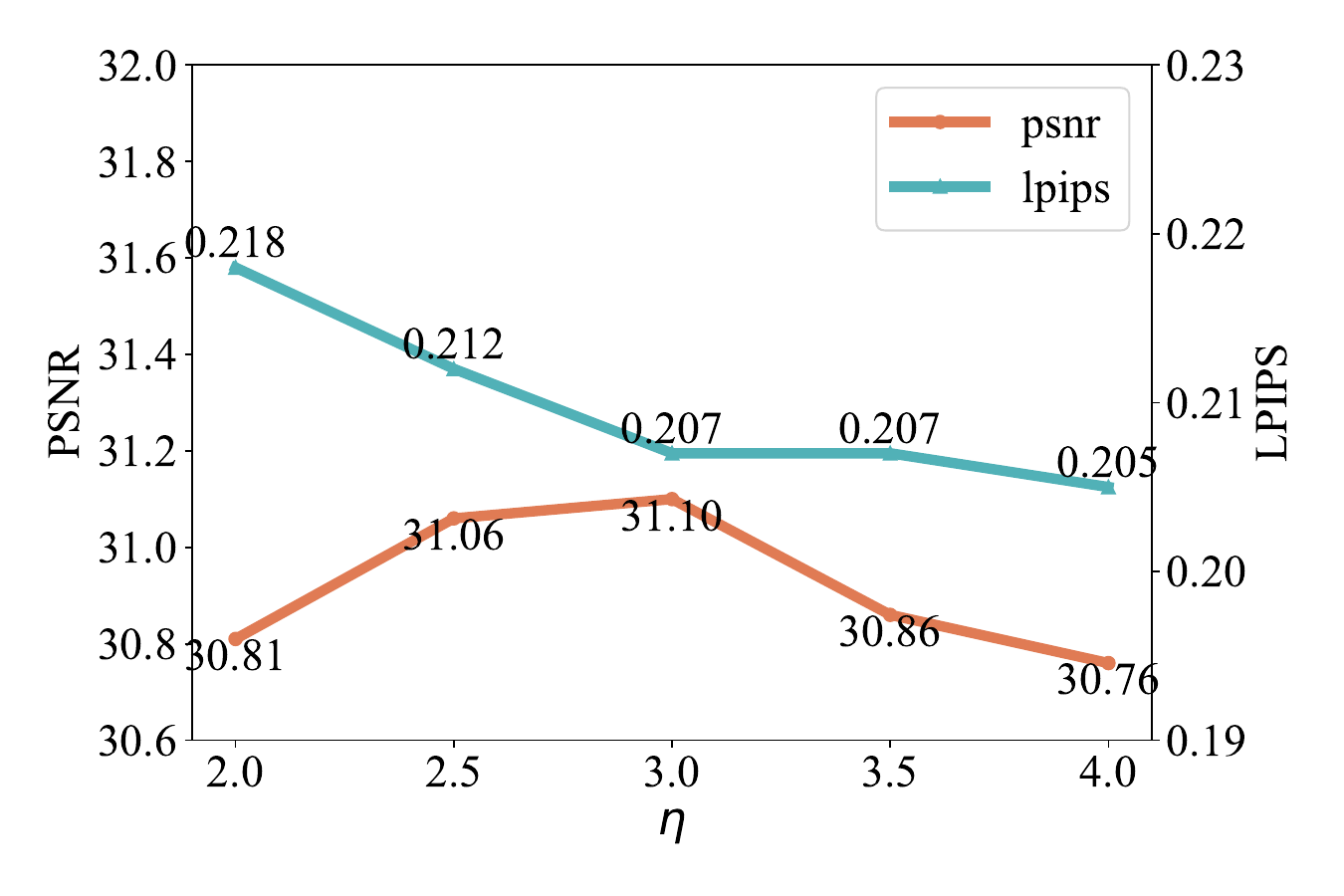}
		\end{minipage}\vspace{-0.5cm}
        \caption{The robustness analysis results for Inpainting (Box) on the FFHQ 256×256-1k and the CelebA-HQ 256×256-1k validation sets with different parameters. Columns 1-3 are the plots of PSNR (blue) and LPIPS (orange) values versus the changes of parameters $q_1$, $q_2$, and $\eta$, with the other two fixed.}
		\label{fig:Box}
	\end{center}
\end{figure}

\begin{figure}[H]
	\begin{center}
			\begin{minipage}[b]{0.32\linewidth}
				\includegraphics[width=1.05\textwidth]{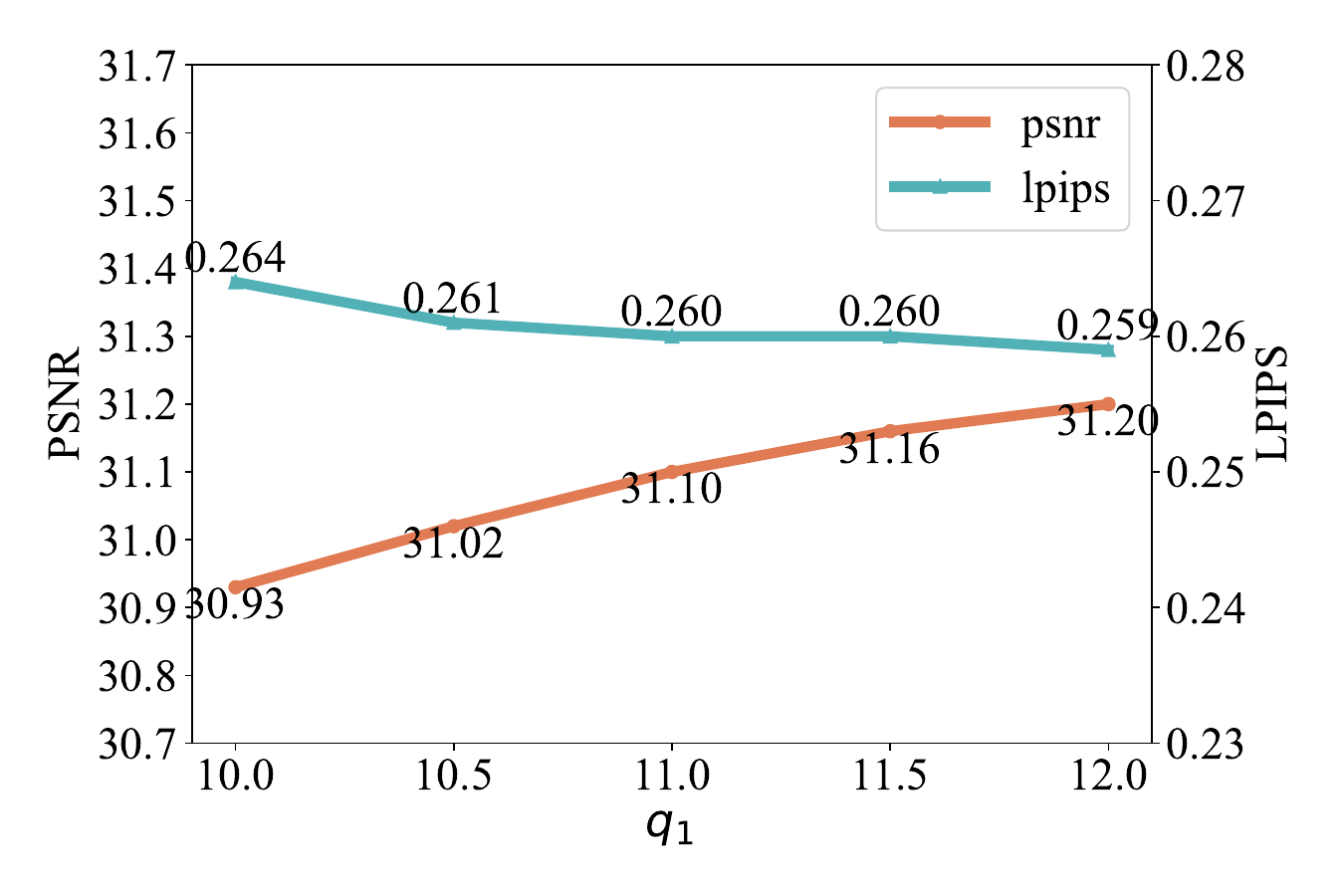}
				\includegraphics[width=1.05\textwidth]{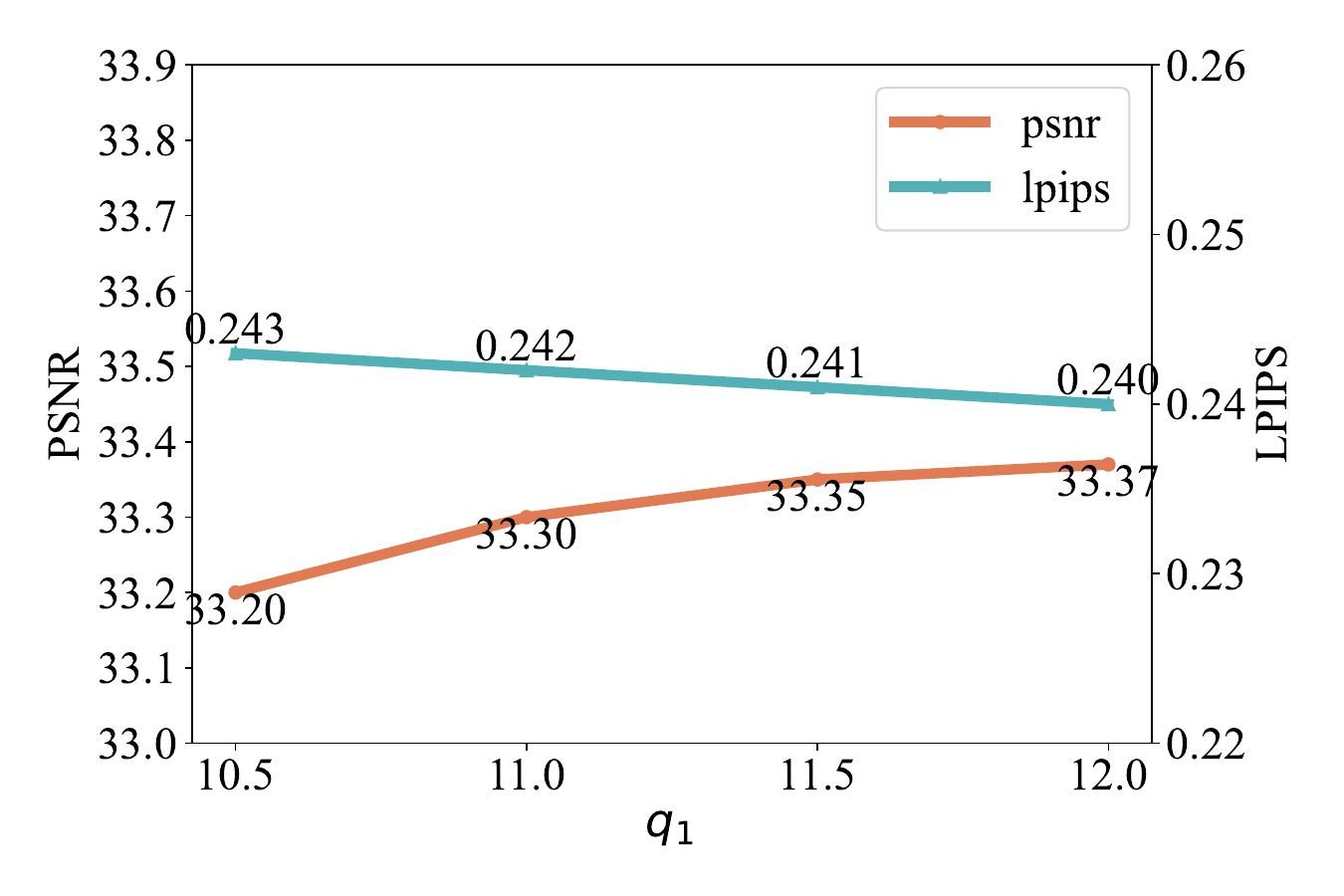}
		\end{minipage}
			\begin{minipage}[b]{0.32\linewidth}
				\includegraphics[width=1.05\textwidth]{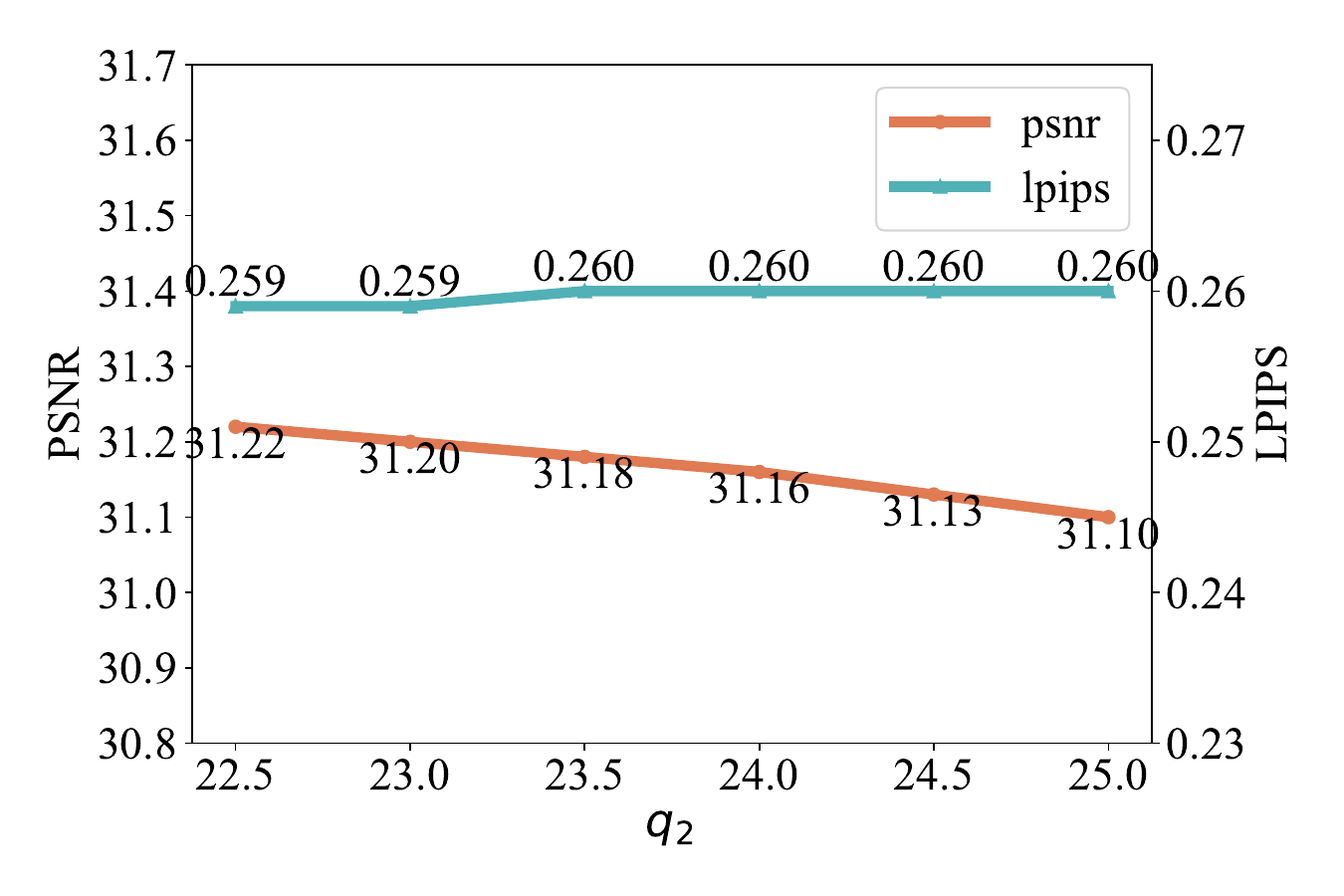}
				\includegraphics[width=1.05\textwidth]{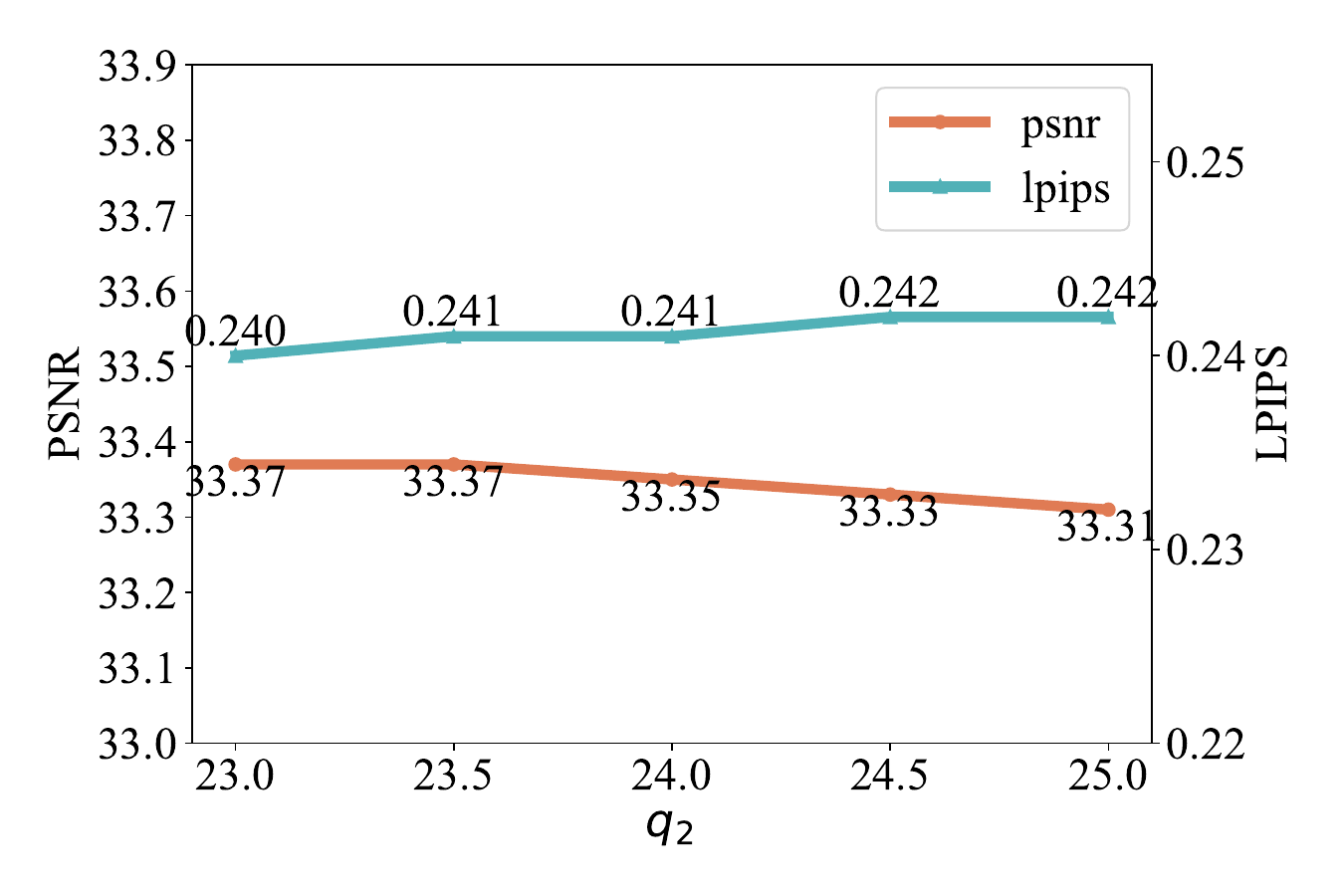}
		\end{minipage}
			\begin{minipage}[b]{0.32\linewidth}
				\includegraphics[width=1.05\textwidth]{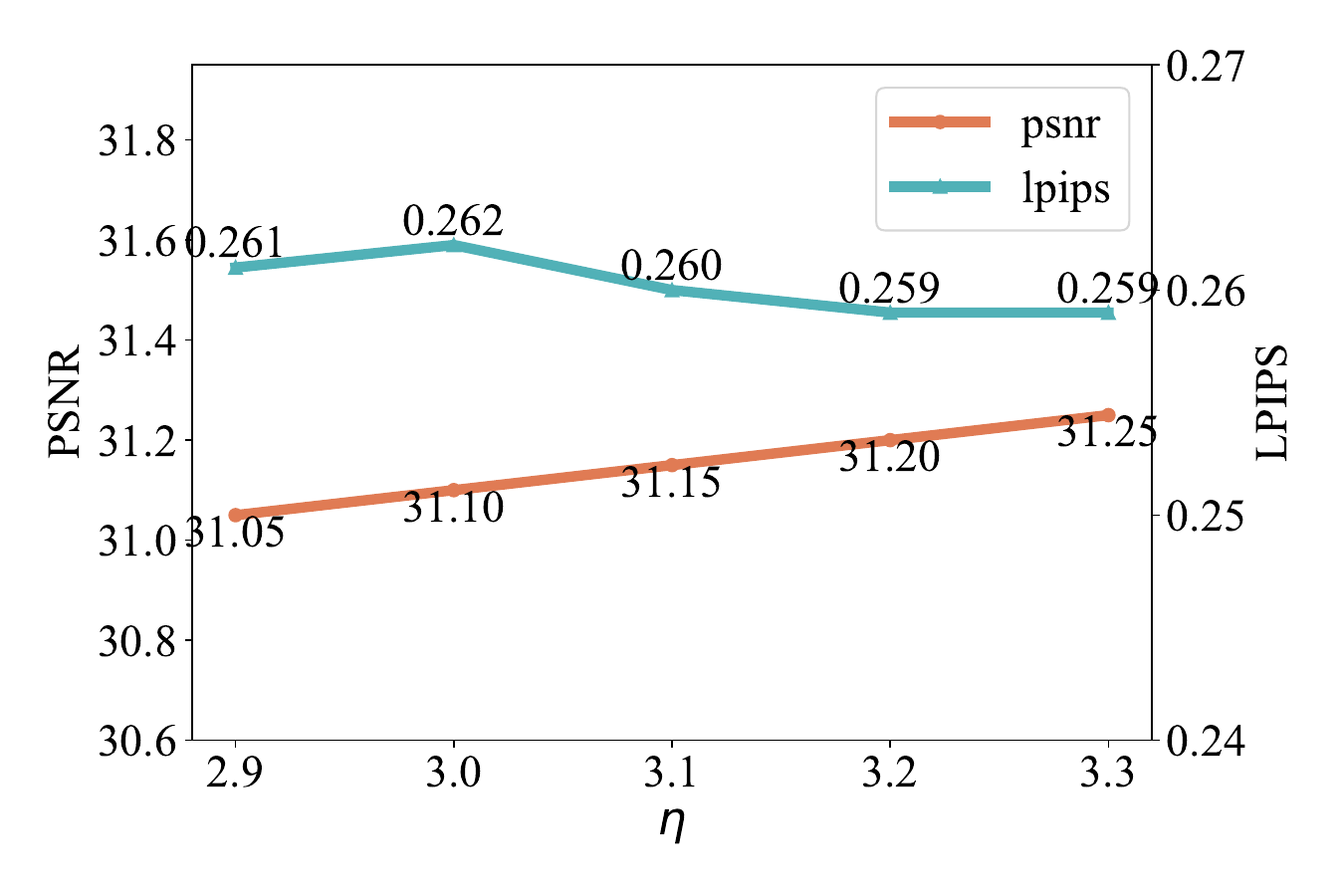}
				\includegraphics[width=1.05\textwidth]{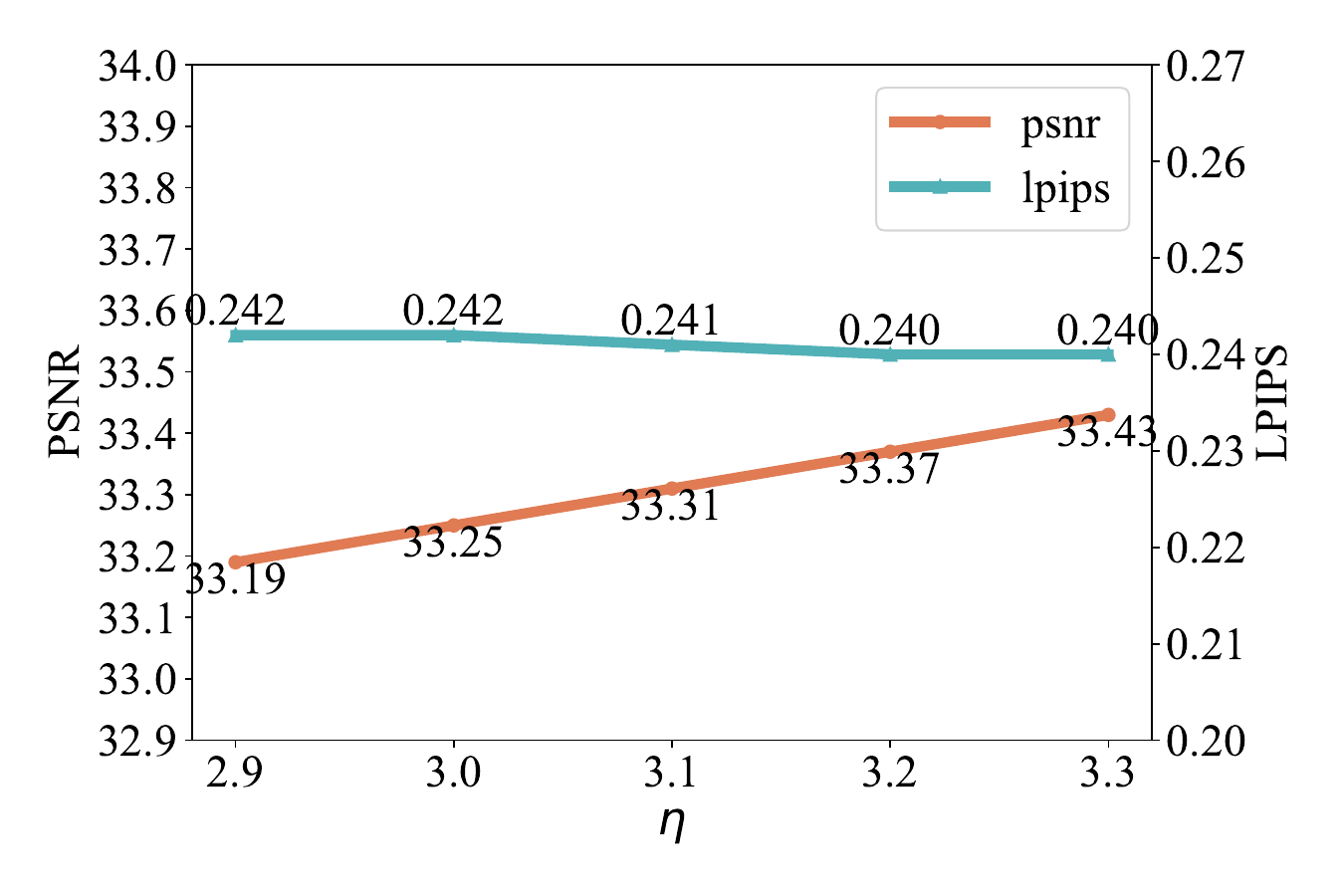}
		\end{minipage}\vspace{-5mm}
        \caption{The robustness analysis results for Inpainting (Lolcat) on the FFHQ 256×256-1k and the CelebA-HQ 256×256-1k validation sets with different parameters. Columns 1-3 are the plots of PSNR (blue) and LPIPS (orange) values versus the changes of parameters $q_1$, $q_2$, and $\eta$, with the other two fixed.}		\label{fig:Lolcat}
	\end{center}
\end{figure}

\begin{figure}[H]
	\begin{center}
			\begin{minipage}[b]{0.32\linewidth}
				\includegraphics[width=1.05\textwidth]{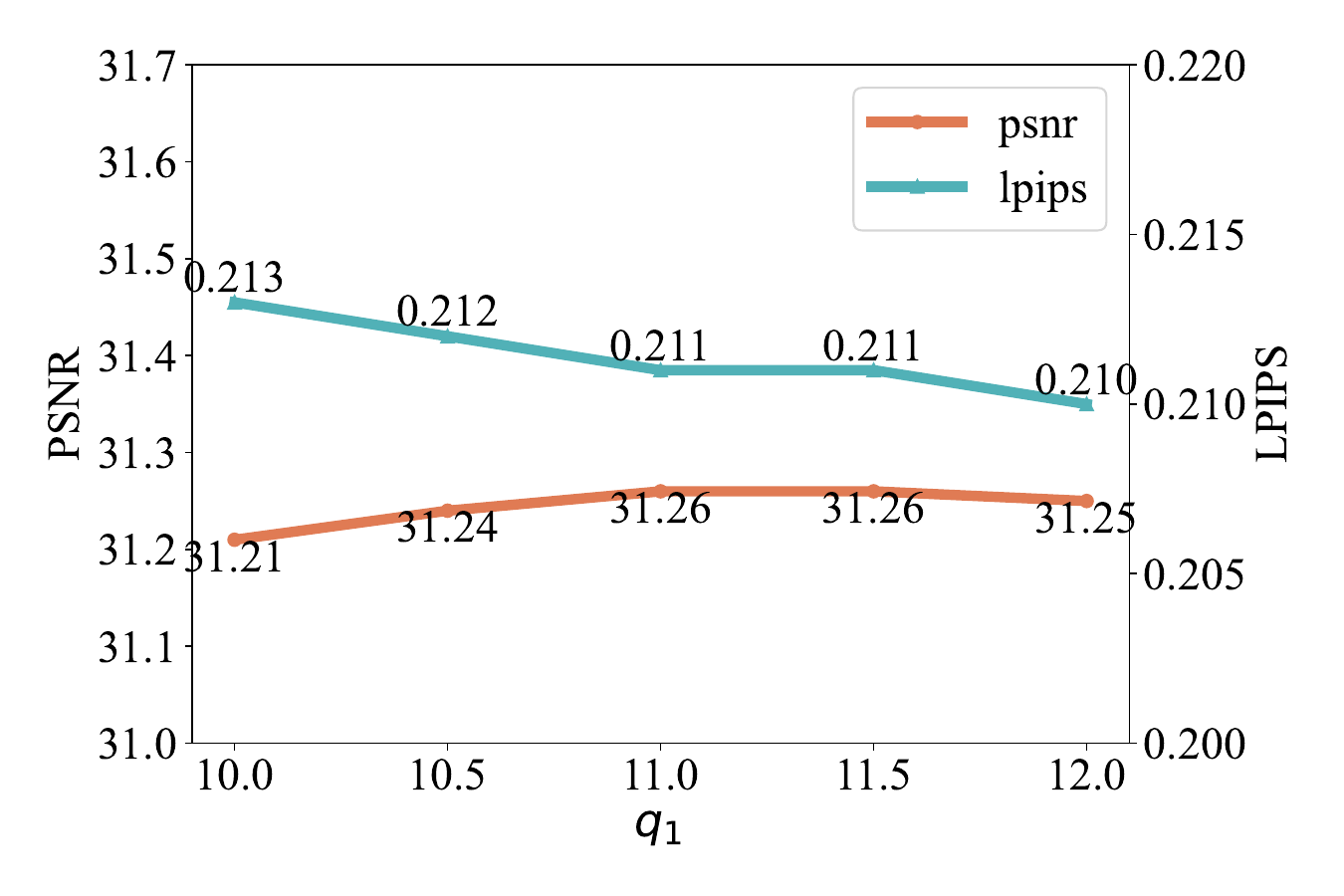}\vspace{4pt}
				\includegraphics[width=1.05\textwidth]{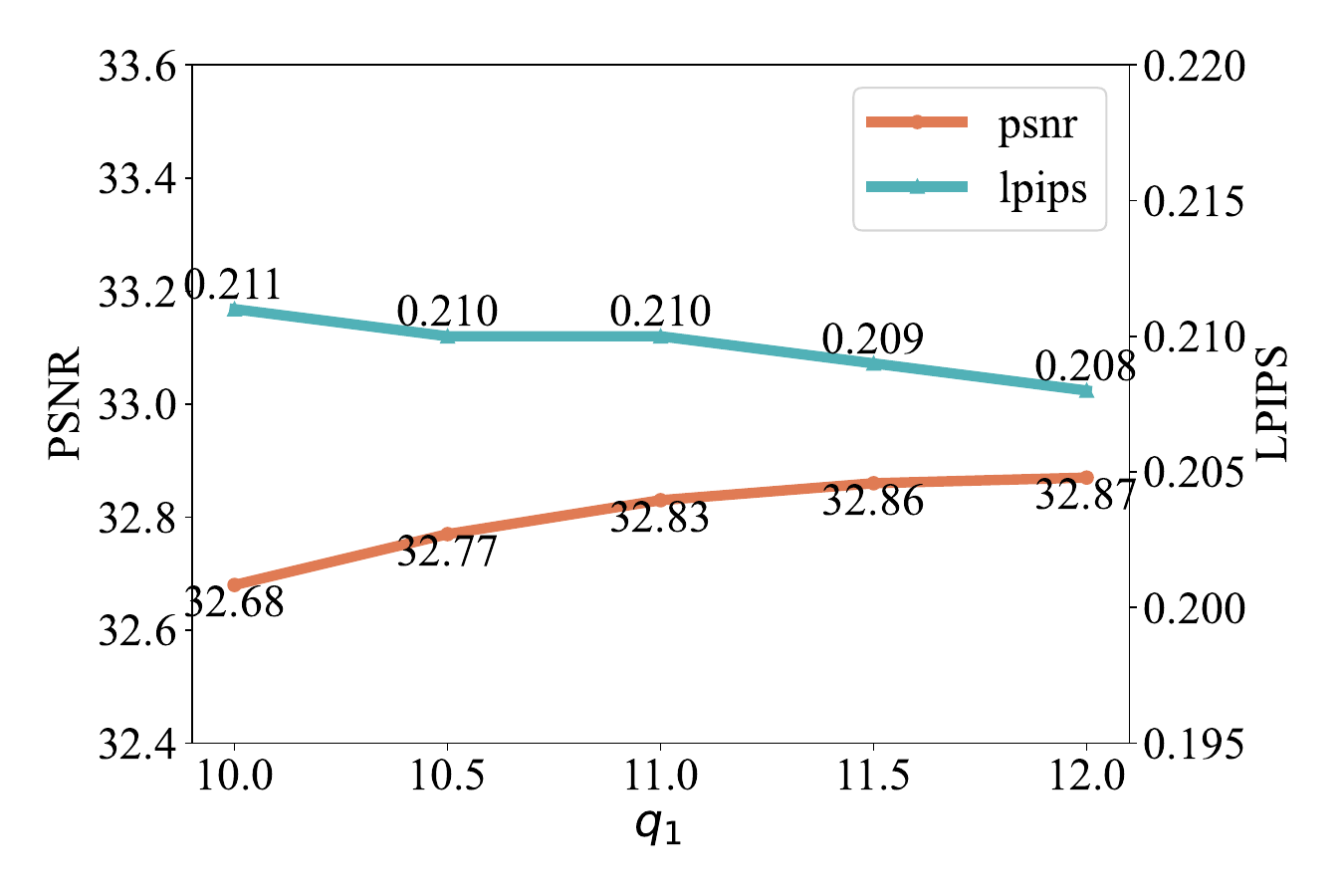}
		\end{minipage}
			\begin{minipage}[b]{0.32\linewidth}
				\includegraphics[width=1.05\textwidth]{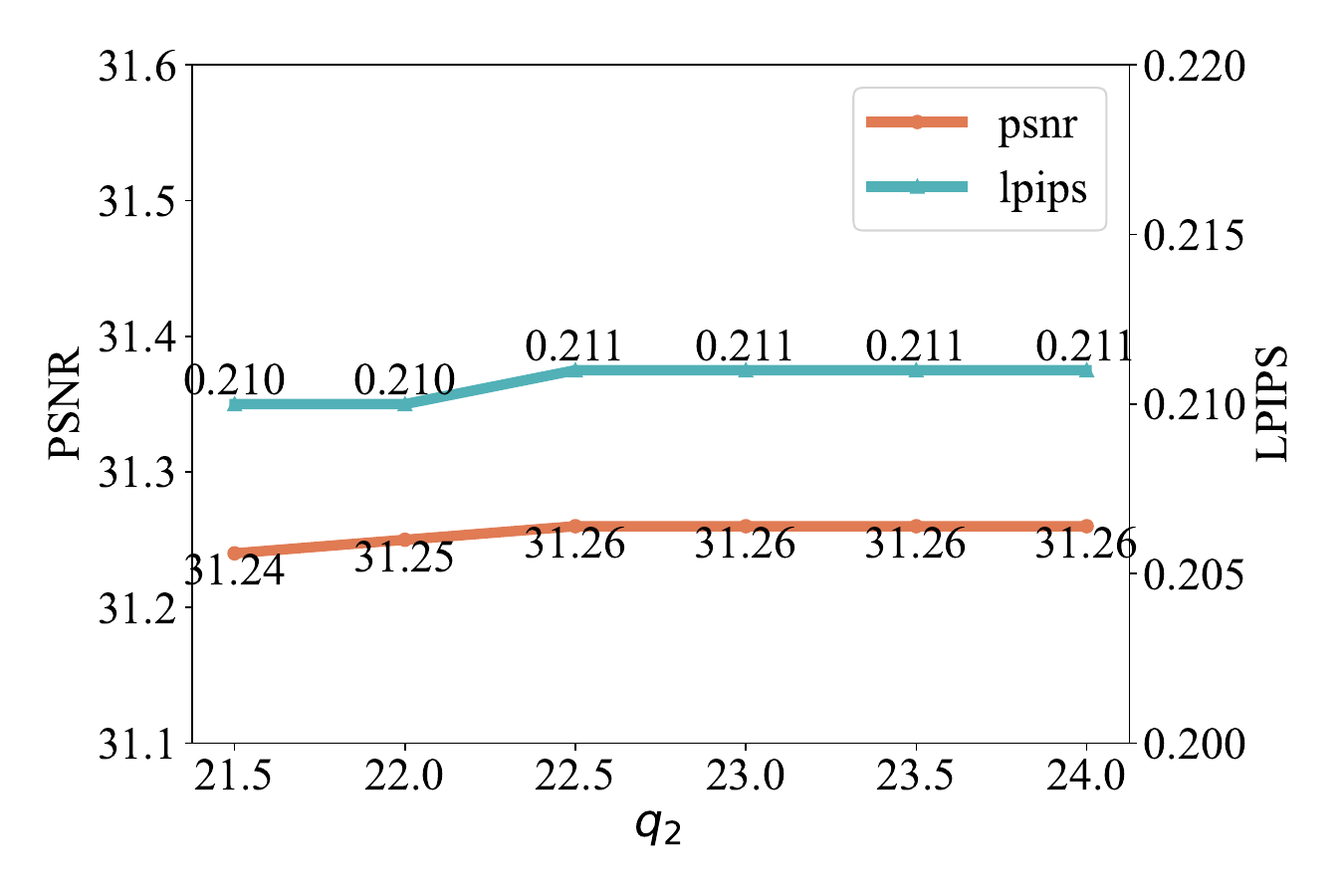}\vspace{4pt}
				\includegraphics[width=1.05\textwidth]{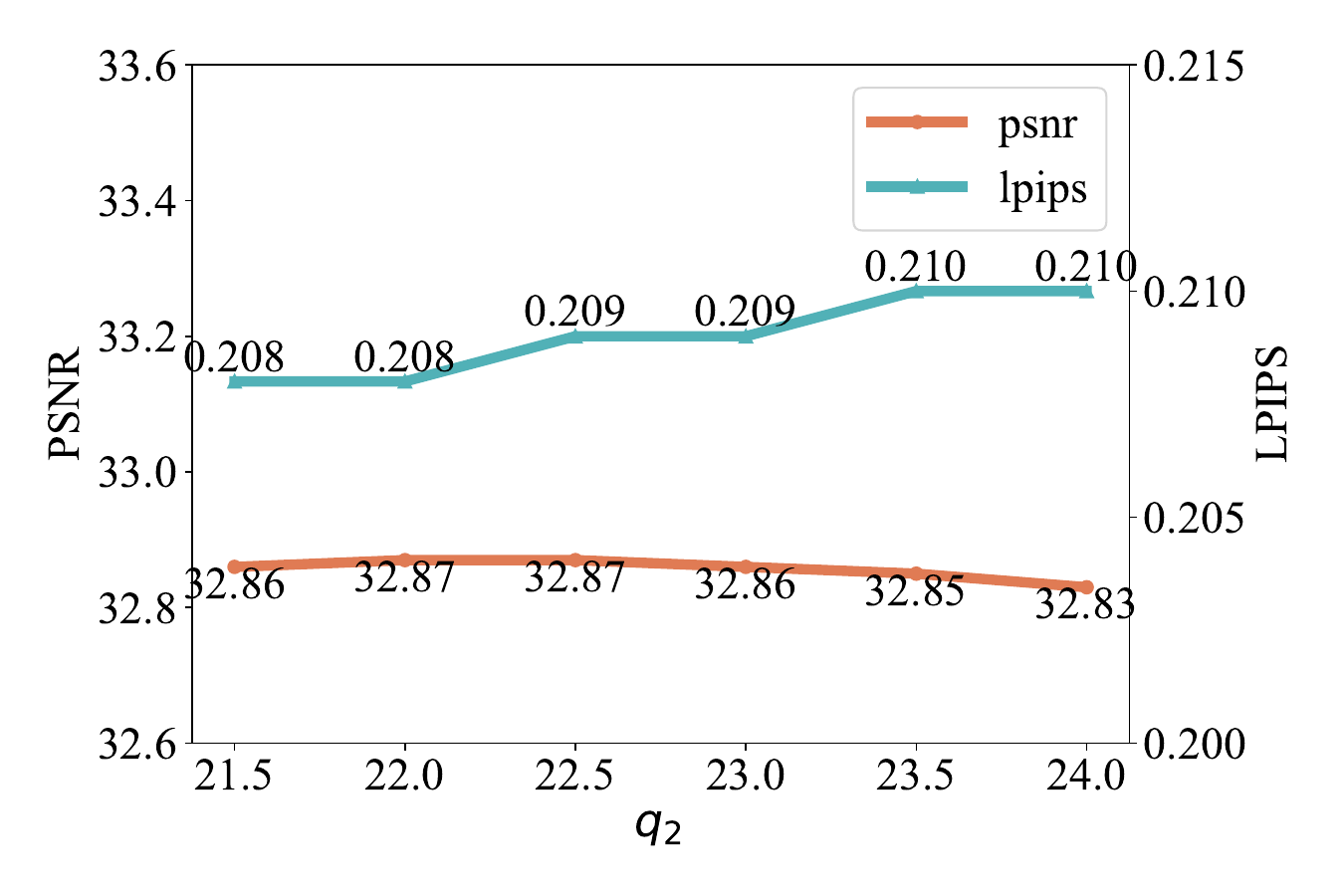}
		\end{minipage}
			\begin{minipage}[b]{0.32\linewidth}
				\includegraphics[width=1.05\textwidth]{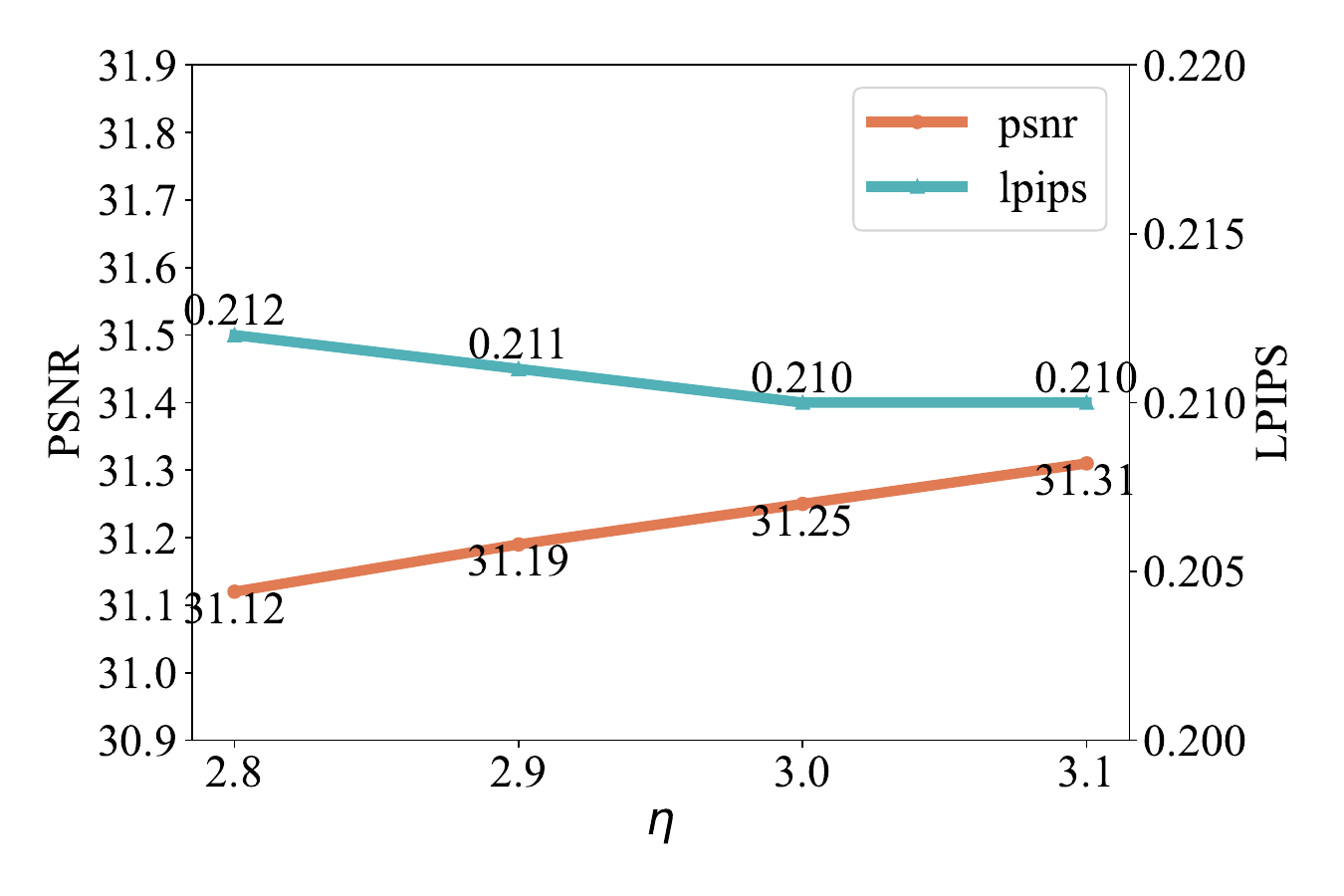}\vspace{4pt}
				\includegraphics[width=1.05\textwidth]{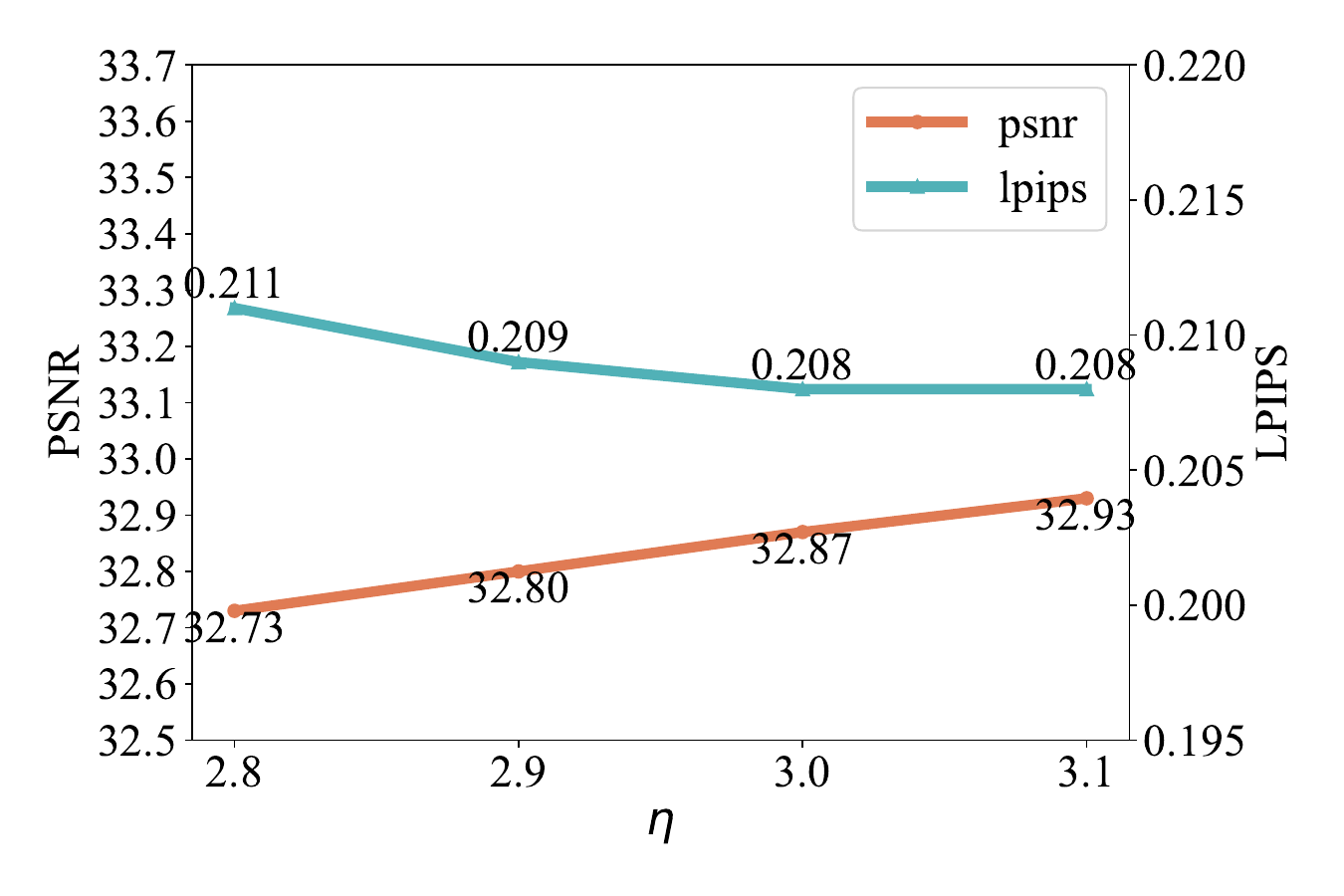}
		\end{minipage}\vspace{-3mm}
        \caption{The robustness analysis results for Inpainting (Lorem) on the FFHQ 256×256-1k and the CelebA-HQ 256×256-1k validation sets with different parameters.  Columns 1-3 are the plots of PSNR (blue) and LPIPS (orange) values versus the changes of parameters $q_1$, $q_2$, and $\eta$, with the other two fixed.}
		\label{fig:Lorem}
	\end{center}
\end{figure}

In this section, we focus on the robust analysis on the tasks of denoising, inpainting, whose results are illustrated in Figures \ref{fig:Denoise1}, \ref{fig:Box}, \ref{fig:Lolcat}, \ref{fig:Lorem}. 

\end{document}